%% file: mind_the_gaps.tex
\documentclass[fleqn,usenatbib]{mnras}

\usepackage{newtxtext,newtxmath}

\usepackage[T1]{fontenc}

\DeclareRobustCommand{\VAN}[3]{#2}
\let\VANthebibliography\thebibliography
\def\thebibliography{\DeclareRobustCommand{\VAN}[3]{##3}\VANthebibliography}

\usepackage{graphicx}	
\usepackage{amsmath}	
\usepackage{enumitem} 
\usepackage{multirow}
\usepackage[symbol]{footmisc}

\newcommand{\omegabend}{$\omega_\mathrm{bend}$}
\newcommand{\matern}{Mat\'ern-3/2}
\newcommand{\xmm}{\textit{XMM--Newton}}
\newcommand{\nicer}{\textit{NICER}}
\newcommand{\swift}{\textit{Swift}}

\newcommand{\granulation}{SHO$_\mathrm{Q=1/\sqrt{2}}$}

\newcommand{\celerite}{\texttt{celerite}}

\newcommand{\tlrt}{$T_\mathrm{LRT}$}
\newcommand{\rms}{$F_\mathrm{var}$}
\newcommand{\interval}{\Delta t_{ij}}

\newcommand{\nsimsfalsenegatives}{100}

\title[Mind the gaps]{Mind the gaps: improved methods for the detection of periodicities in unevenly-sampled data}

\author[A. G\'urpide et al.]{
Andr\'es G\'urpide,$^{1}$\thanks{E-mail: A.Gurpide-Lasheras@soton.ac.uk}
Matthew Middleton$^{1}$
\\
$^{1}$School of Physics \& Astronomy, University of Southampton, Southampton, Southampton SO17 1BJ, UK\\
}

\date{Accepted XXX. Received YYY; in original form ZZZ}

\pubyear{2015}

\begin{document}
\label{firstpage}
\pagerange{\pageref{firstpage}--\pageref{lastpage}}
\maketitle

\begin{abstract}
The detection of periodic signals in irregularly-sampled time series is a problem commonly encountered in astronomy. Traditional tools used for periodic searches, such as the periodogram, have poorly defined statistical properties under irregular sampling, which complicate inferring the underlying aperiodic variability used for hypothesis testing. The problem is exacerbated in the presence of stochastic variability, which can be easily mistaken by genuine periodic behaviour, particularly in the case of poorly sampled lightcurves. Here we present a method based on Gaussian Processes (GPs) modelling for period searches and characterization, specifically developed to overcome these problems. We argue that in cases of irregularly-sampled time series, GPs offer an appealing alternative to traditional periodograms, because the known distribution of the data (correlated Gaussian) allows a well-defined likelihood to be constructed. We exploit this property and draw from existing statistical methods to perform traditional likelihood ratio tests for an additional, (quasi-)periodic component, using the aperiodic variability inferred from the data as the null hypothesis. Inferring the noise from the data allows the method to be fully generalizable, with the only condition that the data can be described as a Gaussian process. We demonstrate the method by applying it to a variety of objects showing varying levels of noise and data quality. Limitations of the method are discussed and a package implementing the proposed methodology is made publicly available.


\end{abstract}

\begin{keywords}
Galaxies: active -- 
                Accretion --
                Stars: neutron -- Stars: black holes
                Methods: data analysis -- Methods: statistical
               
\end{keywords}



\section{Introduction}
The identification of periodic/quasi-periodic signals in irregularly-sampled time series is a common problem in astronomy due to the predominance of interrupted or intermittent observing. While uninterrupted, regularly-sampled observations may be achieved for periods of up to around a day at most, for timescales extending to the tens or hundreds of days this becomes impractical, particularly for faint sources which require the most sensitive of instruments. Such `long' timescales are however of great interest in the study of many phenomena such as superorbital periods in X-ray binaries (XRBs, e.g. \citealt{kotze_characterizing_2012, vasilopoulos_m51_2020}) and binary supermassive black hole (SMBH) signals (e.g. \citealt{graham_possible_2015}).

Arguably the most widely used technique to search for periodicities in time series is the periodogram, which involves calculating the modulus-squared of the discrete Fourier transform. \citet{lomb_least-squares_1976} and \citet{scargle_studies_1982} extended the periodogram to the case of irregularly-sampled time series, a technique known today as the Lomb-Scargle periodogram \citep[see][for a review of this technique]{vanderplas_understanding_2018}. Periodic signals appear as peaks or `outliers' in power, allowing the frequency of the repeating signal to be estimated. If the distribution of the powers in the absence of a signal is known, the chance probability of generating such an outlier can then be calculated, providing an estimate of the significance of the candidate period. 

In the absence of source variability other than the repeating signal, i.e. when the sole source of additional variance in the lightcurve is due to Poisson noise, the problem is somewhat straightforward. In regularly-sampled time series, the problem can be tackled analytically as the powers in the periodogram are independent and follow a $\chi^2$ distribution with 2 degrees of freedom \citep[$\chi^2_2$; e.g.][]{van_der_klis_fourier_1988}. In the case of irregularly-sampled data, the powers are no longer independent, but the problem can be tackled easily by randomizing the time series \citep{frescura_significance_2008, vanderplas_understanding_2018}.

Searching for periods is made considerably harder when systems show \textit{intrinsic} aperiodic or stochastic (i.e. non deterministic) variability, as is universally observed in both accreting systems (\citealt{vaughan_characterizing_2003}) and stars \citep{bowman_photometric_2022}. These types of source have steep power spectral densities (PSDs), commonly referred to as `red noise'. Failing to account for this background noise tends to overestimate the significance of peaks in the periodogram \citep{vaughan_simple_2005}. For this reason, sources showing stochastic variability are more prone to misidentified periods, exacerbated in the case of uneven sampling \citep[e.g.][]{vaughan_false_2016}. While this problem was tackled by \citet{israel_new_1996} and \citet{vaughan_simple_2005, vaughan_bayesian_2010} in the case of regularly-sampled time series, there is as yet no standard procedure for the case of irregularly-sampled time series. 

In this paper, we present a recipe aimed at detecting periodicities in irregularly-sampled time series, with particular focus on cases where the systems under study show additional aperiodic variability \citep[as it is the case in e.g. AGN;][]{gonzalez-martin_x-ray_2012}, although the method is completely generalizable. We argue that in such cases, Gaussian Process (GP) modelling offers a clear advantage over traditional (Lomb-Scargle) periodograms, because the likelihood of the data is known. This allows us to constrain the underlying, aperiodic variability using GP modelling, and use well-established statistical techniques to determine the candidate period significance and a goodness-of-fit.

This paper is structured as follows: in Section~\ref{sec:method} we review the case of period-detection in regularly-sampled time series and describe our proposed methodology for the irregularly-sampled case. In Section~\ref{sec:application} we demonstrate the method, applying it to both simulated and real data. Finally in Section~\ref{sec:discussion} we discuss the advantages of the methodology over more traditional Fourier-based techniques and outline certain limitations and caveats of the proposed methodology.

\section{Searching for a period}\label{sec:method}
\subsection{Regularly-sampled time series}

The standard methodology to test for the presence of a narrow peak associated with a periodic/quasi-periodic signal
in a periodogram generally involves estimating the broadband noise (or continuum) and using that estimate -- and its uncertainties -- as the null hypothesis \citep[e.g.][see also \citealt{gierlinski_periodicity_2008, pasham_loud_2019, ashton_searching_2021} for an example of the application of such a methodology]{israel_new_1996, vaughan_simple_2005, vaughan_bayesian_2010}. In the case of regularly-sampled time series, the powers in the periodogram can be considered independent and their distribution is well known \citep[scattered as a $\chi^2_2$ around the underlying PSD; e.g.][]{van_der_klis_fourier_1988}. Knowing the distribution of powers allows the construction of a well defined likelihood function \citep[e.g.][]{stella_continuum_1994} \citep[commonly known as the Whittle likehood; see][and references therein]{vaughan_bayesian_2010}. This allows forward-fitting of the periodogram\footnote{Notably if aliasing and red noise leakage effects are negligible.} and models to be rejected based on the data alone, as it is customarily done when fitting using $\chi^2$ statistics. 

Knowledge of the likelihood not only allows models to be rejected based on the data alone, but also to test for the presence of additional components (e.g. quasi-period oscillations; QPOs) by performing a likelihood ratio test (LRT) \citep{protassov_statistics_2002}. In particular, \citet{vaughan_bayesian_2010} proposed to follow \citet{protassov_statistics_2002} and perform a LRT using the Whittle likelihood function from fits to the periodogram:
\begin{equation}
    T_\mathrm{LRT} = -2 \ln \frac{\mathit{L}_\mathrm{0}}{\mathit{L}_{1}}
\end{equation}
\noindent Here $\mathit{L}_{0}$ and $\mathit{L}_{1}$ are the maximum of the likelihood functions for the null hypothesis and the alternative model respectively. Subsequently, one would simulate periodograms drawn from the null hypothesis \citep[and its uncertainties, see e.g.][]{ashton_searching_2021} and derive the same quantity for each of the simulated periodograms. As stated earlier, as the distribution of powers is known in the case of evenly sampled data (scattered as $\chi^2_2$ around the PSD) one can avoid the additional step of simulating lightcurves (so long as aliasing and red noise leakage effects are not important). Finally, a comparison of the {\it observed} \tlrt\ against the reference distribution derived from the simulated datasets allows the probability of rejecting the null-hypothesis model to be assessed (through the derived $p$-value), thereby providing an estimate of the significance of the putative signal. Note that, because the null hypothesis is derived from the data, the method makes no assumptions about the underlying noise, and is completely generalizable. We seek to replicate this process in the case of unevenly-sampled data. 

\subsection{Irregularly-sampled time series}\label{sub:irregular_sampling}
In the case of irregular sampling, there is no straightforward way to model the broadband noise as for the regularly-sampled case. If we were to replicate the standard procedure outlined above using the Lomb-Scargle periodogram, we would encounter a variety of problems. First, the powers in the Lomb-Scargle periodogram are known to {\it not} be statistically independent \citep[e.g.][]{lomb_least-squares_1976} and their distribution is therefore unknown and dependent on the underlying (also unknown) PSD. Secondly, the irregular sampling implies that there is no well-defined set of frequencies over which to evaluate the periodogram (e.g. \citealt{frescura_significance_2008}). Finally, as the Nyquist frequency is ill-defined or nonexistent \citep{vanderplas_understanding_2018}, aliasing effects are exacerbated. The combination of these problems typically precludes forward-fitting of the (Lomb-Scargle) periodogram or at the very least, forward-fitting will lead to biased estimates. 

\begin{figure}
    \centering
    \includegraphics[width=0.48\textwidth]{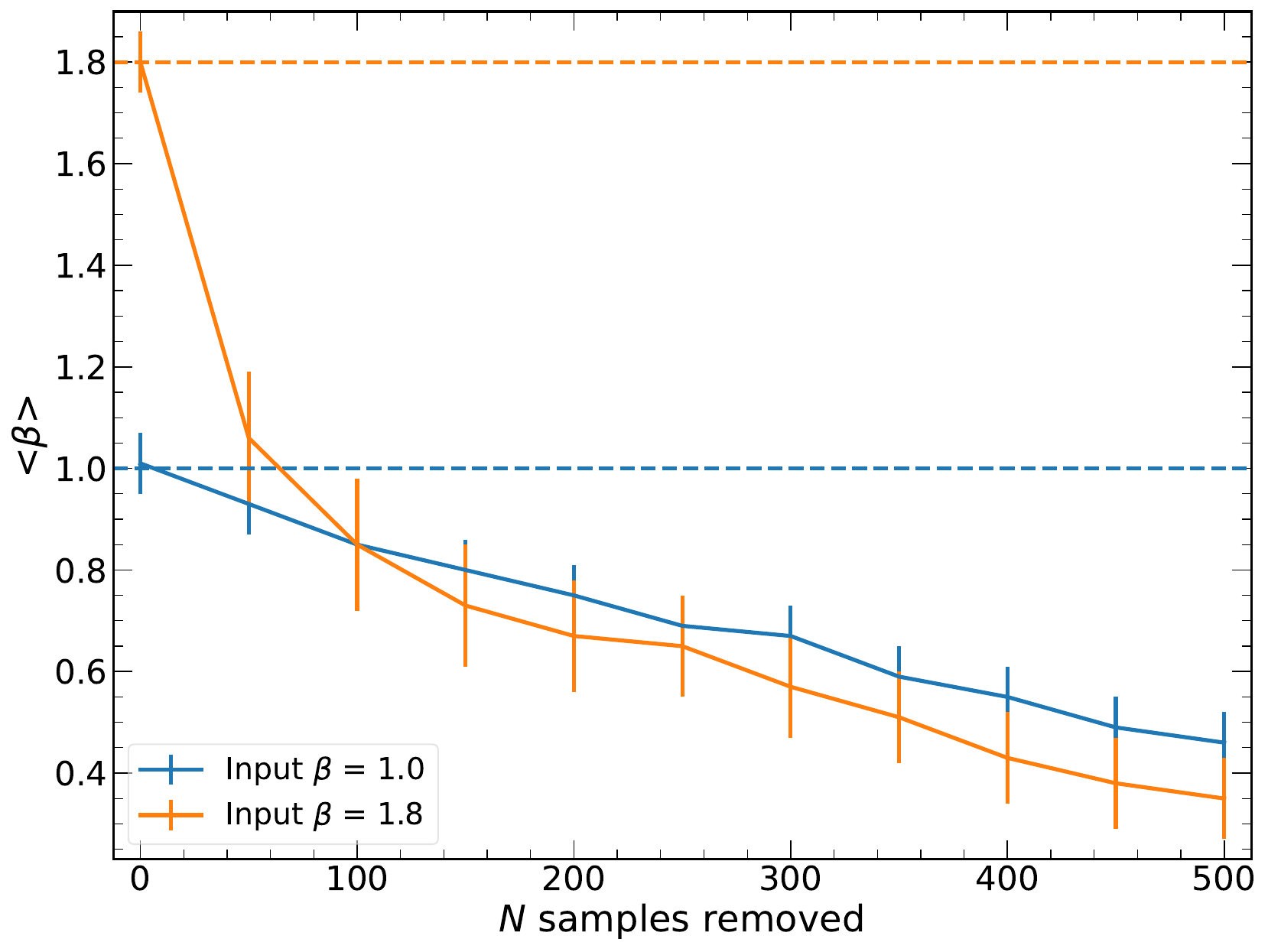}
    \caption{Mean best-fit $\beta$ for an ensemble of 1,000 Lomb-Scargle periodograms of lightcurves generated with a PSD following a powerlaw $S(f) \sim f^{-\beta}$ with $\beta = 1$ (blue solid line) and $1.8$ (orange solid line). The periodograms were fitted with a linear function in log-log space (i.e. assuming the powers follow a $\chi^2_2$ as for the regularly-sampled case) as we progressively removed datapoints. The best-fit $\beta$ quickly deviates from the input $\beta$ value (dashed horizontal lines), showing how the Lomb-Scargle periodogram becomes a biased estimator as the sampling regularity decreases.}
    \label{fig:lombscargle_bias}
\end{figure}

To illustrate this, we simulated 1,000 lightcurves using the method proposed by \cite{timmer_generating_1995}, initially with $N = 1,000$ evenly sampled datapoints
using an input PSD where the power ($S(f)$) follows a powerlaw $S(f) \propto f^{-\beta}$ with $\beta = 1$, and $1.8$ respectively. The lightcurves were initially simulated to be ten times longer to introduce red-noise leakage effects and then truncated into the aforementioned length. We then randomly removed 50 datapoints from each lightcurve and computed Lomb-Scargle periodograms from the resulting lightcurves. We then fit the periodograms in log space with a linear function \citep[i.e. assuming the powers follow a $\chi^2_2$;][]{vaughan_simple_2005} and retrieve the best-fit slope ($\beta$) in each case. We then progressively removed a further 50 datapoints, until 500 datapoints had been removed (but always keeping the last and the first datapoint to maintain the same lightcurve length), recording the mean best-fit $\beta$ for the ensemble of the 1,000 lightcurves. The mean best-fitting $\beta$ as a function of number of datapoints removed is shown in Figure~\ref{fig:lombscargle_bias} for both $\beta$ values.

Figure 1 shows that, for $\beta = 1$, when $\gtrsim$200 datapoints have been removed, the best-fit $\beta$ is underestimated by $\sim$20\%, illustrating the inherent issues in fitting to the Lomb-Scargle periodogram. The case of $\beta = 1.8$ shows the bias is more dramatic for steeper PSDs. The situation becomes even worse in a real-case scenario, where the frequency-grid will be unknown (here we can at least assume the frequency-grid given by the initially, evenly-sampled lightcurves) and there will be no way to know whether the fit is an adequate description of the data. Note also that the biases will affect any parameter fitted, including the normalization of the powerlaw, which we have not shown here.  


One way to circumvent the above problems is to rely on Monte Carlo simulations of lightcurves, as pioneered by \citet{done_x-ray_1992} and later refined by \citet{uttley_measuring_2002}. This approach attempts to find the power spectral model that, when convolved with the observing window, best matches the (Lomb-Scargle) periodogram of the real data, so that all distorting effects are taken into account. However such methods still run into some problems, particularly when dealing with irregularly-sampled time series. Once again, the unknown distribution of the powers and their lack of independence implies that the choice of the fit-statistic will not be straightforward. One could aim to rebin the periodogram, hoping that enough samples will converge to Gaussianity and independence, but again the lack of a well-defined frequency-grid complicates this approach \citep[whilst the number of averages needed to reach Gaussianity is unclear and depends on the underlying, unknown PSD;][]{ingram_physical_2011}. In addition, if binning cannot be avoided, any periodic component and its structure due to the observing window will be smeared out, which will affect the estimate of the continuum. For that reason, tests relying on simulated Lomb-Scargle periodogram peaks often have to excise the frequency of the candidate period in order to determine the putative underlying noise \citep[e.g.][]{pasham_lense-thirring_2024}, thereby making a priori assumption about the presence of any QPO. This is because if the feature is real and not removed, the broadband continuum estimate used for the simulations will be biased, often towards steeper indexes if the QPO is at the low-frequency end. This overestimate of the amount of aperiodic variability will therefore underestimate the significance of the periodic component.

The method of Monte Carlo simulations also becomes quickly computationally expensive as it relies heavily on Monte Carlo simulations for estimation of the best-fit parameters, and additional simulations are often needed to obtain parameter uncertainties (e.g. \citealt{mueller_parameter_2009, markowitz_x-ray_2010}). An appealing aspect of such a method however is that one can obtain the goodness-of-fit through the simulated lightcurves, using them to derive the empirical distribution of fit-statistic from which the goodness-of-fit \citep[or `rejection confidence' as per ][]{uttley_measuring_2002} can be derived. We note that, as pointed out by \citet{mueller_parameter_2009}, one should re-fit the simulated lightcurves in the same manner as for the observed dataset in order to obtain the empirical distribution of the fit-statistic, which would again dramatically increase the computational time. We return to this point in Section~\ref{sub:goodness}.

An alternative approach to the above is to use time-domain fitting methods such as GP, where the irregular sampling and measurement (heteroskedastic) errors are fully accounted for, and are less susceptible to the distorting effects inherent in a Fourier-domain approach \citep[e.g.][]{kelly_stochastic_2011}. 
The covariance functions or `kernels', when the data is stationary (when they depend only on the $\interval = |t_i - t_j|$ interval between any two datapoints), describe the autocorrelation function, which can be Fourier-transformed to obtain the PSD \citep{rasmussen_gaussian_2006}. Therefore GP modelling offers an equally-flexible but frequency-distortion-free access to the PSD, while maximising data usage by making full account of the measurement uncertainties and avoiding binning. Moreover, complications arising from the unknown distribution of powers in the case of the Lomb-Scargle periodogram are avoided.

Beyond the computational demand, which generally scales as $N^3$ (although \citealt{foreman-mackey_fast_2017} showed that the computational time can be reduced to scale as $\sim J^2N$ -- where $J$ is the number of model components -- for a restricted set of kernels), a more general drawback of GP modelling compared to traditional methods is that there is no measure of goodness-of-fit. As a result, models cannot be rejected solely on the basis of the data, and only model comparison (e.g. using an information criterion) is possible. In addition, although QPO-searches have been performed using GP \citep[e.g.][]{hubner_searching_2022}, establishing the significance of such signals remains challenging. In particular, it is important to quantify the chance probability of generating a fit-improvement (or any other metric such as the Bayes factor) when including a QPO/periodic component (hereafter we will refer to this simply as the `signal')
given the specific sampling, priors, fitting technique, and other factors involved in the analysis. This is particularly important if such methods are to be extended to include non-stationary kernels, where the time window becomes a parameter of the model \citep{hubner_searching_2022}. In such cases, one needs to account for the additional sets of free-trials or model-flexibility introduced in allowing signals to be transient.

\subsection{The method}\label{sub:testing_qpo}

Our procedure can be considered equivalent to the LRT approach proposed by \citet{vaughan_bayesian_2010}, but adapted to deal with irregularly sampled data. First, to circumvent the issues related to use of the Lomb-Scargle periodogram, we obtain the likelihood directly from the GP modelling in the time domain. We then make a comparison between a continuum-only model (the null hypothesis) to a more complex model that includes the signal, obtaining a fit improvement (quantified through $L_\mathrm{1} - L_\mathrm{0} = \Delta L$ or \tlrt). Next, from the posteriors of the null hypothesis modelling, we draw kernel parameter samples and then use the PSD of these kernels to generate a number of simulated lightcurves via inverse-Fourier transform (the full methodology employed to simulate the lightcurves is described in Appendix~\ref{sec:lightcurve_simulations}). We finally perform the same GP modelling on the synthetic lightcurves to derive the reference distribution for the LRT. While throughout this work we employ uninformative (uniform) priors, Bayesian priors could easily be incorporated, provided the same priors are also used when fitting the simulated datasets.

Intuitively this method can be understood as follows: if the improvement in fit-statistic provided by the added model component (the putative signal) is due to random noise fluctuations in the original data-set (i.e. the signal is spurious), the fit-improvement obtained in the simulated datasets (which were simulated using the model \textit{without} the additional model component) will be of the same order as that of the real dataset. If the signal is real, then the improvement in fit-statistic provided by the additional model component will be generally larger than any of the values obtained in the simulated datasets. Such a procedure not only inherently accounts for the number of free trials -- as long as the parameter ranges/priors are kept the same as for the original dataset -- but also for the fact that some signals may have a more complex profile than a simple peak in a Lomb-Scargle periodogram (as is often assumed in significance testing). Another advantage of relying on fit-improvements is that it removes the need to make any prior assumptions about the presence of the QPO in the periodogram, which, as stated in Section~\ref{sub:irregular_sampling}, is often the case when relying on periodogram peaks.

\subsection{Kernel functions}
There are naturally a range of possible models one could potentially use to describe the underlying noise and the signal, and we refer to \citet{rasmussen_gaussian_2006} for some examples. In this paper we use the \texttt{celerite} kernels proposed by \cite{foreman-mackey_fast_2017} for GP modelling -- which we note bear many similarities to \texttt{CARMA} \citep{kelly_flexible_2014} -- to reduce the computational burden, but note our method is generalisable to any choice of kernels. These kernels can then be combined through additions or multiplications to achieve more complex covariance matrices. However, as shown by \citet{foreman-mackey_fast_2017}, any multiplication of \texttt{celerite} kernels can always be reformulated as an addition under a new parameter set. Therefore we only explore additions of the kernels described below.

\begin{figure}
    \centering
    \includegraphics[width=0.48\textwidth]{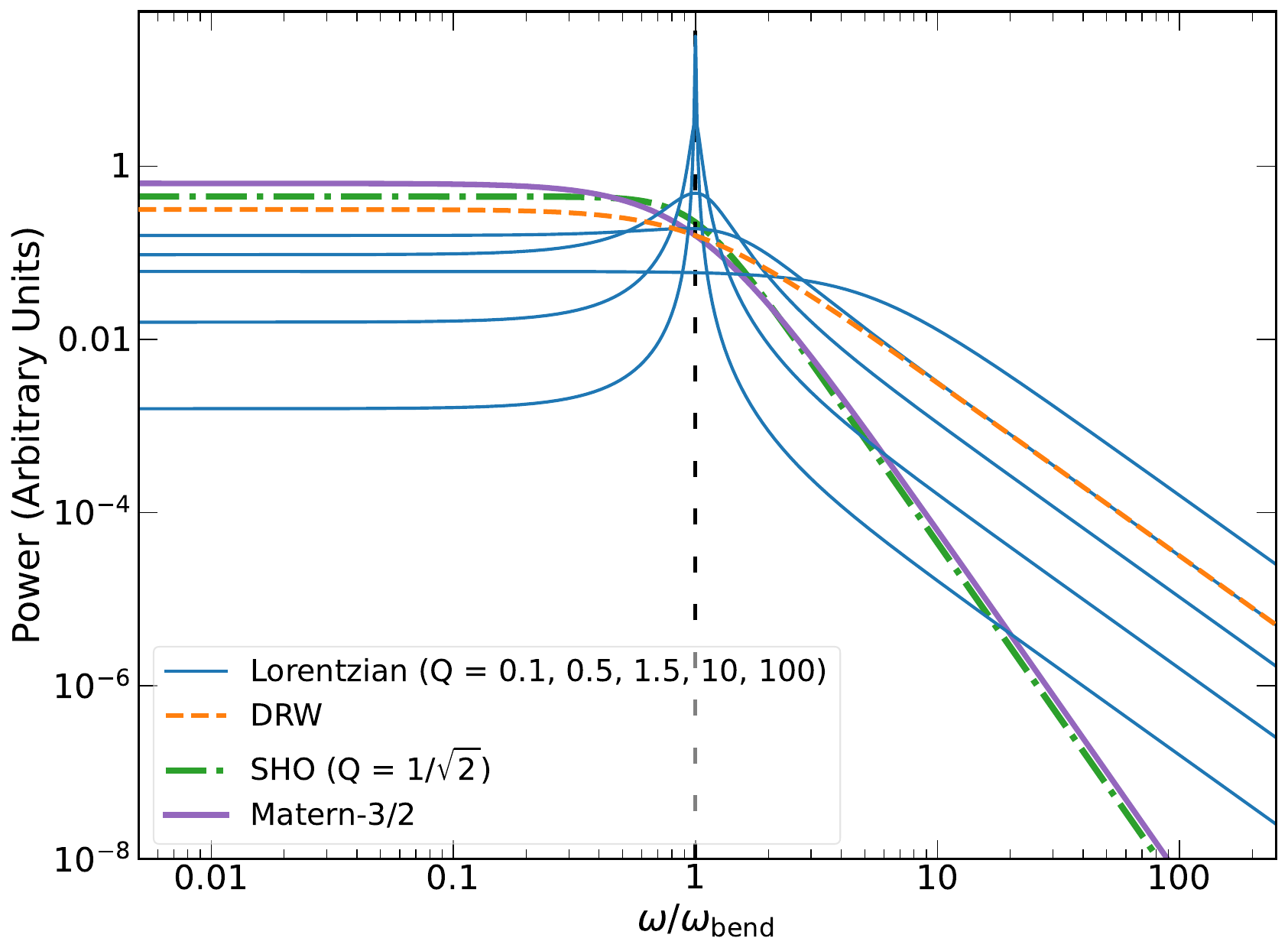}
    \caption{PSDs of the \texttt{celerite} models used in this work. All PSDs are shown with the same integrated variance. The DRW results in a bending powerlaw in Fourier space (dashed orange line), whereas the exponentially decaying sinusoid gives a Lorentzian, which is shown for different values of coherence, $Q$, in blue solid lines. For $Q \lesssim 3/2$, the Lorentzian becomes broad, mimicking a bending powerlaw \citep[see][]{belloni_unified_2002}. The PSD of the SHO for the special case of $Q = 1 / \sqrt{2}$ and $Q = 1/2$ (which yields the Mat\'ern-3/2 kernel approximation) are shown as a dashed-dotted green line and a solid purple line, respectively. The vertical dashed line indicates the central frequency of the Lorentzian ($\omega_\mathrm{0}$) and \omegabend\ for the DRW, \matern\ and \granulation\ kernels. Note that for the \matern\ $\rho = \sqrt{3}$/\omegabend.}
    \label{fig:celerite_models}
\end{figure}

The simplest choice of \texttt{celerite} kernel for modelling aperiodic variability is the Damped Random Walk (DRW), whose kernel is simply a decaying exponential:
\begin{equation}
    k(\interval) = \sigma^2 \exp(-\omega_\text{bend} \interval)
\end{equation}
\noindent the PSD of which is a bending powerlaw (Figure~\ref{fig:celerite_models}):
\begin{equation}\label{eq:bending}
     S(\omega) = \sqrt{\frac{2}{\pi}} \frac{\sigma^2} {\omega_\text{bend}} \frac{1}{1 + \left( \frac{\omega}{\omega_\text{bend}}\right)^2}
\end{equation}
\noindent with an index of --2 for $\omega >> \omega_\text{bend}$ bending smoothly to a flat ($S(\omega) \sim \omega^0$) powerlaw around \omegabend. In the above, $\sigma^2$ is the variance of the process.  

A further possible kernel (as proposed by \citealt{foreman-mackey_fast_2017}) is the stochastically-driven damped harmonic oscillator (SHO), which can model both aperiodic and periodic variability. For the full details of this kernel we refer the reader to \citet{foreman-mackey_fast_2017}; in this work we consider two special cases of this kernel used to model aperiodic noise. The first one is commonly used to model (aperiodic) granular noise in stars:
\begin{equation}
      k(\interval) = S_\text{N} \omega_\text{bend} e^{-\frac{1}{\sqrt{2}} \omega_\text{bend} \interval } \cos \left(\frac{\omega_\text{bend} \interval}{\sqrt{2}} - \frac{\pi}{4} \right)
\end{equation}
with a PSD of the form:
\begin{equation}
    S(\omega) = \sqrt{\frac{2}{\pi}} \frac{S_\mathrm{N}}{\left(1 + \omega/\omega_\text{bend} \right)^4}
\end{equation}


\noindent here $S_\mathrm{N}$ scales the variance of the noise process ($\sigma^2 = \frac{1}{\sqrt{2}}S_\mathrm{N} \omega_\mathrm{bend}$). The PSD of this kernel is similar to the DRW but here the powerlaw has a stepper index of --4 for $\omega >> \omega_\text{bend}$ (Figure~\ref{fig:celerite_models}). Hereafter we refer to this model as \granulation\ as this kernel is obtained for the special case of $Q = 1 /\sqrt{2}$ within the more general SHO \citep[for more details we refer the reader to][]{foreman-mackey_fast_2017} .


The second special case of the SHO we consider is an approximation to the Mat\'ern-3/2 kernel\footnote{In practice we have found the parameter controlling the approximation in \texttt{celerite} to have very small effect on the results and was fixed to the arbitrary small value of 10$^{-7}$. While preparing this manuscript we have learned that an exact state-representation of the Mat\'ern-3/2 has now been derived in \citet{jordan_state-space_2021}.}, which using \texttt{celerite} kernels can be approximated setting $Q = 1/2$ in the SHO:
\begin{equation}\label{eq:mattern}
    k(\interval) = \sigma^2 \left(1 + \frac{\sqrt{3}}{\rho}\interval\right) e^{-\frac{\sqrt{3}\interval}{\rho}}
\end{equation}
\noindent where $\rho$ sets the characteristic timescale in a similar fashion to the DRW. The PSD of this function is only slightly dissimilar to the \granulation\ kernel as can be seen in Figure~\ref{fig:celerite_models}. Hereafter we refer to this kernel as Mat\'ern-3/2 for simplicity.

Finally we also considered a `Jitter' or white-noise kernel to model uncorrelated aperiodic variability, parameterized only by its variance:
\begin{equation} \label{eq:jitter}
    k(\interval) = \sigma^2 \delta_{ij}
\end{equation}
\noindent where $\delta_{ij}$ is the Kronecker delta, indicating this term simply adds a diagonal term to the covariance matrix. This kernel can be interpreted in two ways. The first is that the uncertainties on the data are underestimated; in this case $\sigma^2$ provides the constant, missing contribution to the noise; the second is as an extra white noise term to capture some random variations (e.g. instrumental effects) not captured by the main model. Here we consider it as an independent model to describe cases where the data does not support the use of a different kernel (signaling that white noise as the null hypothesis might be justified).. 

For the periodic component, we have employed only a single exponentially decaying sinusoid:
\begin{equation}
\begin{split}
    k(\interval) = \sigma^2 & \exp(-b \interval) \cos (\omega_\mathrm{0} \interval) = \\ \sigma^2 & \exp\left(-\frac{\omega_\mathrm{0}}{2Q} \interval\right) \cos (\omega_\mathrm{0} \interval)
\end{split}
\end{equation}
\noindent where $b = \omega_\mathrm{0} /2Q$ following the nomenclature of \citet{foreman-mackey_fast_2017}.
The resulting PSD takes the form of a Lorentzian  (Figure~\ref{fig:celerite_models}):
\begin{equation}\label{eq:lor}
    S(\omega) = \frac{1}{\sqrt{2\pi}}\frac{\sigma^2b}{b^2 + (\omega - \omega_\mathrm{0})^2}= \sqrt{\frac{2}{\pi}}  \frac{\sigma^2 Q \omega_\mathrm{0} }{\omega_\mathrm{0}^2 + 4 Q^2 (\omega - \omega_\mathrm{0})^2}
\end{equation}
\noindent which is a phenomenological model commonly used to model QPOs in X-ray binaries \citep[e.g.][]{belloni_unified_2002, vaughan_where_2005}, and is flexible enough to capture strict periodicities (where the coherence is extremely high). The Lorentzian has three parameters: the period  of the oscillation $P = 2\pi / \omega_\mathrm{0}$, the coherence or quality factor $Q$\footnote[1]{Note that our definition is consistent with \cite{belloni_unified_2002} but differs by a factor two compared to other works \citep[e.g.][]{vaughan_where_2005}}, which sets how stable the oscillation amplitude is over time, or how peaked the Lorentzian is, and $\sigma^2$, which is again the variance of the oscillation. Note that unlike periodogram modelling which is agnostic to the underlying mechanism broadening the QPO, our periodic model here can only capture variations in amplitude.

The presence of Poisson (or Gaussian) noise can be included in the usual manner, by adding in quadrature the observation uncertainties to the covariance matrix \citep{rasmussen_gaussian_2006, foreman-mackey_fast_2017}. Here we take the mean function to simply be the mean of the lightcurve. This choice also helps to limit the number of variable parameters, but again, our method can be generalised to include any mean function. 

\subsection{Model Selection}\label{sub:model_selection}

Since we are performing a test for an additional component, our models will always be of the form \textit{underlying noise} + \textit{periodic component(s)}. Before testing for the presence of a signal, it is important to select a suitable null-hypothesis that captures the underlying, stochastic noise. Information criteria (IC), which penalise more complex models if the increase in fit statistic is not deemed `worthy' of the extra parameters, are commonly used for model selection. If priors are included, model selection can be performed using the Bayes factor. Given that we have used non-informative priors throughout this work, we perform model selection using the Akaike Information Criterion \citep[AIC;][]{parzen_information_1998}:
\begin{equation}
   \mathrm{AIC} =  2k - 2 \ln \mathit{L}_\mathrm{*}
\end{equation}
\noindent where models with higher AIC values are considered to have excessive complexity with respect to the quality of the data. Here $k$ is the number of model parameters and $L_\mathrm{*}$ for a particular model is the maximum of the likelihood function. The AIC is only correct asymptotically (i.e. for large sample sizes) but it can be corrected for finite sample sizes as shown in \citet{hurvich_regression_1989}:
\begin{equation}
    \text{AICc} = \text{AIC} + \frac{2k(k+1)}{N-k-1}
\end{equation}  

In order to find the best-fit model, we have implemented a small iterative routine in which we start by testing each of the single-kernel models (Jitter, DRW, Mat\'ern-3/2, \granulation\ and Lorentzian) on the data. From these 5 fits, we selected the one yielding the lowest AICc$_\mathrm{min}$ and those within $\Delta$AICc = 2 from AICc$_\mathrm{min}$. Subsequently, we tested each of the selected kernels in combination with any of the other five. From this second stage we again retained the lowest overall model and those within the above $\Delta$AICc, and repeated the process until adding an extra component no longer resulted in a decrease in the AICc.
 
We have compared the results of this routine with brute-forcing testing all possible model combinations and while we have found this routine yields the correct model in most instances, this was not the case in two of the objects tested here. As an example, there might be instances where a combination of a Lorentzian + DRW might be a better overall combination than a Lorentzian + Mat\'ern-3/2, even if in isolation a Mat\'ern-3/2 may be preferred over a DRW and Lorentzian. Nevertheless, we have found the routine useful in performing a preliminary triage and establishing the number of components required. Therefore after the model minimizing AICc was found, we have tested alternations keeping the same model components to refine the final selection. We leave developing a more refined search process when focusing on large-scale survey searches, where we will calibrate the method against specific datasets.

Once the model (or combination of model components) has been selected for the alternative and null hypothesis models, the posteriors derived from the null hypothesis (the noise-only model) can be used to calibrate the reference LRT distribution as proposed by \citet{protassov_statistics_2002} and derive the posterior predictive $p$-value (PPP). In doing so, we are able to map $\Delta$IC changes to $p$-values. 

Note it may not always be possible to establish a unique pair of alternative and null hypothesis models when the differences between two models are small (typically $\Delta$AICc $\lesssim$2). In such instances hypothesis testing may be repeated using the various competing models in order to assess the robustness of the results to the choice of null hypothesis. This situation is akin to the regularly-sampled case \citep[e.g.][]{alston_detection_2014} and practical examples will be discussed in Section~\ref{sub:xmm_seyfert} and Section~\ref{sub:agn_qpo}. Note that we refer here to differences in $\Delta$AICc between two alternative models, as small differences in $\Delta$AICc between null and alternative models are precisely the type of situation our method is designed to address.

\subsection{Goodness-of-fit}\label{sub:goodness}

The goodness-of-fit is one of the main statistical quantities lacking in GP modelling. As opposed to the commonly used $\chi^2$ statistic (whose value can be mapped to a $p$-value, indicating the likelihood that the data was generated by the model), the maximum of the GP likelihood $\mathit{L}_\mathrm{*}$ alone tells us nothing about whether the model is a good description of the data or whether the data can be described by a GP. Regarding the latter, there may be concern the lightcurves of accreting compact objects cannot be described using GPs, because the fluxes are observed to follow a lognormal probability density function (PDF), suggesting a multiplicative process generates the variability \citep{uttley_non-linear_2005}, which we should not be able to describe using GPs. In the Appendix (Section~\ref{sec:gp_lognormal}) we discuss this aspect, and show through simulations that, despite the lightcurves having a lognormal distribution, GPs are able to recover the underlying process generating the variability (the PSD), indicating that their applicability might be broader than originally thought.

Despite the likelihood telling us nothing about whether the model is an appropriate description of the data, there are still several diagnostics that can be employed to test whether the model describes the data (or whether the data can be described by a GP). Following \citet{kelly_flexible_2014}, we derive two diagnostics for model testing. First, we assess whether the standarized residuals follow a standard normal distribution ($\mu = 0$, $\sigma = 1)$ by performing a Kolmogorov-Smirnof (KS) test. In practice, if the data cannot be described by a GP, the standarized residuals will be narrower than a standard normal distribution. This indicates the GP is overcompensating by assuming the entire variability in the time series is just random noise. In other words, the GP will indicate excess variance with respect to the data. On the other hand, residuals broader than a standard normal distribution indicate a deficiency in the chosen model. The other quantity we derive verifies there are no remaining trends by computing the autocorrelation function (ACF) of the standarized residuals. Any deviations from white noise will indicate the GP has not captured the full variability present in the lightcurve.

As described above (Section~\ref{sub:irregular_sampling}), in cases where the underlying distribution of the fit-statistic is unknown, the reference distribution can be built empirically using numerical simulations. This general method \citep[e.g.][]{waller_monte_2003, kaastra_use_2017} involves simulating realistic datasets from the best-fit model parameters that yielded $\mathit{L}_\mathrm{*}$, applying the same fitting procedures and retrieving the reference distribution of $\mathit{L}_\text{sim}$ values to compare to $\mathit{L}_\mathrm{*}$. While a similar approach was also suggested by \cite{kelly_stochastic_2011}, who proposed to simulate lightcurves from the best-fit-derived PSD and compare their periodograms to the periodogram of the data, here we avoid the Fourier domain entirely by fitting the simulated lightcurves in time domain too.

This approach can be understood as follows: if the best-fit parameters are truly representative of the data, then the simulated datasets (lightcurves in this case) will yield values of $L$ close to $\mathit{L}_\mathrm{*}$ when fitted, and so $\mathit{L}_\mathrm{*}$ will sit roughly at the median of the $\mathit{L}_\text{sim}$ distribution. If the best-fit parameters are \textit{not} representative of the data, then the value of $\mathit{L}_\mathrm{*}$ will be an outlier in the distribution of $\mathit{L}_\text{sim}$, i.e. $\mathit{L}_\mathrm{*}$ will in general be much lower than each of the $\mathit{L}_\text{sim}$ values from the synthetic datasets; the model can then be statistically rejected (typically $p \lesssim 0.05$). If the data is over-fitted, then the value of $\mathit{L}_\mathrm{*}$ will be towards the higher end of the $\mathit{L}_\text{sim}$ distribution, implying the model has captured the data beyond the statistical noise which is injected into the simulations (the model is deemed "too good", which may also occur where the errors have been overestimated). 

Finally, we note that, as opposed to more traditional $\chi^2$-fitting, where more complex models always lead to lower $\chi^2$, in GP modelling this is not necessarily the case. As opposed to $\chi^2$, where the likelihood depends exclusively on the fit residuals, in GP modelling, the likelihood depends on the residuals \textit{and} a term depending on the kernel (or model) through the determinant of the covariance matrix. Therefore the best-fit is determined from a trade-off between the residuals and the part of the likelihood that depends on the model alone. This makes it possible for less complex models to actually have more flexibility than models involving more hyperparameters, yielding better fits even if the complexity of the model is reduced.




\subsection{Recipe}

As a summary of the above, we outline the proposed steps of our method:
\begin{itemize}
    \item Chose a periodic kernel (or set of kernels) and a set of models for the underlying noise. 
    \item Fit the models (and combinations of) to the data and rank them using one of the widely used IC (e.g. AICc, BIC).
    \item Ensure that the model with the lowest IC provides a good fit (e.g. via standarized residuals and their ACF or deriving the reference distribution for $L_\mathrm{*}$).
    \item Compare the maximum of the likelihood function $\mathit{L}_{1}$ of the best-fit signal + underlying noise model (the alternative model) to the maximum of the likelihood function $\mathit{L}_{0}$ of the model {\it without} the signal (the null-hypothesis) and retrieve the fit-improvement, quantified as $T_\mathrm{LRT}$.
    \item Use the posteriors of the null model (the stochastic noise-only model) to generate synthetic datasets.
    \item Fit the synthetic datasets with the alternative and null-hypothesis models, derive the reference distribution for the LRT, and obtain the PPP by locating $T_\mathrm{LRT}$ in the distribution.
    \item Based on the significance of the signal, decide whether the component should be added to the null hypothesis (i.e. whether the signal is present in the data).
\end{itemize}
A \texttt{python} package that implements the proposed methodology has been made available at \url{https://github.com/andresgur/mind_the_gaps} and was employed throughout this work.
\section{Application}\label{sec:application}
We initially apply the recipe above to simulated data to explore the sensitivity of our method to variations in cadence and observing baseline. We present two sets of tests to examine the robustness of our approach to false negatives (failure to detect a signal) and false positives (identification of spurious signals).


\subsection{Application to simulated data}\label{sec:proof_concept}

In order to examine the sensitivity of our method to false negatives, we start by generating lightcurves (using the method explained in Appendix~\ref{sec:lightcurve_simulations}) with a (quasi)periodic component (Lorentzian) and red noise (a DRW) to mimic the case of a QPO identified in a stochastically-varying lightcurve \citep[e.g.][]{graham_possible_2015}. 

We assumed a period of 100 days for the QPO with a coherence $Q= 200$, a bending timescale of 60 days for the DRW and that the QPO and the DRW contribute equally to a total variance of 6.7$\times10^{-4}$ (ct/s)$^2$. We assumed a mean count rate of 0.1 ct/s for the source, background contribution of 1\% and 2 ks exposure for all observations (these values were motivated by the faintest sources \swift-XRT is capable of monitoring).

\begin{figure*}
    \centering
    \includegraphics[width=0.49\textwidth]{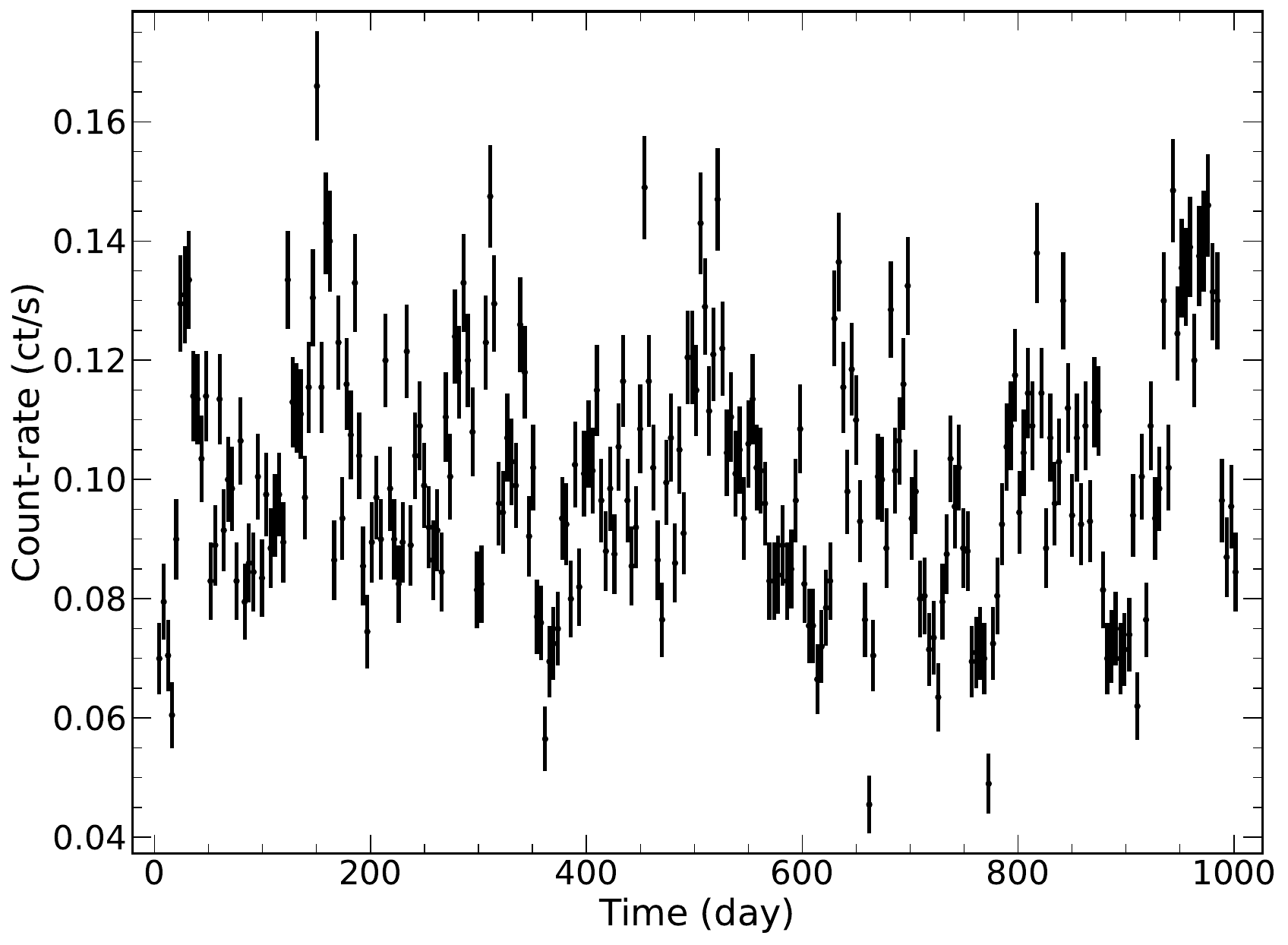}
    \includegraphics[width=0.49\textwidth]{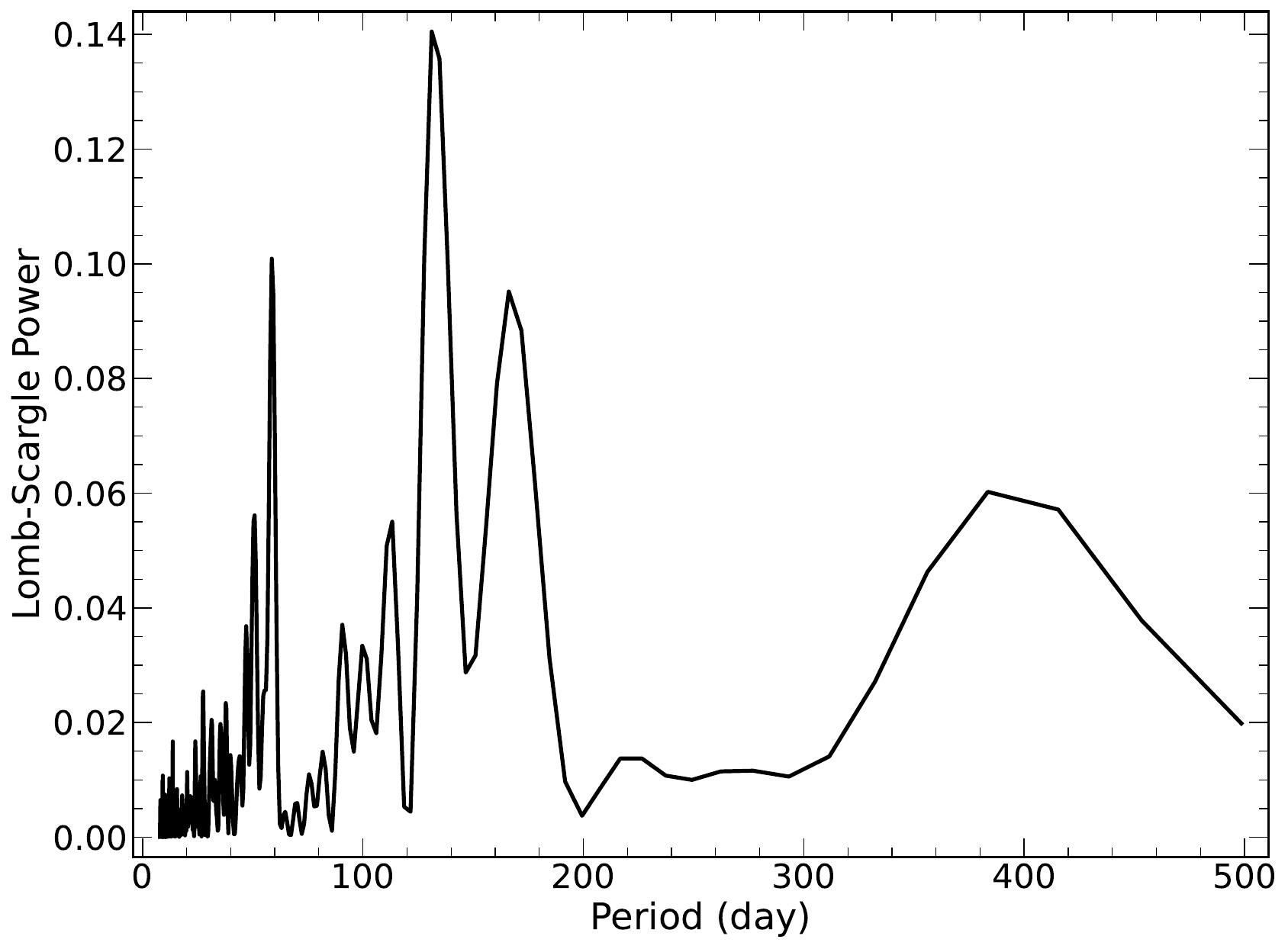}
    \caption{Example of a simulated lightcurve, generated to test the sensitivity of our method to false negatives ($N = 250$ $\Delta t \approx 4$ days). (Left) Lightcurve generated using a Lorentzian + DRW, with bending timescale of 60 days, and period of 100 days. (Right) Corresponding Lomb-Scargle periodogram.}
    \label{fig:lor_test}
\end{figure*}

We performed two types of tests using our input PSD model. First, to test the sensitivity of our method to changes in the sampling frequency, we generated lightcurves with a length of approximately $T$ = 1,000 days,  with a sampling rate $\Delta t$ drawn from a Gaussian distribution of mean = 1, 2, 4 and 10 days, and a standard deviation of 0.2 days, such that the lightcurves had 1000, 500, 250 and 100 datapoints respectively and a realistic, irregular observing cadence. Figure~\ref{fig:lor_test} shows a test lightcurve ($N = 250$ and $\Delta t$ $\approx$ 4 days) and its corresponding (Lomb-Scargle) periodogram. Secondly, to test the effects of having a shorter baseline, we fixed $\Delta t$ (mean and standard deviation of 1 and 0.2 days respectively) but progressively reduced the number of datapoints to generate lightcurves of shorter duration. In addition to the 1,000 day lightcurve, we also simulated lightcurves spanning approximately $T$ = 500, 400 and 300 days, respectively. 

For a given cadence/baseline combination, we simulated \nsimsfalsenegatives\ lightcurves and carried out the PPP method described in Section~\ref{sub:testing_qpo}, i.e. \nsimsfalsenegatives\ lightcurves were fitted with the DRW and the DRW + Lorentzian models and the LRT reference distribution was built using 2,000 simulations from the DRW posteriors. We chose to simulate \nsimsfalsenegatives\ lightcurves as a trade off between computational time and having roughly a representative sample for each cadence/baseline. Similarly, the rather low number of 2,000 simulations was set by computational constraints.  

\begin{figure*}
    \centering
\includegraphics[width=0.98\textwidth]{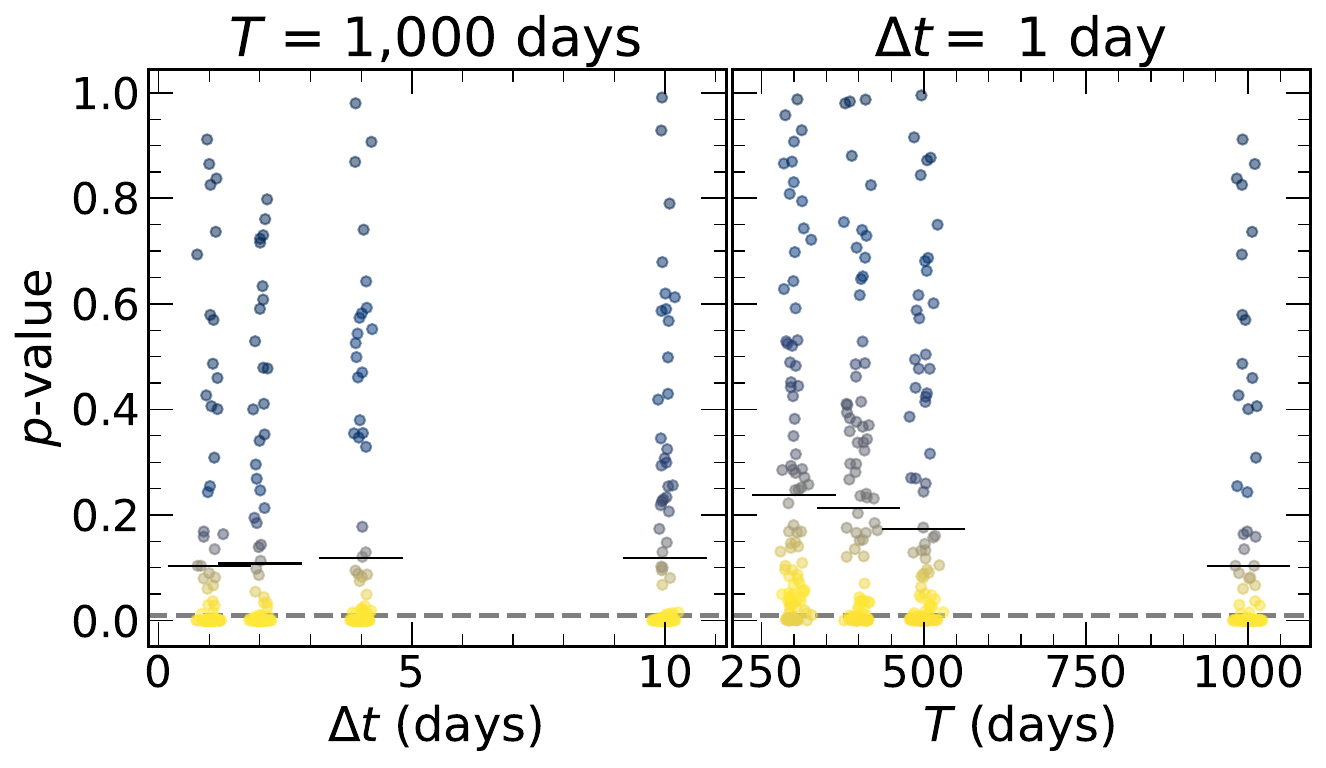}
    \caption{Distribution of $p$-values (with colors indicating the density of values) obtained from the application of our PPP method to \nsimsfalsenegatives\ simulated lightcurves using a Lorentzian (QPO) + DRW with varying cadence (left) and baseline (right). Dashed colored horizontal lines show the mean $p$-values for each cadence/baseline strategy. The horizontal grey dashed line shows the 99\% significance detection level ($p$ = 0.01).}
    \label{fig:proof_concept_results}
\end{figure*}

Figure~\ref{fig:proof_concept_results} show the distribution of the retrieved $p$-values for the \nsimsfalsenegatives\ lightcurves for the case of varying cadence (left panel) and varying baseline (right panel). Table~\ref{tab:false_negatives} shows the mean retrieved $p$-values and the number of significant ($p\lesssim$0.01) detections per cadence/baseline combination.

\begin{table}
    \centering
        \caption{Summary of the analysis carried out to test the sensitivity of our method to false negatives. \nsimsfalsenegatives\ lightcurves were simulated per baseline/cadence combination from a Lorentzian + DRW PSD and we examined whether we could detect the additional Lorentzian (QPO) component over the DRW.} 
    \label{tab:false_negatives}
    \begin{tabular}{ccccc}
    \hline
    \noalign{\smallskip}

    $N^a$ & $<\Delta t >^b$ &  $T^c$ & $<p>^d$ & $n_{p < 0.01}^e$\\
        &   days       &  days &    &    \\
        \noalign{\smallskip}
        \hline 
        \hline
        \noalign{\smallskip}
100 & 10  & 1,000 & 0.12 & 65 \\ 
250 & 4   & 1,000 & 0.12 & 62 \\ 
500 & 2  &  1,000 & 0.11 & 68 \\ 
1000 & 1 & 1,000& 0.10 & 68 \\ 
300 & 1 & 300 & 0.24 & 21 \\ 
400 & 1  & 400 &  0.21 & 39 \\ 
500 & 1  &  500 & 0.17 & 44 \\ 
\noalign{\smallskip} 
         \hline
    \end{tabular}
    \\
        \begin{flushleft}
        \noindent\textbf{Notes.}  \\
    $^{a}$ Number of datapoints of the generated lightcurves. \\
    $^{b}$ Mean cadence. \\
    $^{c}$ Observing baseline. \\
    $^{d}$ Mean retrieved PPP value of \nsimsfalsenegatives\ lightcurves for the presence of the QPO (Lorentzian) component. \\
    $^{e}$ Number of lightcurves for which the Lorentzian was significantly detected ($p < 0.01$). 
    \end{flushleft}
\end{table}

From Figure~\ref{fig:proof_concept_results}, we can see that, despite the rather low count rates of the lightcurves, we are able to recover the period in more than half of the instances (in $\geq$ 62/\nsimsfalsenegatives) as long as enough cycles are observed. In particular, there is little improvement in detection rates in the lightcurves with fixed baseline (left panel). This is partially due to the fact that the parameter space (namely $P$ and \omegabend) is accommodated with the sampling (as the lowest cadence in the lightcurve sets the minimum allowed $P$ and bend timescale) but it suggests the number of cycles might be the most important metric when attempting to detect a periodicity. This simple result is consistent with the requirement identified by \cite{vaughan_false_2016} when looking for periodicities in stochastically-varying systems.

Interestingly, our results show that lightcurves spread over time with lower-cadence sampling prove more advantageous for detecting periodicities compared to shorter lightcurves with a higher number of datapoints (at least as long the timescale of the period is much longer than the cadence). For instance, in the case of a $\Delta t \approx$ 2 days and $N = 500$ (left panel), we are able to recover the period in 68/\nsimsfalsenegatives\ instances, whereas for  $\Delta t \approx$ 1 day and the same number of datapoints (right panel), only in 44/\nsimsfalsenegatives\ instances we are able to recover the period (Table~\ref{tab:false_negatives}). Our results suggest that, in the presence of stochastic variability, it will hard to reliably confirm periodicities in a lightcurve covering five or less cycles of the putative period, in agreement with \citet{vaughan_false_2016}.

\begin{table}
    \centering
        \caption{Summary of the analysis carried out to test the sensitivity of our method to false positives. We tested for the presence of an additional Lorentzian (QPO) component in 50 lightcurves, simulated from a DRW PSD per baseline/cadence combination.}  
    \label{tab:false_positives}
    \begin{tabular}{ccccc}
    \hline
    \noalign{\smallskip}
    $N^a$ & $<\Delta t >^b$ &$T^c$& $p_\mathrm{uniform}^d$\\
        &   days       &  days &   \\
        \noalign{\smallskip}
        \hline 
        \hline
        \noalign{\smallskip}
100 & 10 & 1,000 & 0.35  \\ 
250 & 4  & 1,000 & 0.56  \\ 
500 & 2  & 1,000 & 0.16  \\
1000 & 1  & 1,000& 0.35 \\ 
300 & 1  & 300 & 0.91  \\ 
400 & 1  & 400 & 0.30  \\ 
500 & 1  & 500 & 0.98  \\ 
\noalign{\smallskip}
         \hline
    \end{tabular}
                 \\
        \begin{flushleft}
        \noindent\textbf{Notes.}  \\
    $^{a}$ Number of datapoints of the generated lightcurves. \\
    $^{b}$ Mean cadence. \\
    $^{c}$ Observing baseline. \\
    $^{d}$ $p$-values for the distribution of the 50 retrieved PPPs following a uniform distribution between 0 and 1, as expected when the null hypothesis is true. \\
    \end{flushleft}
\end{table}

\begin{figure*}
    \centering
    \includegraphics[width=0.49\textwidth]{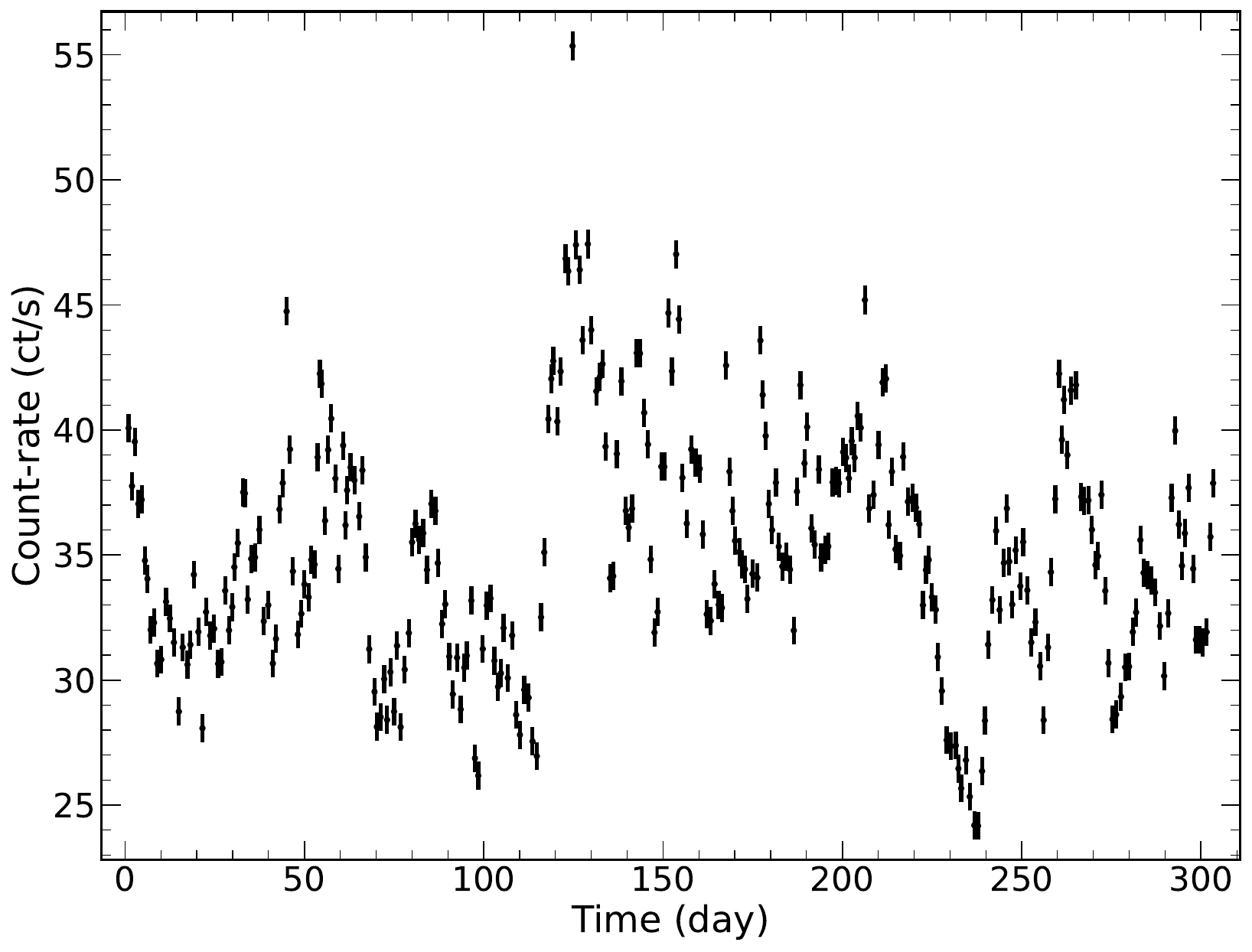}
     \includegraphics[width=0.49\textwidth]{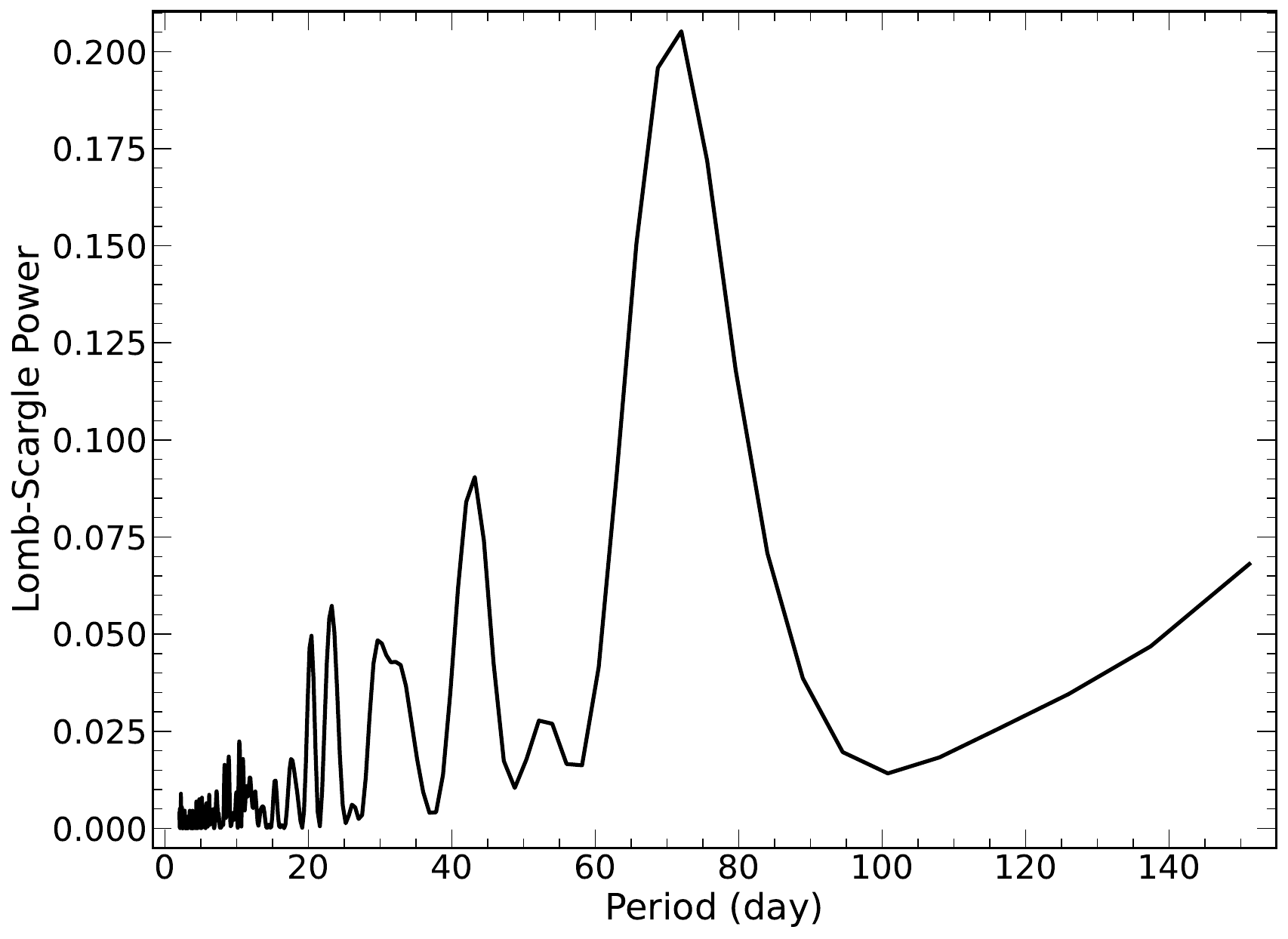}
         \includegraphics[width=0.49\textwidth]{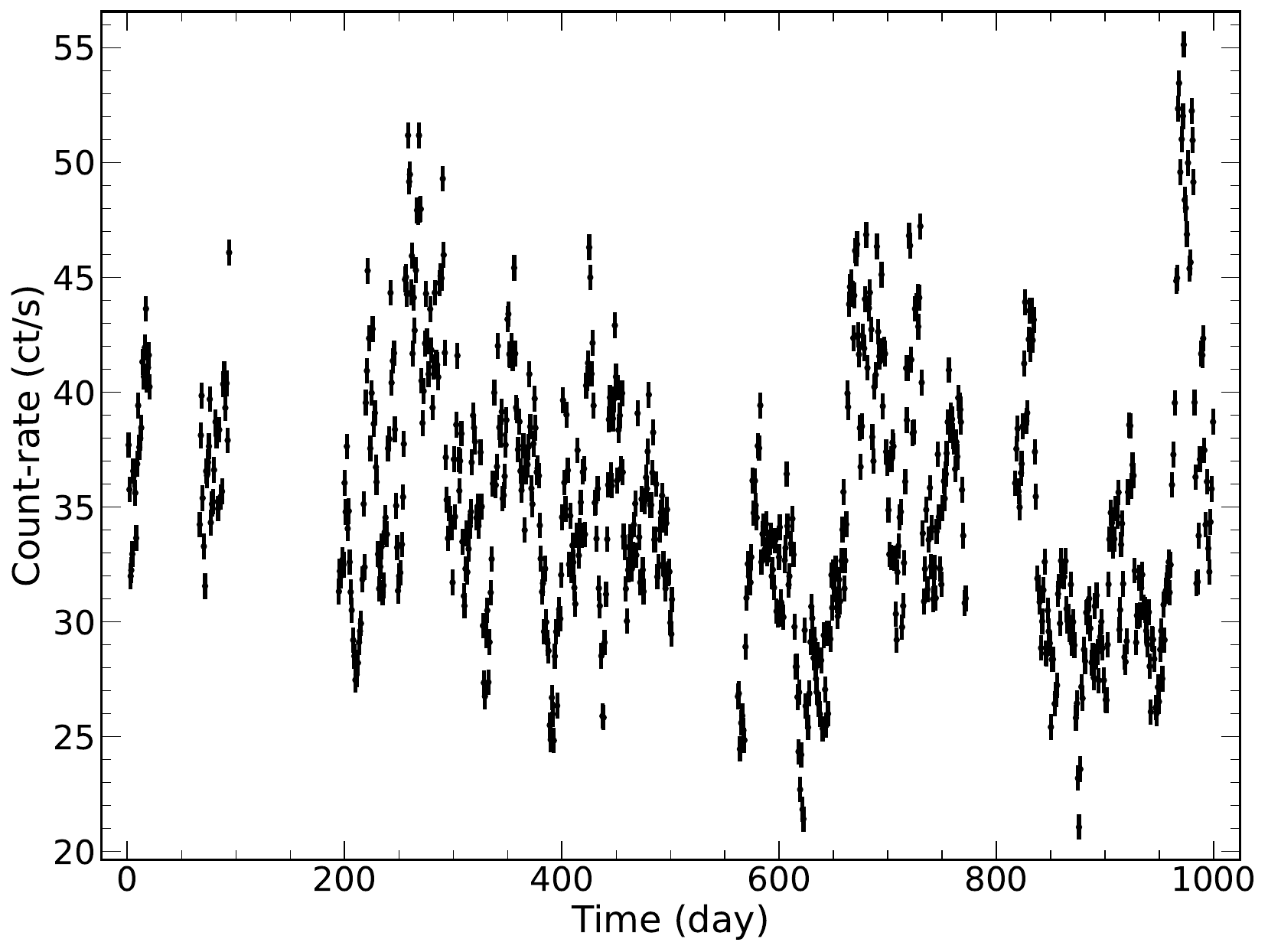}
     \includegraphics[width=0.49\textwidth]{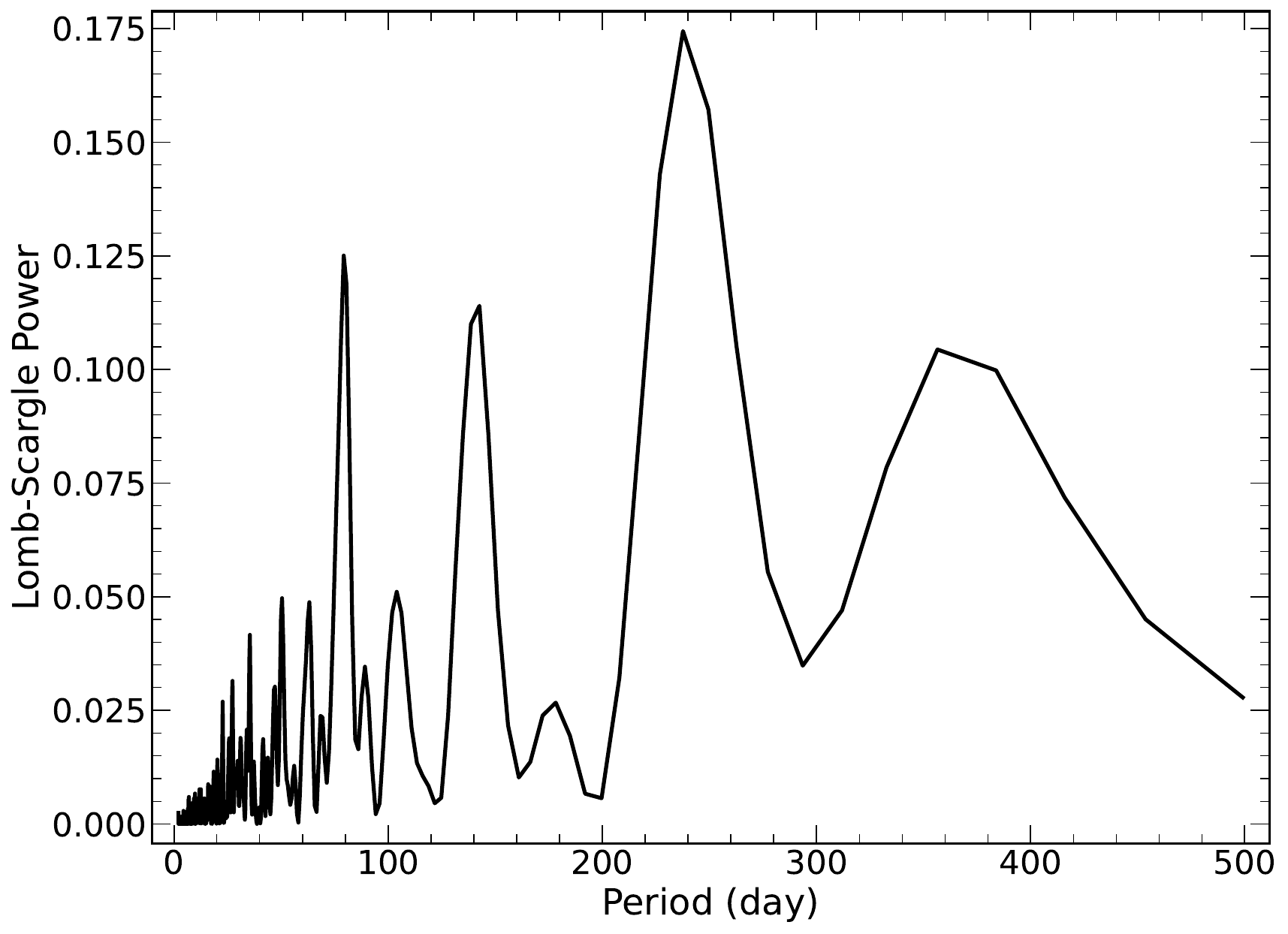}
    \caption{Example of simulated lightcurves generated to test the sensitivity of our method to false positives. (Top left) Lightcurve generated using a DRW with bending timescale of 65 days ($N = 300$, cadence roughly every 1 days). (Top right) Corresponding Lomb-Scargle periodogram. (Bottom left) As before, but initially with $N = $1,000 and then including two gaps of 45 days, one of 60 days and another of 100 days. (Bottom right) Corresponding Lomb-Scargle periodogram.}
    \label{fig:test_drw}
\end{figure*}

Next we turn to examine the robustness of our method to false positives (i.e. misidentifying aperiodic variability as periodic), one of the aspects that motivated us to devise better methods for period detection. To this end, we performed a second series of simulations using a simple DRW with $\Delta t$ again drawn from the same Gaussian distributions. Here we use a mean count rate of 35 ct/s, a background contribution of 300 ct/s and a variance of 36 (ct/s)$^2$ and a break at 65 days for the DRW. These parameters are similar to those observed in the  in the TESS lightcurves of Blazars (see Section~\ref{sub:qso}) and were chosen to generate lightcurves which {\it appear} periodic. Figure~\ref{fig:test_drw} (top panel) shows an example lightcurve ($N= 300$, median $\Delta t$ = 1 day) with the corresponding periodogram; clearly naive inspection of the periodogram may lead to the conclusion that some genuine periodicity is present in the lightcurve.

We ran our PPP method again as above, this time creating 50 lightcurves per cadence/baseline combination and simulating 2,000 lightcurves from the DRW posteriors, and comparing the fit improvements when adding a Lorentzian. Under the absence of the signal, the distribution of retrieved PPP values from the \tlrt\ tests is expected to be uniformly distributed from 0 to 1. We therefore tested using a KS test whether the 50 retrieved PPP values per cadence/baseline combination followed the aforementioned uniform distribution. Table~\ref{tab:false_positives} shows all $p_\mathrm{uniform}$-values are consistent with the expected uniform distribution, regardless of the observing strategy, indicating false positives are unlikely, at least for the cadences explored here. 

Finally, to test our ability to avoid false-positives when gaps are introduced, we have repeated the method above, keeping the same PSD and using $N = 1000$, samples taken roughly at 1 day intervals but then adding three gaps: one of 45 days, one of 60 days and another of 100 days. Figure~\ref{fig:test_drw} (bottom panels) shows an example lightcurve and corresponding periodogram. We then ran the PPP method for 50 sample lightcurves and tested whether the recovered PPPs followed a uniform distribution as expected for the absence of a signal, finding a $p$-value of 0.37. This indicates that, at least for the cadence/variability timescales explored here, our method can robustly avoid false-positives as long as the noise is well described. We will present an exhaustive exploration of period detectability under different combinations of observing strategies in a forthcoming publication.

\subsection{Application to real data}\label{sub:celerite_results}

Our method has been developed for instances where irregular sampling hampers obtaining the PSD in a straightforward manner. Such a scenario is routinely encountered in many studies of AGN \citep[e.g.][]{jiang_tick-tock_2022} and other accreting systems. 

Nevertheless, to show our method is not restricted to irregularly-sampled time series, we first apply our method to a recent claim of a QPO in {\it XMM-Newton} data of a Seyfert galaxy (Section~\ref{sub:xmm_seyfert}). We then explore claims of periodicites in \swift\ data \citep{gehrels_swift_2004} (both UVOT and XRT) of a ULX (Section~\ref{sub:ngc7793}), a QPO in an AGN in RXTE data (Section~\ref{sub:agn_qpo}) and finally revisit a recent claim of a QPO in the Transiting Exoplanet Survey Satellite \citep[TESS;][]{ricker_transiting_2014} lightcurve of a Blazar (Section~\ref{sub:qso}). The results of the analysis are then discussed in Section~\ref{sub:analysis_results}. The choice of priors and the procedure used to derive the best-fit parameters and their posteriors is described in Appendix~\ref{sub:mcmc_sampling} and unless stated otherwise, we perform 10,000 simulations to derive the LRT reference distribution in the calculation of the PPP value. 
\subsubsection{A high-frequency QPO in the Seyfert NGC 1365}\label{sub:xmm_seyfert}
\input{ngc1365.tex}
\subsubsection{The Pulsating ULX NGC 7793 P13}\label{sub:ngc7793}


Since the discovery of its $\sim$63\,d period \citep{motch_mass_2014}, the pulsating neutron star ULX NGC 7793 P13 (\citealt{furst_discovery_2016,israel_discovery_2017}, P13 hereafter) has been intensively monitored by \swift. Being among the brightest ULXs in the optical bands with a $V$ magnitude of around $\sim$20.2 \citep{motch_mass_2014}, it is one of only a small number of ULXs where the long-term variability can be studied by both the \swift-UVOT and \swift-XRT.
The irregular sampling of the monitoring of this source has revealed two closely but significantly different periods: a $\sim$64-day period in the $U$ band and a $\sim$65-day period in the X-rays \citep{hu_swift_2017, furst_tale_2018}. 
An advantage of using GP for period searching is that uncertainties are well-defined as we can marginalise over the noise parameters. Therefore we can asses both the significance of the claimed periodicities and also the difference between them. 

\begin{figure*}
    \centering
    \includegraphics[width=0.49\textwidth]{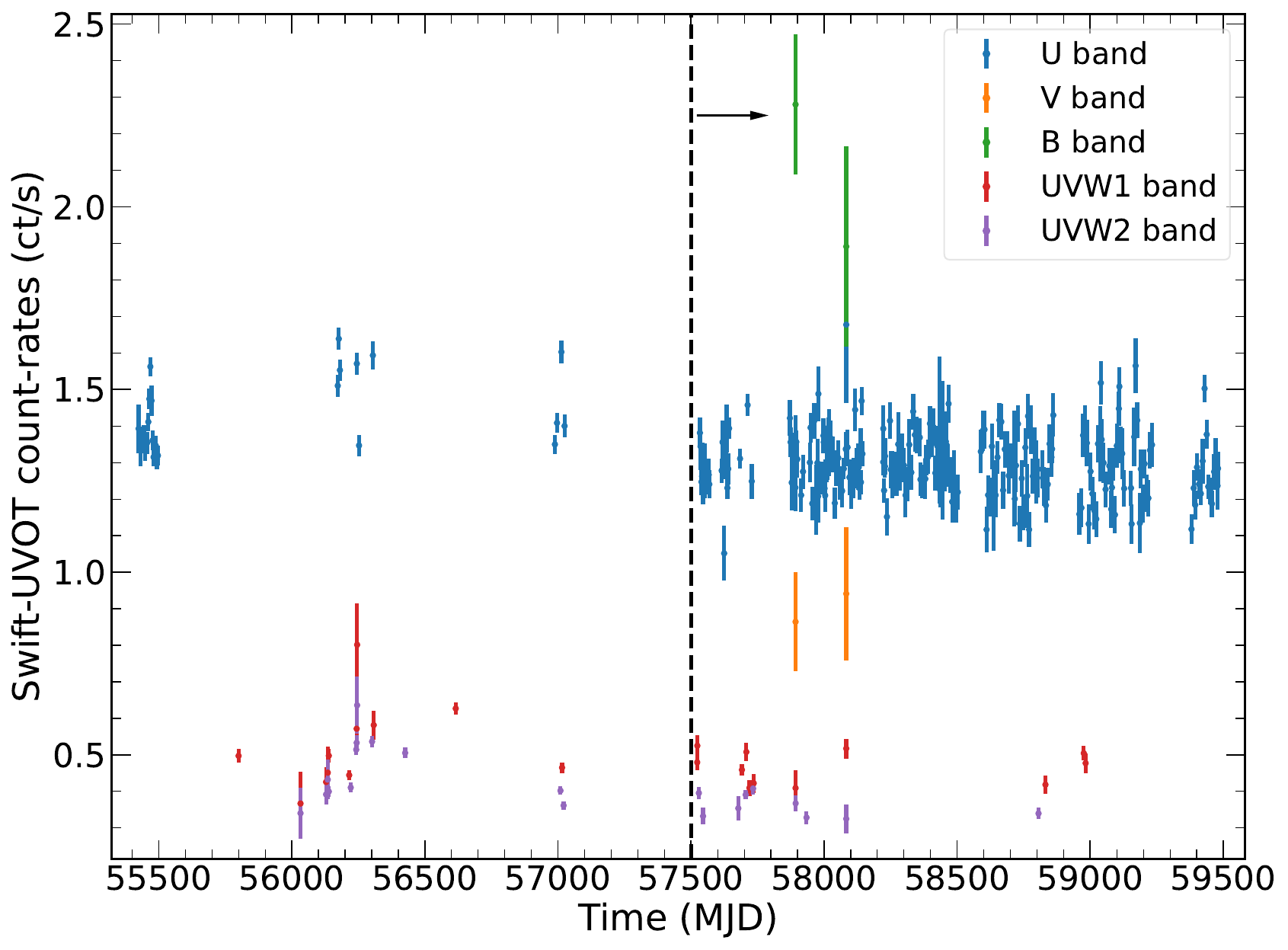}
    \includegraphics[width=0.49\textwidth]{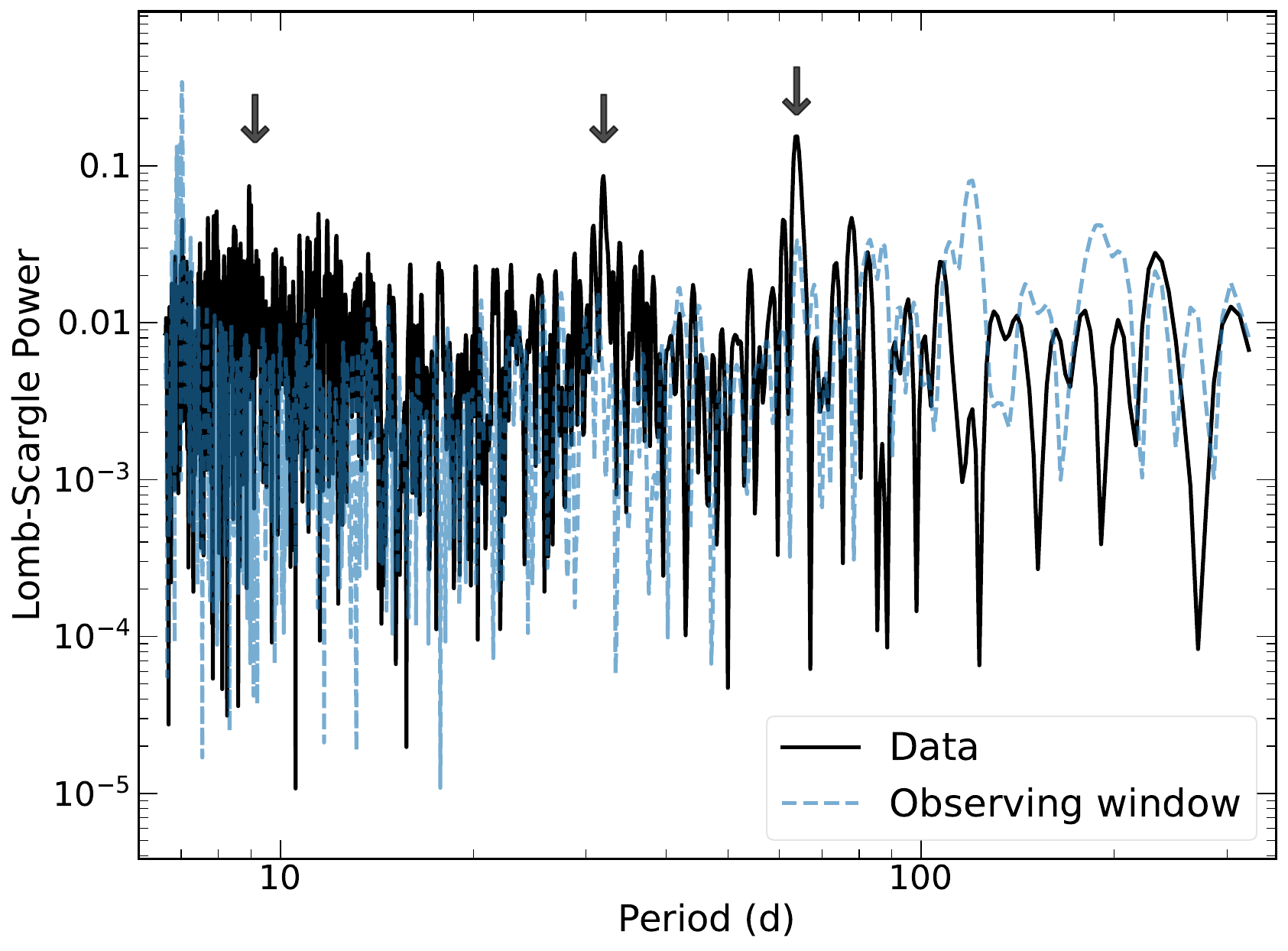}\caption{NGC~7793~P13 \swift-UVOT lightcurve, with the segment considered for analysis indicated with a vertical dashed line and an arrow. (Right) Lomb-Scargle periodogram of the $U$ band lightcurve segment indicated in the left-hand panel. The black vertical arrows indicate harmonics ($P/n$) at $n= 2$ and $n=7$ from the fundamental at $\sim$ 64 days. The dashed blue line shows the power spectrum of the observing window.}
    \label{fig:uvot_P13_lc}
\end{figure*}

The UVOT data were kindly provided by \citet{khan_long-term_2023}, to which we refer the reader for the data reduction details. The $U$ band contained the largest amount of observations (Figure~\ref{fig:uvot_P13_lc}; 260 observations compared to $\lesssim$20 in other bands); we therefore analysed only this band. While an advantage of GP modelling is that more data, regardless of the gaps, should lead to tighter constraints, here the few additional and largely spaced datapoints at the beginning of the monitoring increase the computational cost dramatically for a small gain in accuracy, particularly in our false-alarm probability calculation. Therefore we only considered the data after MJD 57,500 where the monitoring is denser (Figure~\ref{fig:uvot_P13_lc}).

\begin{table}
    \centering
    \begin{tabular}{lccc}
\hline
 Model & AICc & $\Delta$AICc & $p$-value \\ 
 \hline \hline 
3$\times$Lorentzian + Jitter & --556.0 & 0.0 & 0.32 \\
3$\times$Lorentzian  + \granulation & --555.9 & 0.2 & 0.71 \\ 
3$\times$Lorentzian + Mat\'ern-3/2  & --554.1 & 1.9 & 0.27\\
3$\times$Lorentzian + DRW & --553.6 & 2.4 & 0.08 \\
2$\times$Lorentzian + Jitter & --552.2 & 3.8 & 0.43 \\
2$\times$Lorentzian + DRW & --549.5 & 6.5 & 0.35 \\
2$\times$Lorentzian + \granulation & --549.0 & 7.0 & 0.38 \\
2$\times$Lorentzian + Jitter + \granulation & --548.4 & 7.6 & 0.008 \\
2$\times$Lorentzian + Jitter + DRW & --548.1 & 7.9 & 0.02 \\
2$\times$Lorentzian + Jitter + \matern & --547.7 & 8.3 & 0.02 \\
2$\times$Lorentzian + \matern & --540.1 & 15.9 & 0.18 \\
2$\times$Lorentzian & --531.5 & 24.5 & 0.52 \\
Lorentzian + Jitter & --529.2 & 26.8 & 0.44 \\
 Lorentzian + DRW & --526.5 & 29.5 & 0.365 \\
 Lorentzian + \granulation & -525.9 & 30.1 & 0.45 \\
 \granulation\ + 2$\times$DRW & -522.6 & 33.5 & 0.44 \\
 2$\times$\granulation\ + DRW & -516.5 & 39.5 & 0.78 \\
 DRW & --495.5 & 60.5 & 0.178 \\
 Jitter & --495.1 & 60.9 & 0.05 \\
 Mat\'ern-3/2 & --494.3 & 61.8 & 0.28 \\
 \granulation & --493.2 & 62.8 & 0.10 \\
 2$\times$DRW & --492.9 & 63.1 & 0.11 \\
 Lorentzian & --492.6 & 63.4 & 0.69 \\
 \granulation\ & --492.1 & 63.9 & 0.21 \\
 \hline 
 \hline 
\end{tabular}
\caption{As per Table~\ref{tab:ngc1365} but now showing the AICc, $\Delta$AICc and $p$-values for the standarised residuals following a Gaussian distribution for the different models tested against the \swift-UVOT data of P13.}
\label{tab:uvot_dbic}
\end{table}

Table~\ref{tab:uvot_dbic} lists the models tested in our fit to the data, ranked by AICc value. Part of the modelling was guided by a visual inspection of the Lomb-Scargle periodogram of the lightcurve segment, which we show in Figure~\ref{fig:uvot_P13_lc}. We can see the main peak at $P\sim$64 days and some harmonics at 32 days ($P$/2) and $\sim$9 days ($P$/7), indicating the periodicity -- if real -- is not a pure sinusoid.

\begin{figure*}
    \centering
    \includegraphics[width=0.49\textwidth]{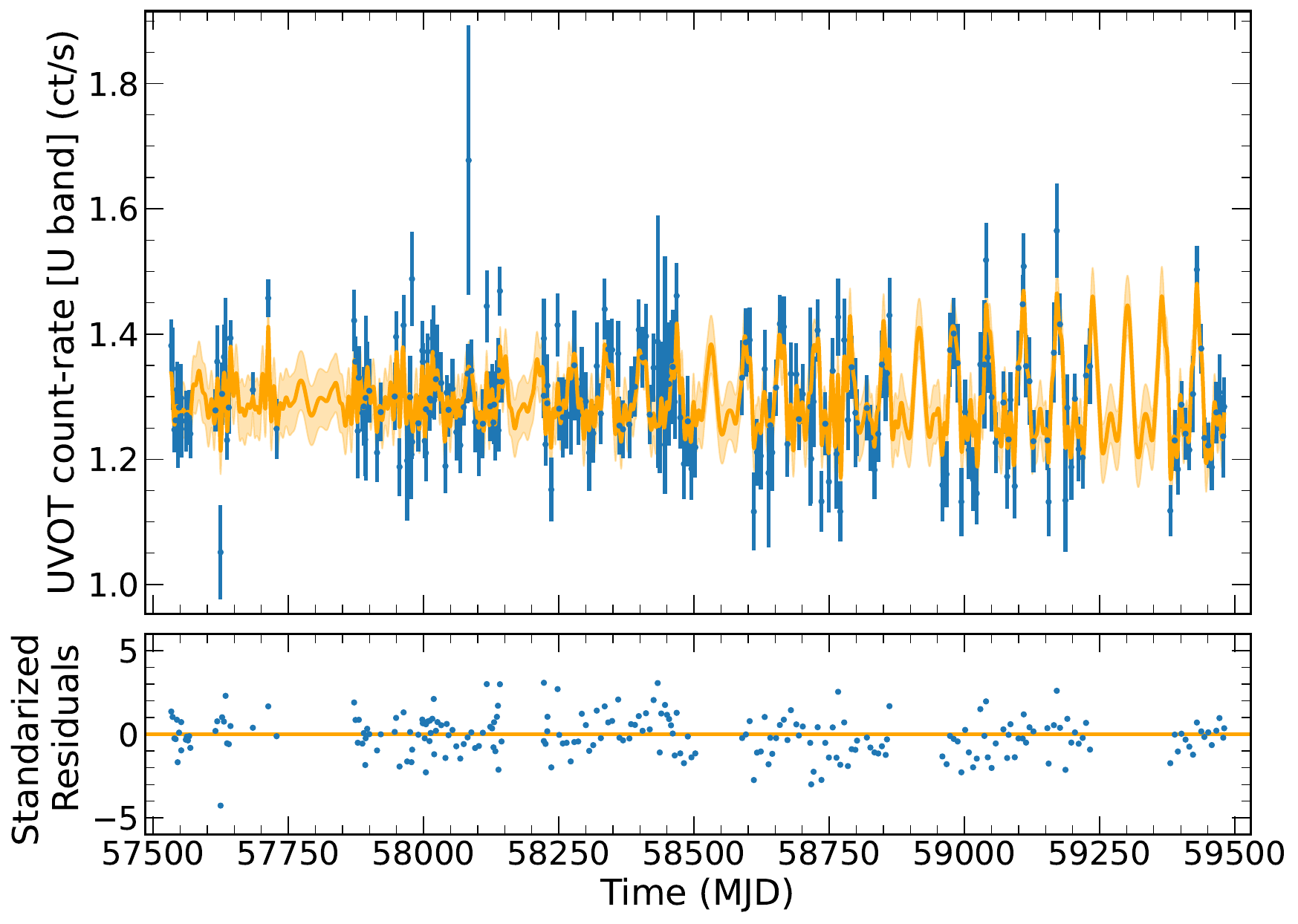}
    \includegraphics[width=0.49\textwidth]{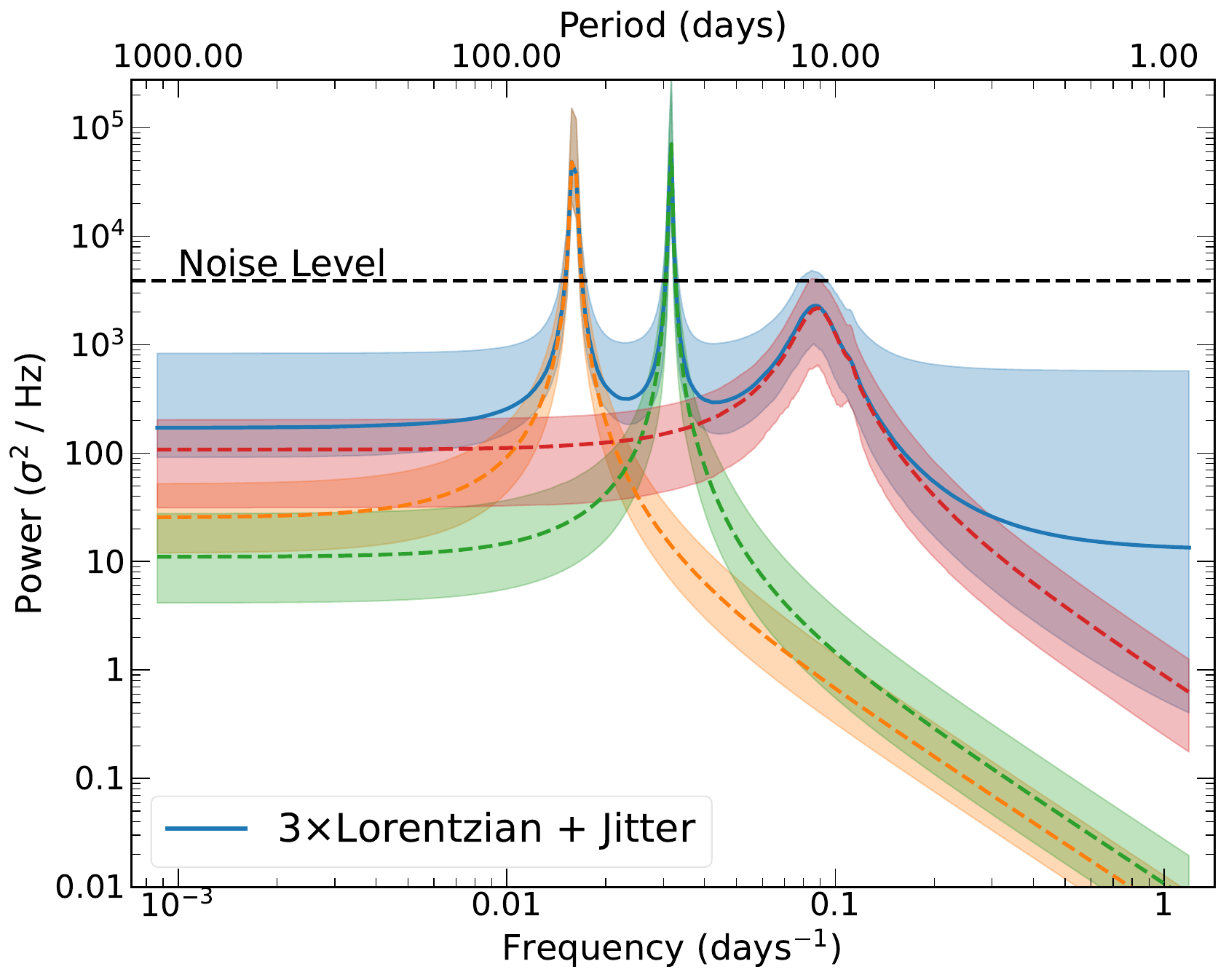}
     \includegraphics[width=0.49\textwidth]{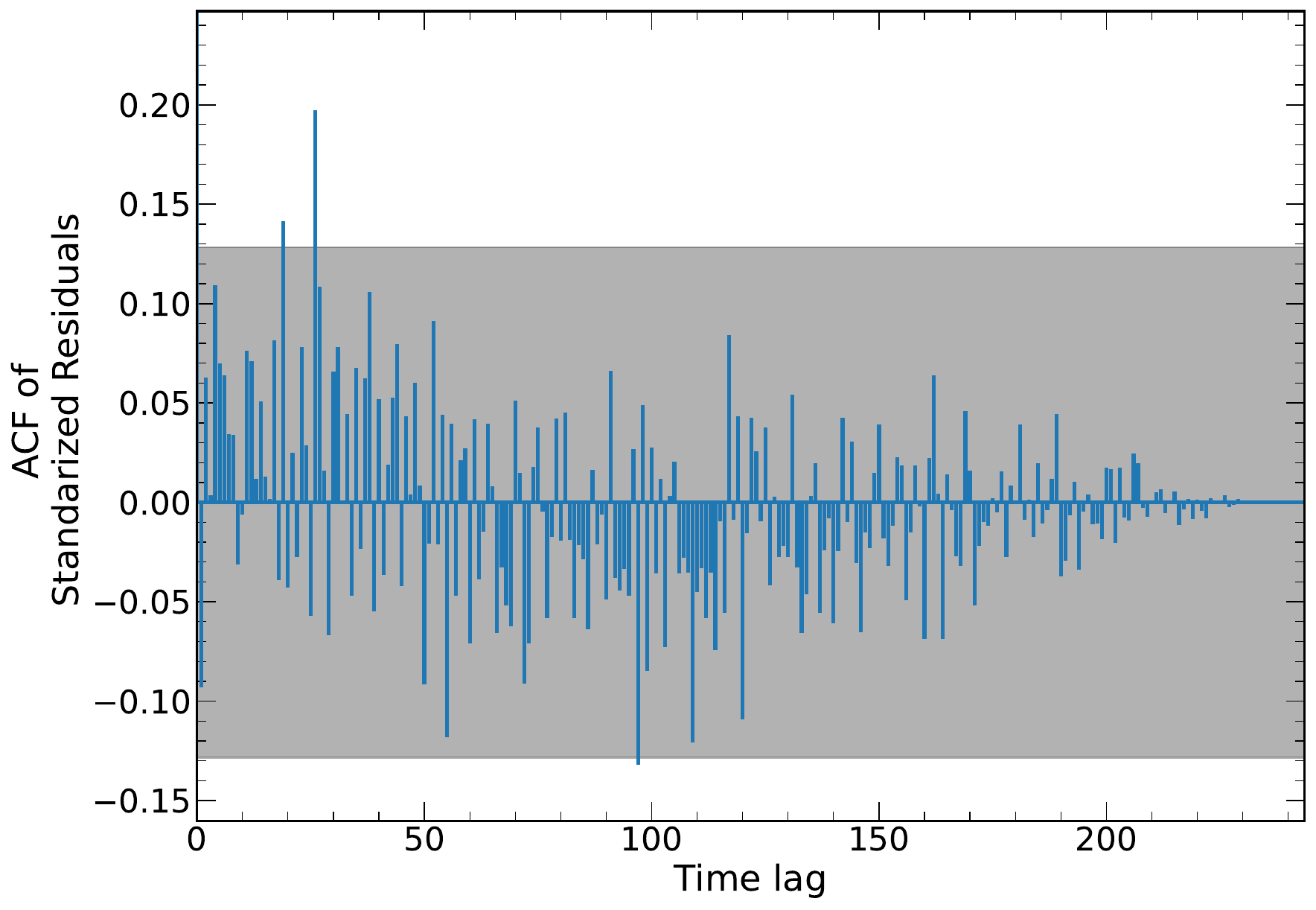}
    \includegraphics[width=0.49\textwidth]{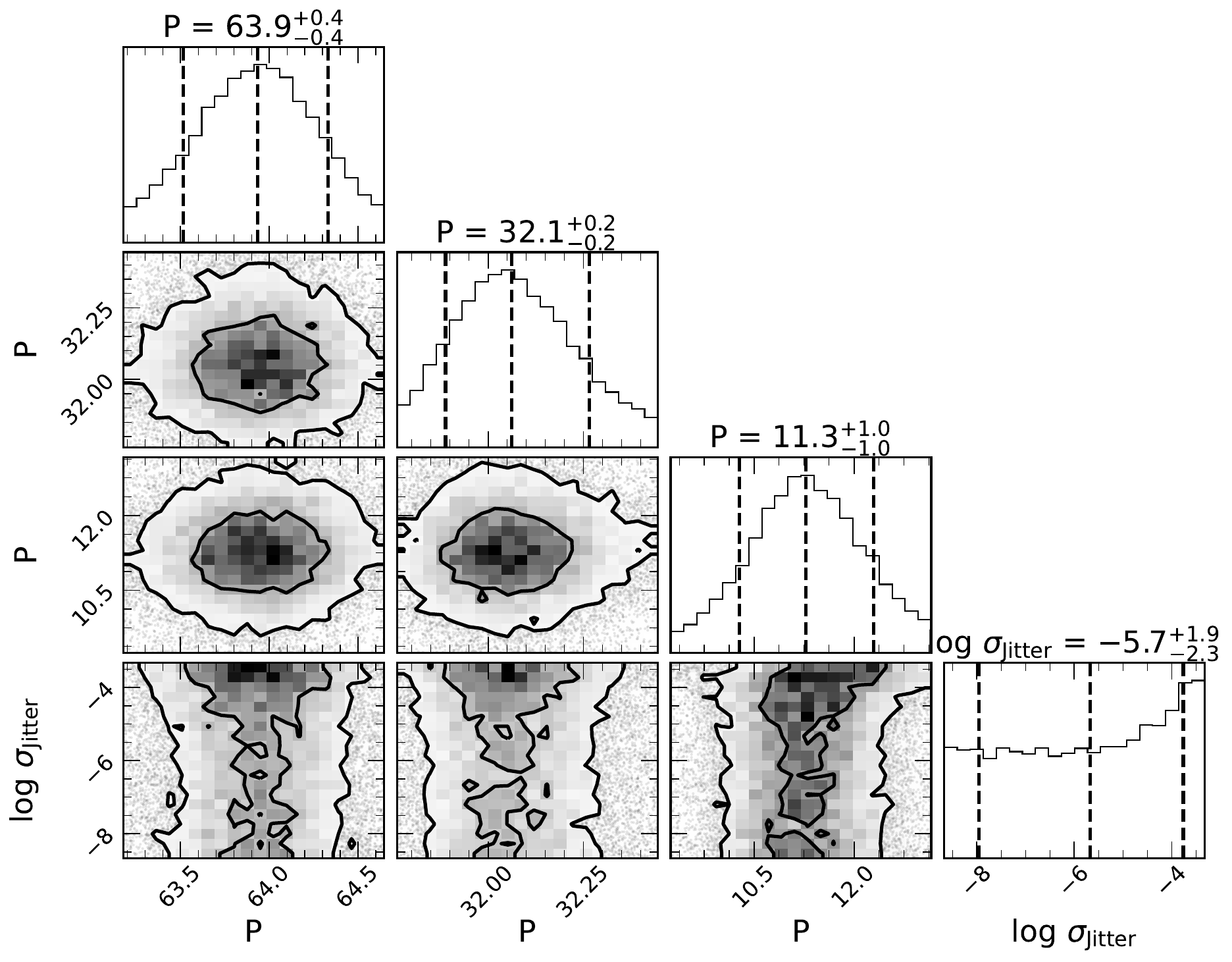}
    \caption{GP modelling results of the \swift-UVOT lightcurve segment (Figure~\ref{fig:uvot_P13_lc}) of the pulsating ULX NGC~7793~P13. (Top left) Best-fitting 3$\times$Lorentzian + Jitter model. (Top right) PSD of the best-fit model. The contribution of the Lorentzians to the PSD (blue solid line) are shown as a orange, green, and red dashed lines. (Bottom left) ACF of the standarized residuals of the best-fit model. (Bottom right) Posteriors of the periods of the three Lorentzians and the Jitter amplitude (other parameters omitted for clarity). The MCMC sampling run for approximately 49,000 steps until convergence, from which we discarded the first 22,000 as burn-in. Symbols as per Figure~\ref{fig:ngc1365_posteriors}.}
    \label{fig:uvot_P13}
\end{figure*}

The harmonics are also reflected in the GP modelling: we can see from Table~\ref{tab:uvot_dbic} that the preferred model consists of three Lorentzians + a Jitter component for the underlying noise. From Table~\ref{tab:uvot_dbic} we can also see that this model is preferred over one where the underlying noise is instead described by a DRW ($\Delta$AICc = 2.4), suggesting that white noise is the statistically preferred null hypothesis. 

The standarised residuals of the best-fit model are fully consistent with a Gaussian distribution, indicating the variability is well-described by a GP, whereas the overall variability is also well-captured, as indicated by the ACF of the standarized residuals (Figure~\ref{fig:uvot_P13} bottom left panel). 

Having established white noise to be a good representation for the underlying noise, we proceed to test for the presence of the Lorentizan(s) components in a hierarchical manner. First, we test for the first Lorentzian over the Jitter-only model, using the posteriors of the Jitter model. If significant, we subsequently use the posteriors of the Lorentzian + Jitter to test for an additional Lorentzian until the new added Lorentzian is no longer significant. 


For the first Lorentzian, we found that none of the simulations showed a \tlrt\ as high as that observed in the data. Fitting the LRT distribution with a lognormal, we estimate the period to be highly significant (99.999\% or $\sim6\sigma$). For the second harmonic at $\approx$32 days, we find the significance to be $\approx$99.1\%, while for the third component the significance is 95.8\%.

From our best-fit (Figure~\ref{fig:uvot_P13}) we obtained $P = 63.9\pm0.4$ days, with a coherence $Q = 220^{+727}_{-146}$, indicating the period amplitude is stable over this time period. 

\begin{figure*}
    \centering
    \includegraphics[width=0.49\textwidth]{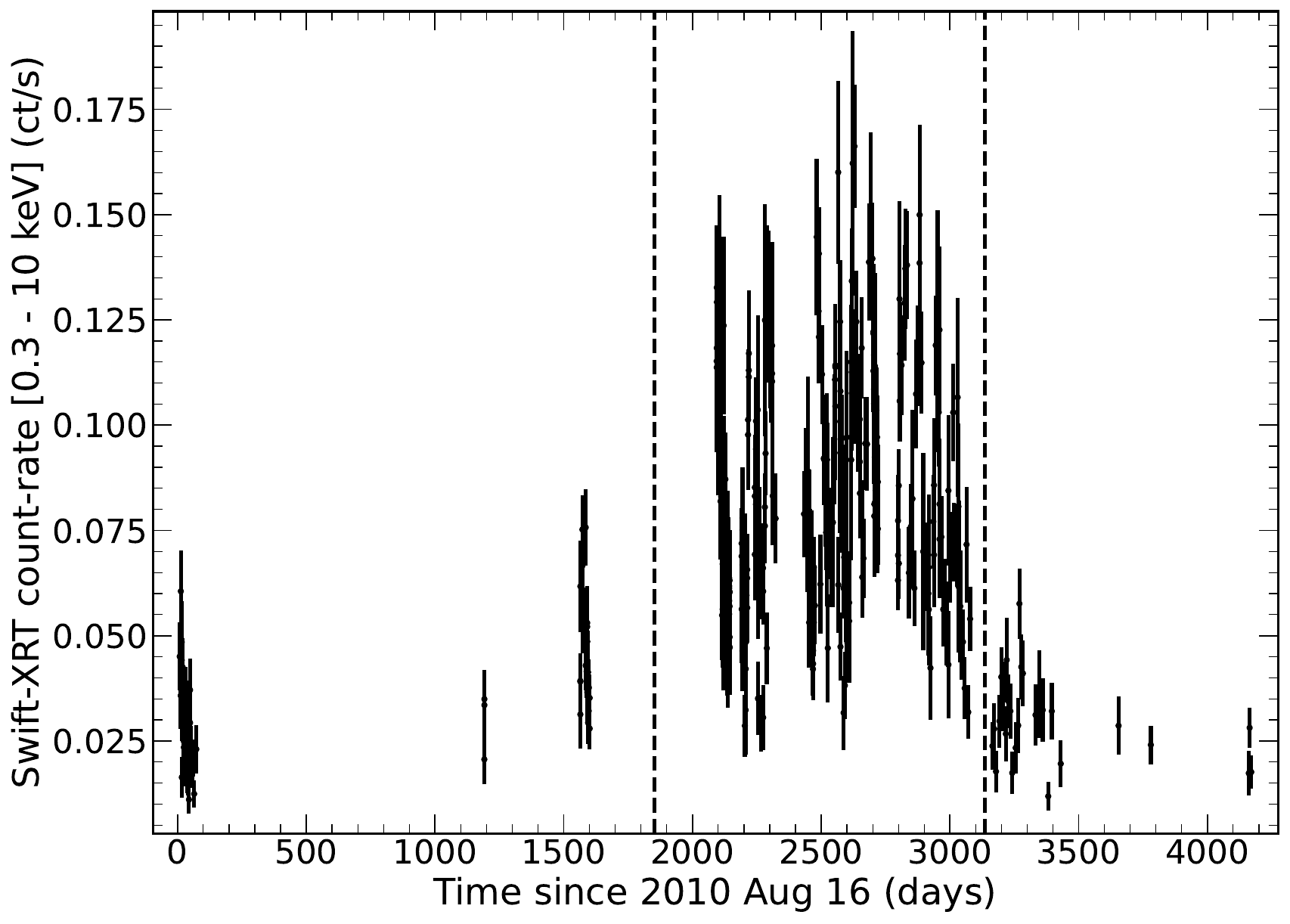}
    \includegraphics[width=0.49\textwidth]{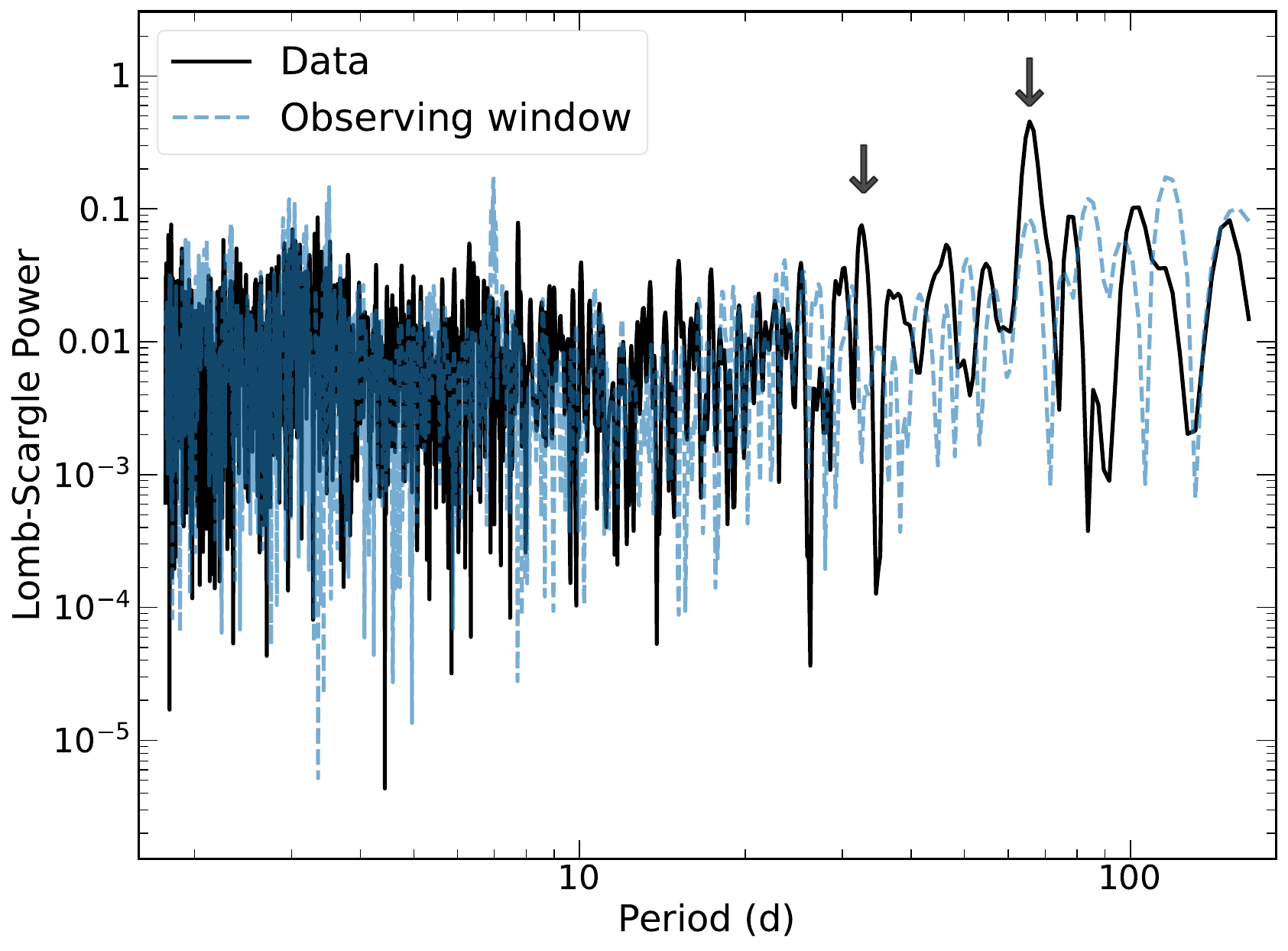}
    \caption{(Left) \swift-XRT lightcurve of NGC~7793~P13, with the segment considered for analysis lying between the two vertical, dashed black lines. (Right) Lomb-Scargle periodogram of the segment indicated in the right-hand panel. Symbols as per Figure~\ref{fig:uvot_P13_lc}. The black vertical arrows show the fundamental period frequency ($P\sim$65.6 days) and an harmonic at $n = 2$.}
    \label{fig:xrt_P13_lc}
\end{figure*}

We now examine the \swift-XRT data and the claimed $\sim$65-day period. The full \swift-XRT lightcurve is shown in Figure~\ref{fig:xrt_P13_lc} and was extracted using the online tools \citet{evans_online_2007, evans_methods_2009}, keeping all snapshots with a detection significance of $\ge$2 $\sigma$. Modelling the full lightcurve would add additional complexity due to potential deviations from stationarity and more complex fine-tuning of the mean function. Additionally, the few largely spaced datapoints would again add little gain in constraining power at the expense of significant computational time. Hence we analysed the indicated segment in Fig.~\ref{fig:xrt_P13_lc} where the variability appears stationary and the monitoring is densest. This segment lasts 984.3\,days with a mean observing cadence of 3.1\,days. 

\begin{figure*}
    \centering
    \includegraphics[width=0.49\textwidth]{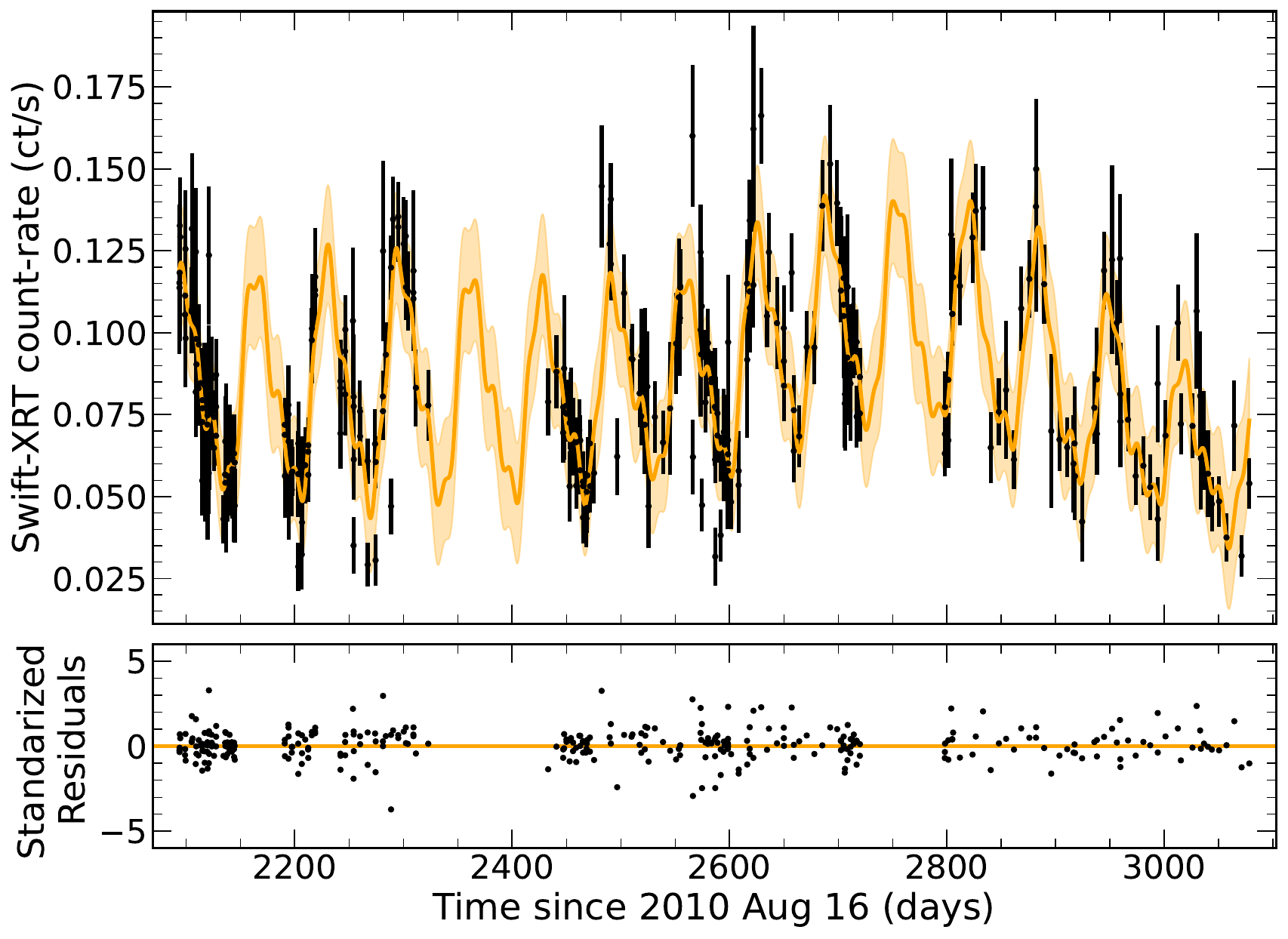}
    \includegraphics[width=0.49\textwidth]{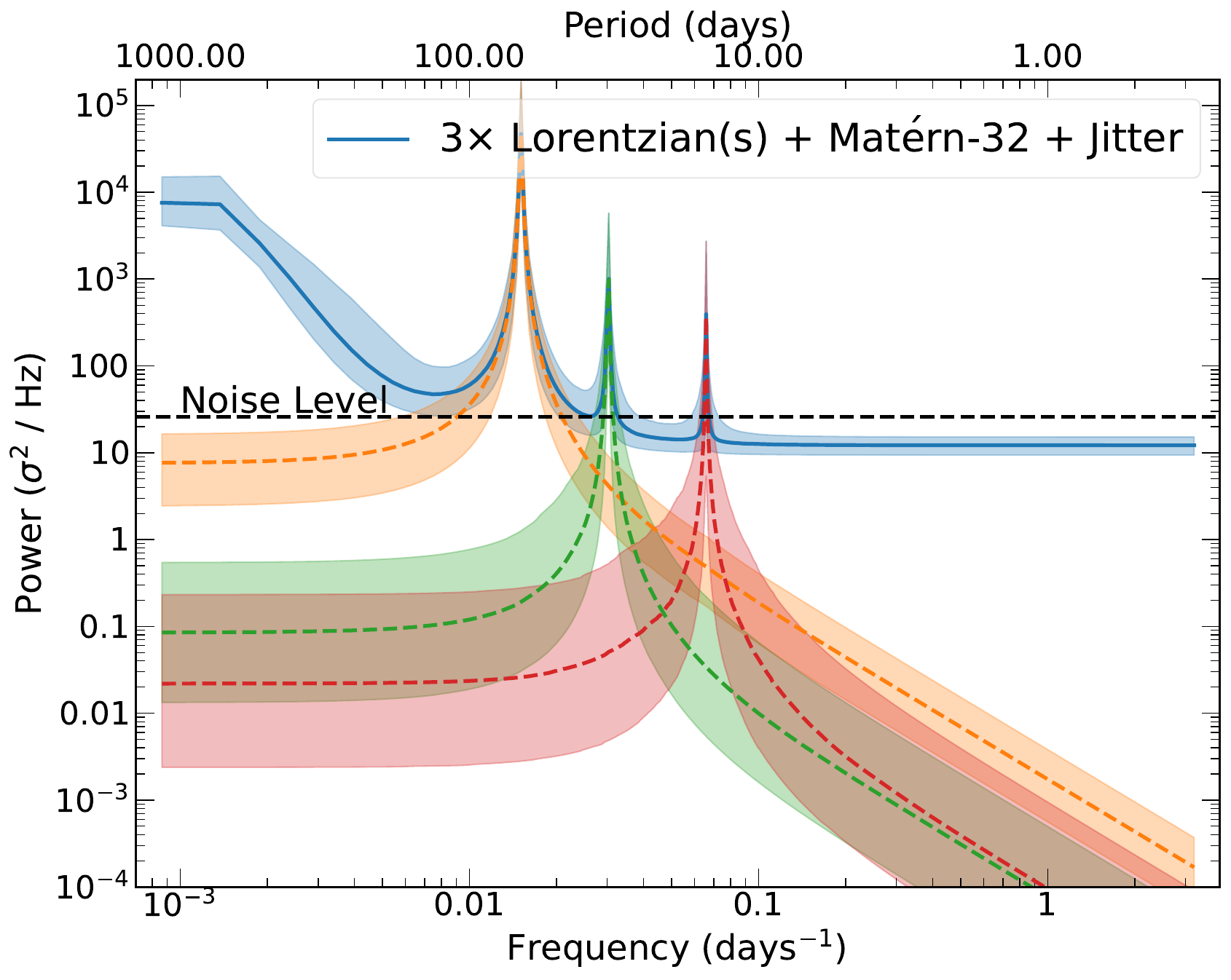}
    \includegraphics[width=0.49\textwidth]{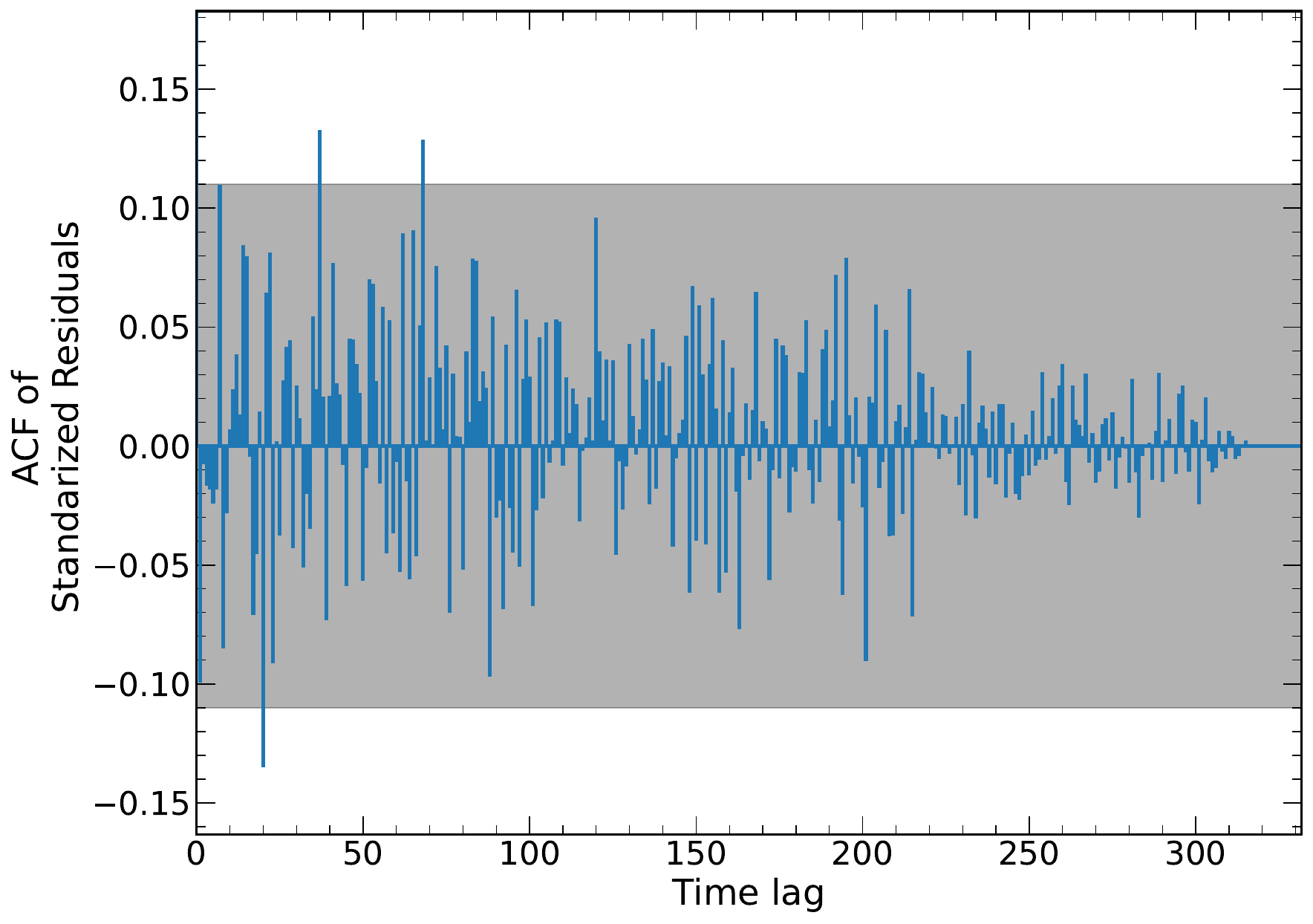}
    \includegraphics[width=0.49\textwidth]{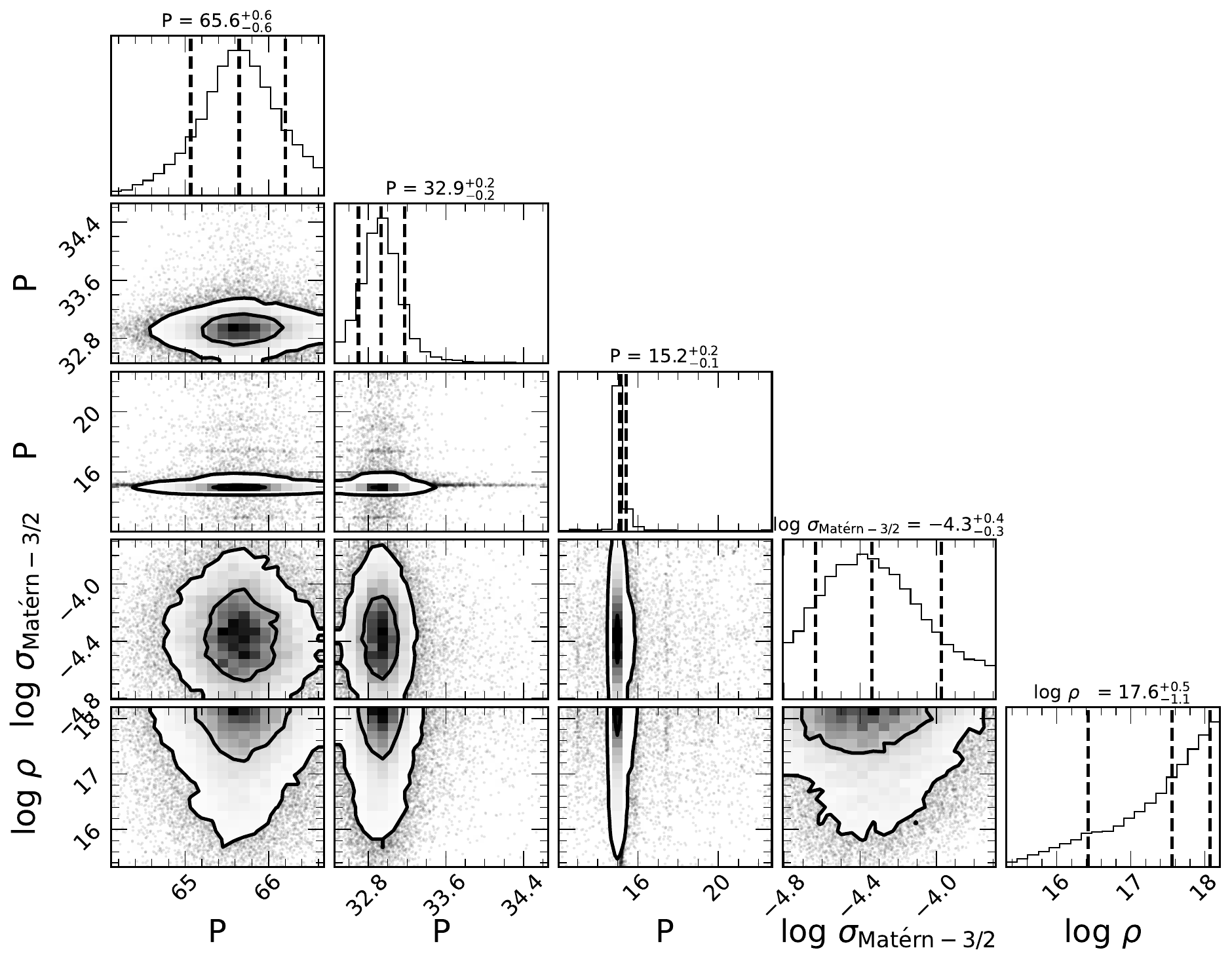}
    \caption{GP modelling results of the \swift-XRT data of the pulsating ULX NGC~7793~P13. (Top Left) Best-fit 3$\times$Lorentzian + Mat\'ern-3/2 + Jitter model to the \swift-XRT 0.3$-$10 keV lightcurve segment of NGC~7793~P13 shown in Figure~\ref{fig:xrt_P13}. (Top right) PSD of the best-fit model. (Bottom left) ACF of the standarized residuals of the best-fit model. (Bottom right) Posteriors of the periods of the three Lorentzians and the Mat\'ern-3/2 (the other parameters are omitted for clarity). The MCMC run for approximately 620,000 steps until convergence and about 124,000 were discarded for the burn-in. Symbols as per Figure~\ref{fig:ngc1365_posteriors}.}
    \label{fig:xrt_P13}
\end{figure*}

Table~\ref{tab:xrt_dbic} shows the $\Delta$AICc with the various models tested along with their $p$-values for the standarized residuals following a standard normal distribution. The best-fitting model is a combination of three Lorentzians + a Mat\'ern-3/2, where a Jitter term is needed as there is additional white-noise variability (jittering) that cannot be captured by any of the kernels. Indeed, we find that none of the models adequately describes the data based on the $p$-values of the residuals following a standard normal distribution. This may not be surprising as the distribution of count-rates is itself non Gaussian ($p = 0.008$ for rejecting a Gaussian distribution based on a KS test). Nevertheless, from Figure~\ref{fig:xrt_P13}, we can see that the failure to describe the data is mostly due to a few datapoints strongly deviating from the model. This is clearly seen in the ACF (Figure~\ref{fig:xrt_P13} bottom left panel), which confirms the lack of trends in the standarized residuals. Thus, while the model may not capture the full complexity of the data, we can at least ascertain that the variability is well represented by the combination of the three Lorentzians + Mat\'ern-32 + Jitter. The PSD of this models is shown in Figure~\ref{fig:xrt_P13} (top right panel). 

As with the UV data, we tested for the addition of the Lorentzians to the noise model in a hierarchical manner. The first Lorentzian with $P\sim65$ days
is found to be significant at the 99.99\% level ($>3\sigma$), the addition of a second is significant at 
the $\approx$99.2\% level and the third Lorentzian is marginally significant, at the $\sim$91.5\% level. 

\begin{table}
    \centering
    \begin{tabular}{lcccc}
    \hline
 Model & AICc & $\Delta$AICc & $p$-value  \\
 \hline
3$\times$Lorentzian + Mat\'ern-3/2 + Jitter        & --1673.4 & 0.0 & 0.001 \\ 
3$\times$Lorentzian + DRW + Jitter                 & --1669.2 & 4.2 & 0.001 \\ 
2$\times$Lorentzian + Mat\'ern-3/2 + Jitter        & --1664.5 & 8.9 & 0.003 \\ 
2$\times$Lorentzian + Mat\'ern-3/2 + DRW + Jitter  & --1662.2 & 11.5 & 0.006 \\ 
2$\times$Lorentzian + DRW + Jitter                 & --1661.2 & 12.2 & 0.003\\ 
2$\times$Lorentzian + DRW + Jitter + \granulation  & --1659.9 & 13.5 & 0.02 \\ 
2$\times$Lorentzian + 2$\times$DRW + Jitter        & --1657.1 & 16.6 & 0.01 \\
Lorentzian + DRW + \matern\ + Jitter               & --1656.1 & 17.3 & 0.003 \\
Lorentzian + DRW + Jitter                          & --1649.4 & 23.9 & 0.003 \\      
Lorentzian + 2$\times$DRW                          & --1647.4 & 25.9 & 0.004\\ 
Lorentzian + Mat\'ern-3/2 + Jitter                 & --1647.1 & 26.3 & 0.00 \\ 
3$\times$Lorentzian + 2$\times$DRW                 & --1644.8 & 28.5 & 0.00 \\ 
Lorentzian + DRW + Mat\'ern-3/2                    & --1641.8 & 31.6 & 0.02 \\ 
3$\times$Lorentzian + Mat\'ern-3/2                 & --1641.6 & 31.8 & 0.00 \\ 
2$\times$Lorentzian + 2$\times$Mat\'ern-3/2        & --1641.5 & 32.3 & 0.00 \\ 
2$\times$Lorentzian + 2$\times$DRW                 & --1640.8 & 32.5 & 0.001 \\ 
3$\times$Lorentzian + \granulation                 & -1639.0 & 34.4  & 0.00  \\ 
3$\times$Lorentzian + 3$\times$DRW                 & --1638.7 & 34.6 & 0.00 \\ 
3$\times$Lorentzian + 2$\times$Mat\'ern-3/2        & --1637.4 & 36.0 & 0.00 \\ 
3$\times$Lorentzian + DRW                          & --1636.9 & 36.8 & 0.00 \\ 
2$\times$Lorentzian + DRW + Mat\'ern-3/2           & --1636.7 & 37.1 & 0.00 \\ 
3$\times$Lorentzian + 3$\times$Mat\'ern-3/2        & --1632.3 & 41.4 & 0.00 \\ 
2$\times$Lorentzian + DRW                          & --1631.1 & 42.7 & 0.00 \\ 
 Lorentzian + DRW                                  & --1627.7 & 46.0 & 0.00 \\ 
 Lorentzian + Jitter                               & --1623.3 & 50.1 & 0.00 \\  
 Mat\'ern-3/2 + Jitter                             & --1619.3 & 54.5 & 0.00 \\ 
 Lorentzian + \granulation + Jitter                & --1619.2 & 54.2 & 0.00 \\ 
 3$\times$Lorentzian + \granulation + Jitter       & --1618.7 & 55.1 & 0.00 \\ 
Lorentzian + \matern\                              & -1617.1 & 56.3  & 0.00 \\
 Lorentzian + \granulation                         & --1611.6 & 61.8 & 0.002 \\ 
 DRW + Jitter                                      & --1607.9 & 65.9 & 0.001 \\ 
 Lorentzian                                        & --1602.3 & 71.4 & 0.00  \\ 
 DRW                                               & --1594.2 & 79.5 & 0.001\\ 
 Mat\'ern-3/2                                      & --1590.5 & 83.2 & 0.00  \\ 
 \granulation                                      & --1588.4 & 85.4 & 0.00  \\  
\hline
\hline
\end{tabular}
\caption{As per Table~\ref{tab:ngc1365} but now showing the AICc, $\Delta$AICc and $p$-values for the standarised residuals following a Gaussian distribution for the different models tested against the \swift-XRT data of the pulsating ULX NGC7793~P13.}
\label{tab:xrt_dbic}
\end{table}

From Figure~\ref{fig:xrt_P13} our final estimate for the period is $P = 65.6\pm0.6$ days. As for the UV data, the high coherence ($Q \gtrsim$ 100) suggests the periodicity is stable throughout the segment.
\subsubsection{NGC~4945} \label{sub:agn_qpo}

\citet{smith_qpo_2020} claimed a $\sim$42-day QPO in the irregularly sampled \textit{RXTE} data of the Type 2 Seyfert, NGC~4945 with a significance of 10.2$\sigma$\footnote{Note that the authors also quote a false-alarm probability of 2.87\%, which corresponds to $\sim2.2\sigma$ only.}. 

\begin{figure*}
    \centering
     \includegraphics[width=0.49\textwidth]{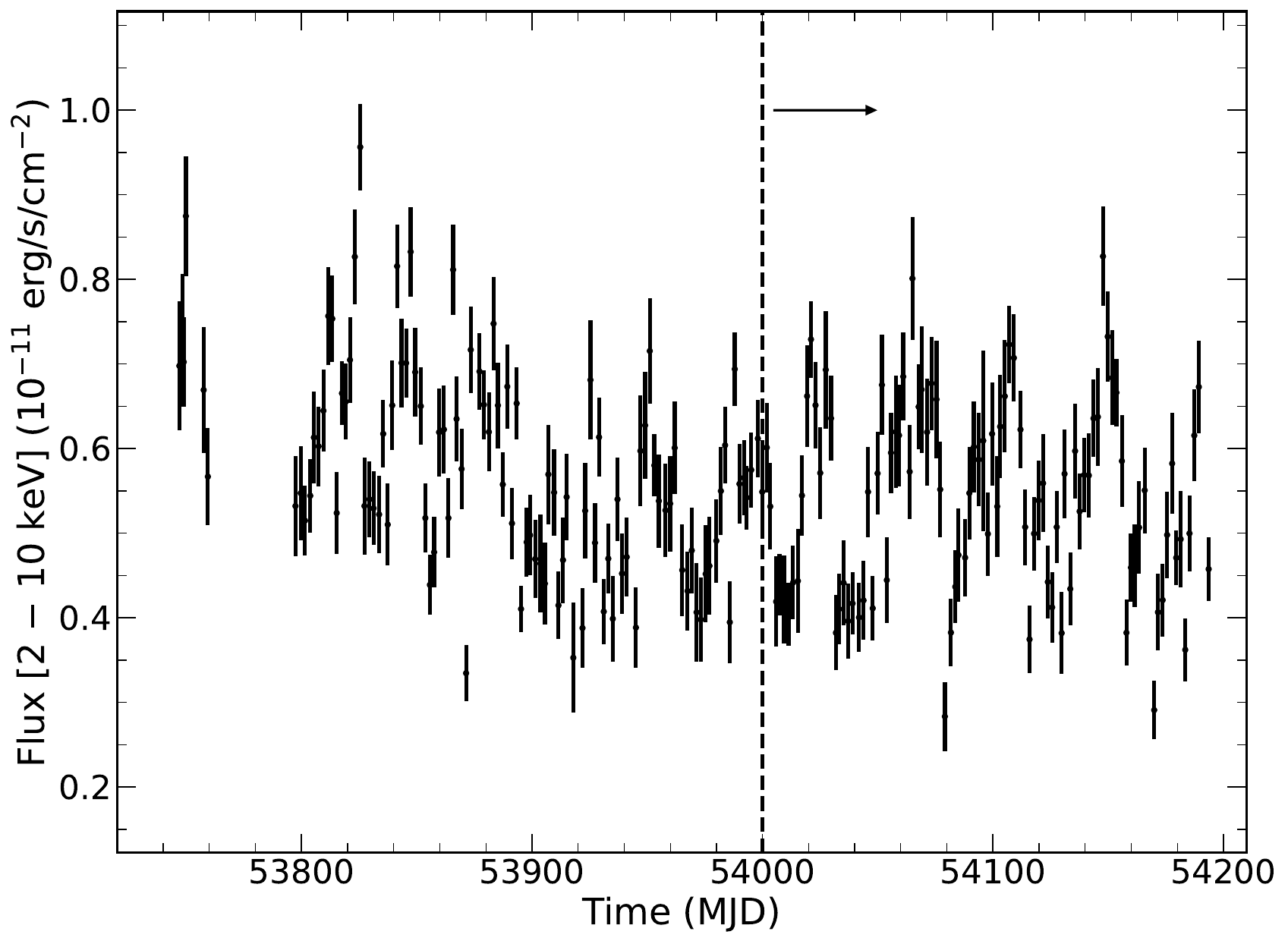}
     \includegraphics[width=0.49\textwidth]{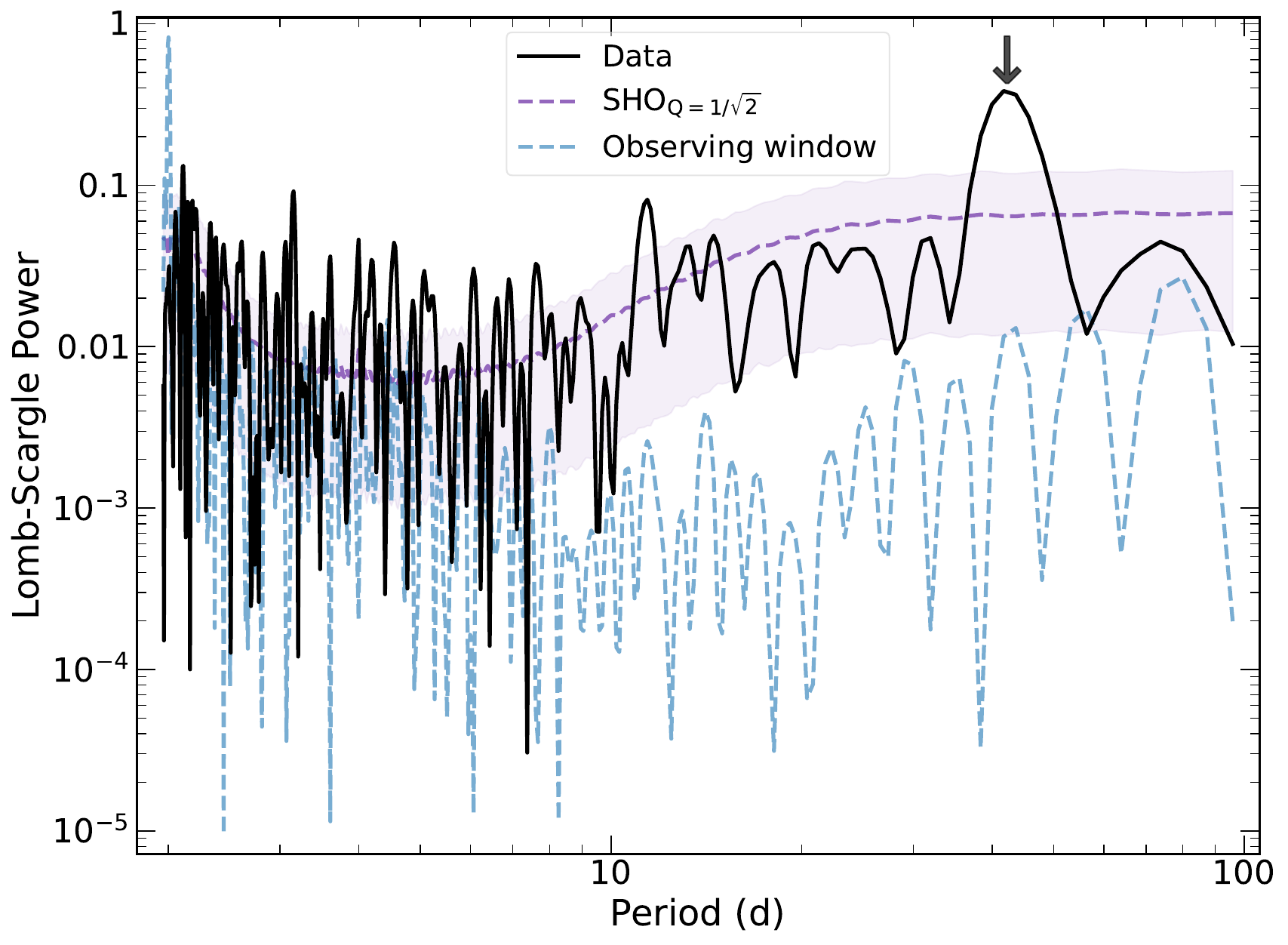}
    \caption{(Left) RXTE lightcurve of the AGN NGC~4945. The segment where \citet{smith_qpo_2020} reported the significance of the QPO to be strongest is highlighted with a dashed line and an arrow. (Right) Lomb-Scargle periodogram of the segment indicated in the right panel. The vertical black arrow shows the QPO reported by \citet{smith_qpo_2020}. The dashed blue line shows the power spectrum of the observing window. The dashed purple line shows the model derived from periodograms of 10,000 lightcurves generated from the posteriors of the \granulation\ model parameters (best--fit model), with the shaded areas indicating the 16\% and 84\% percentiles of the distribution. See text for details.}
    \label{fig:NGC4945_segment}
\end{figure*}

Following \citet{smith_qpo_2020}, we obtained the RXTE/PCA data from the University of California archive\footnote{https://cass.ucsd.edu/~rxteagn/} (for details regarding data filtering criteria, we refer the reader to their website). Figure~\ref{fig:NGC4945_segment} shows a segment of the lightcurve where the monitoring was densest \citep[cf. Figure~2 in][]{smith_qpo_2020}. The full lightcurve spans 442 days, with a median cadence of 2.25 days. The authors found the significance of the QPO to be the strongest in the segment towards the end of the lightcurve after the vertical dashed line in Figure~\ref{fig:NGC4945_segment}. This segment spans 192 days with a mean cadence of 2.04 days.

\begin{figure*}
    \centering
    \includegraphics[width=0.49\textwidth]{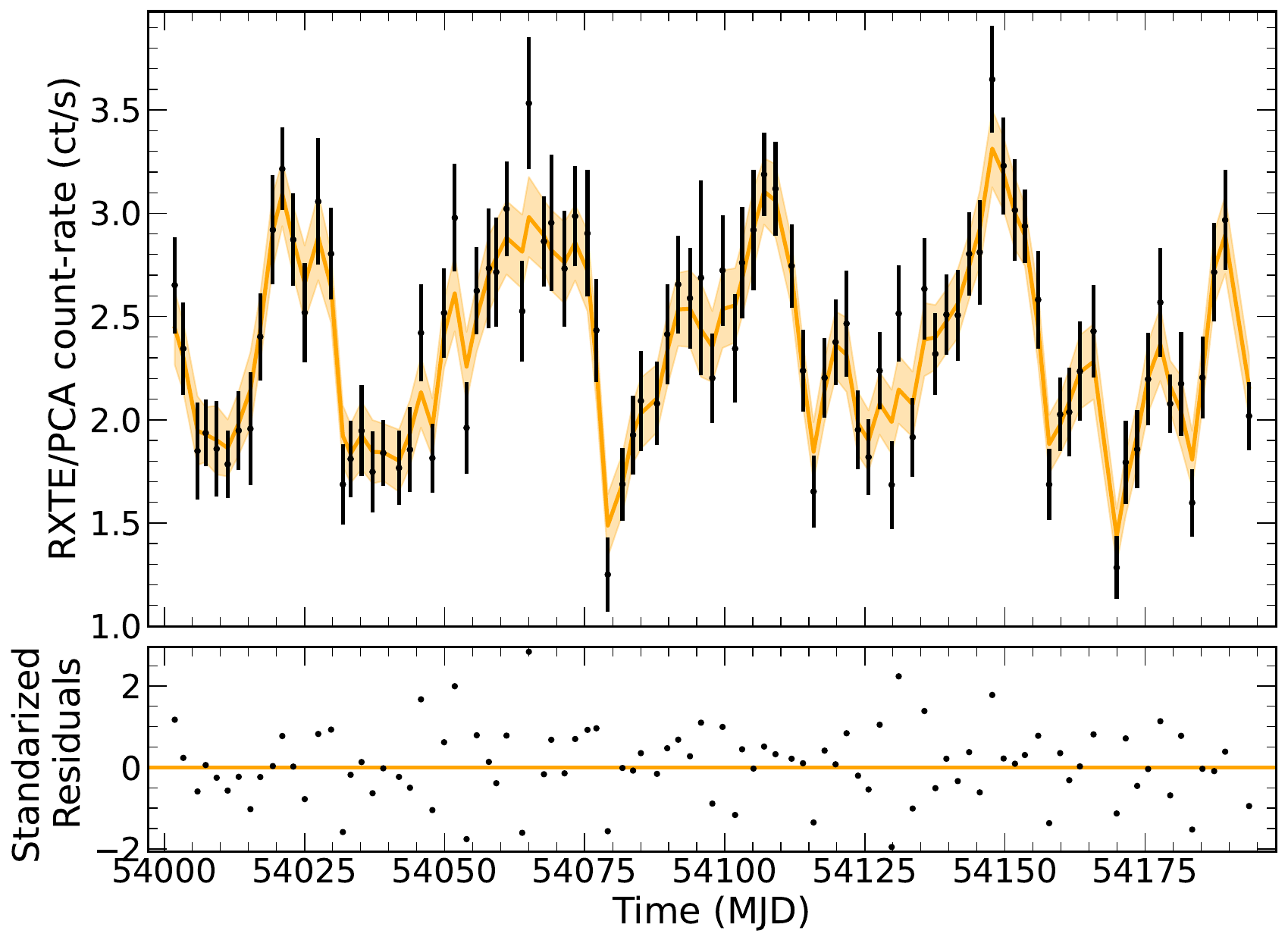}
    \includegraphics[width=0.49\textwidth]{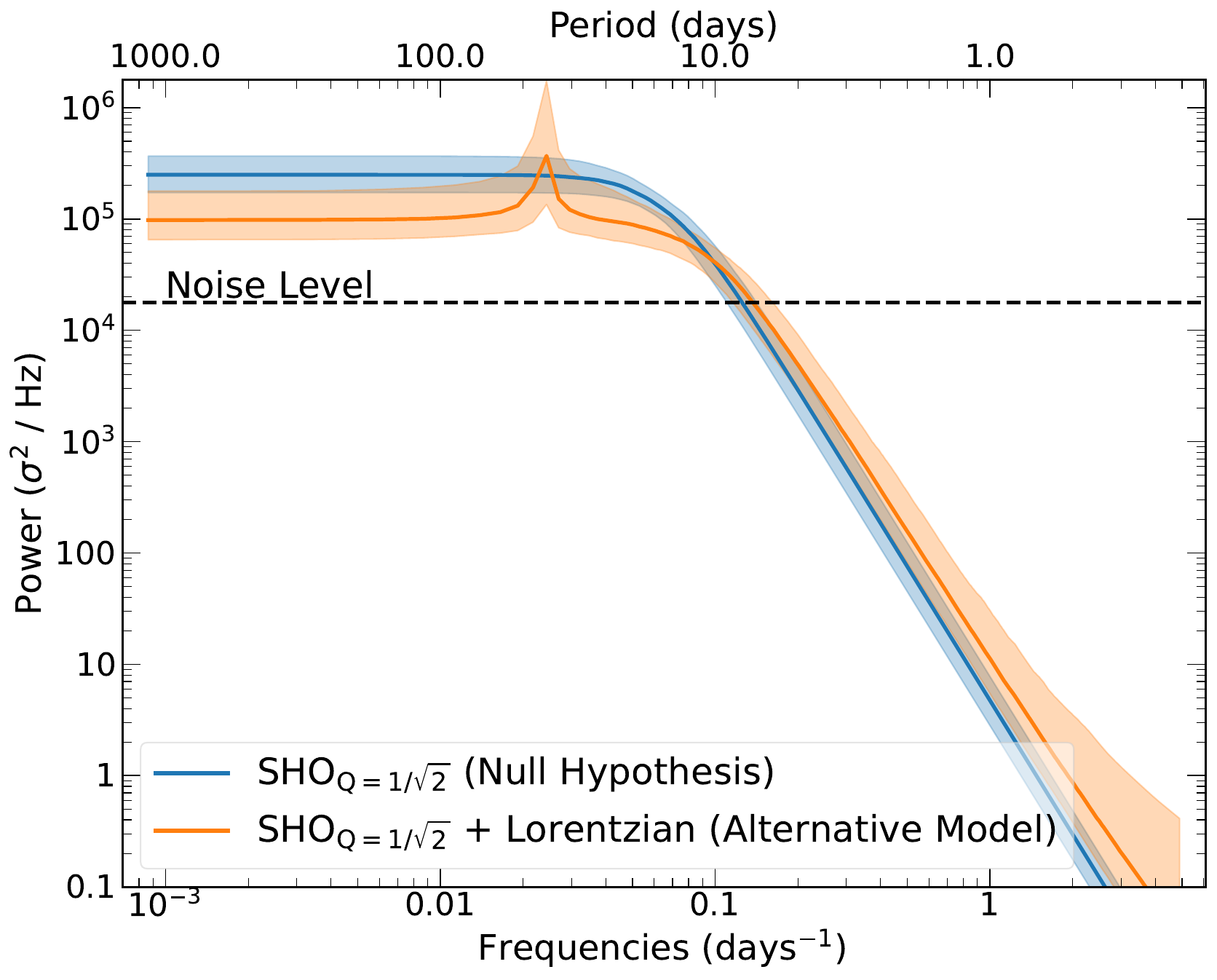}
    \includegraphics[width=0.49\textwidth]{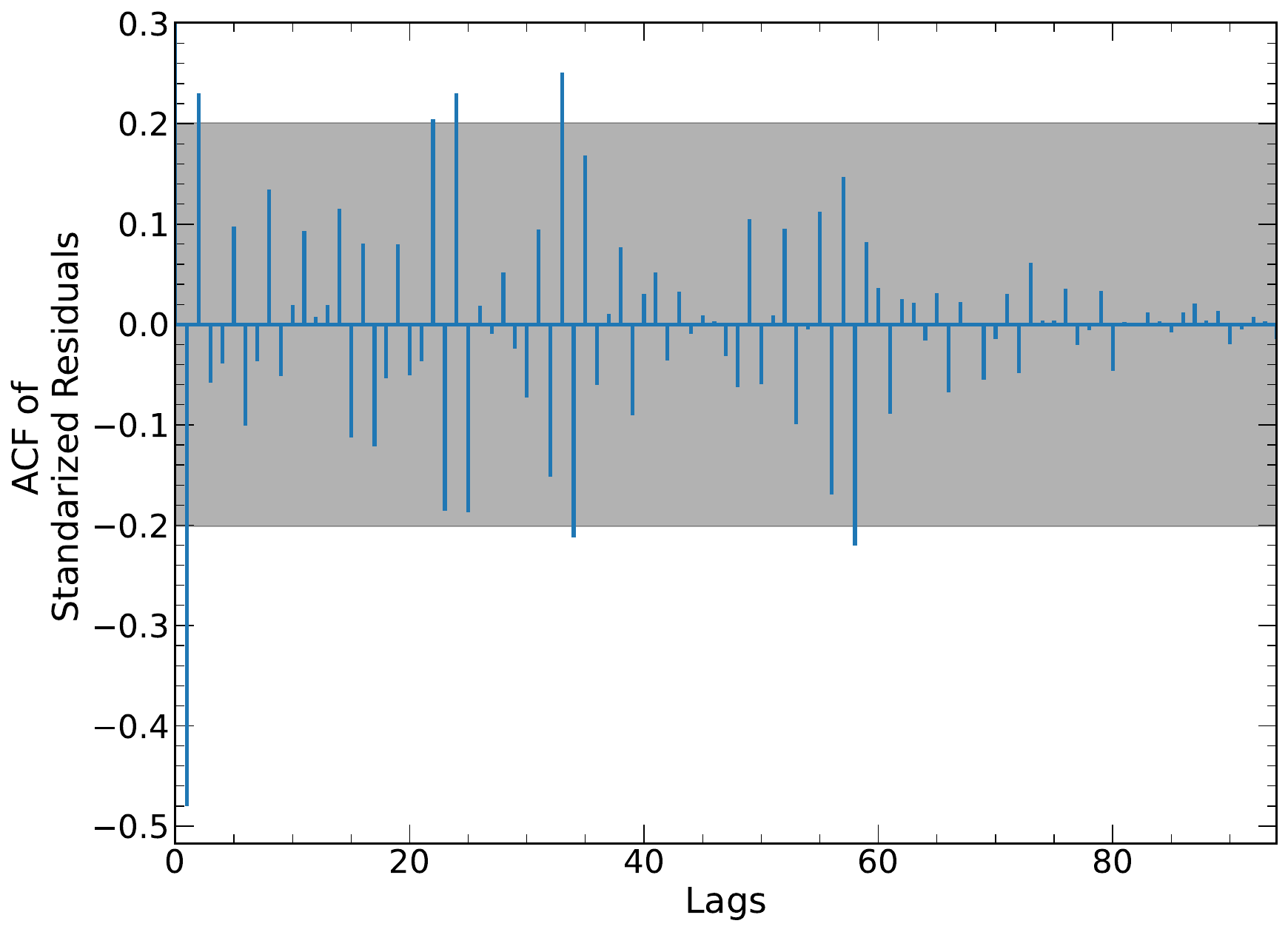}
     \includegraphics[width=0.49\textwidth]{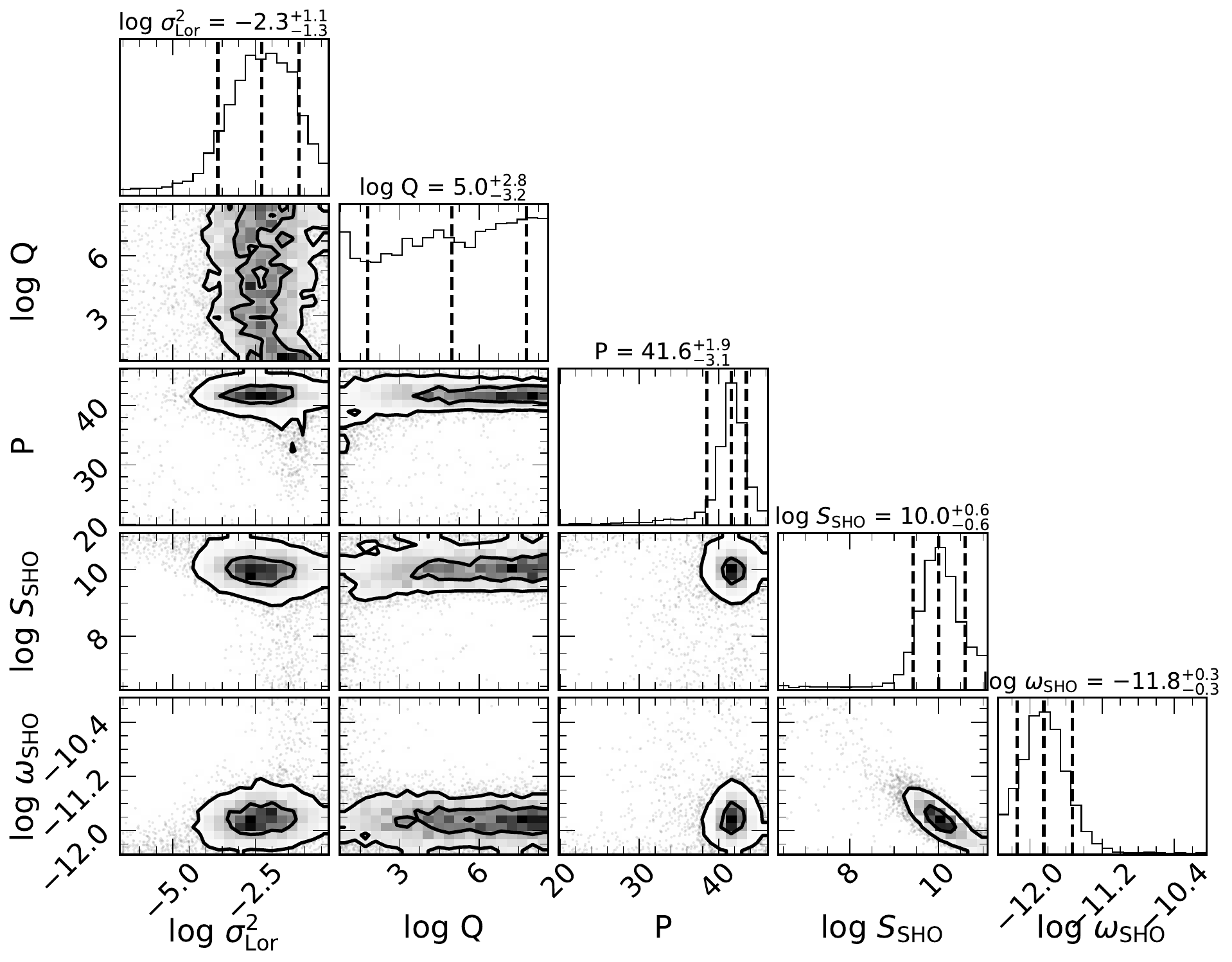}
    \caption{GP modelling results of the RXTE lightcurve segment shown in Figure~\ref{fig:NGC4945_segment} of the AGN NGC~4945. (Top left) Best-fitting Lorentzian + \granulation\ model. (Top right) PSDs of the null hypothesis and alternative models. (Bottom left) ACF of the standarized residuals of the Lorentzian + \granulation\ model. (Bottom right) Posterior parameters for the Lorentzian + \granulation\ model. The MCMC sampler run for approximately 135,000 steps until convergence and about 35,000 steps were discarded for the burn-in. Symbols as per Figure~\ref{fig:ngc1365_posteriors}.}
    \label{fig:NGC4945}
\end{figure*}

\begin{table*} 
 \centering 
\begin{tabular}{lccclccc} 
 \hline 
\multicolumn{4}{c}{Full lightcurve}  & \multicolumn{4}{c}{\citet{smith_qpo_2020}}\\
\cline{1-4} \cline{5-8}
\noalign{\smallskip}
 Model & AICc & $\Delta$AICc & $p$-value & Model & AICc & $\Delta$AICc & $p$-value\\
 \noalign{\smallskip}
 \hline \hline
 \noalign{\smallskip}
DRW                    & 246.1 & 0.00 & 0.00   &  Lorentzian + \granulation       & 97.9  & 0.0 & 0.22 \\
DRW + Jitter           & 247.2 & 1.1  & 0.00    & Lorentzian + Mat\'ern-3/2        & 98.3  & 0.4 & 0.07 \\
DRW + \granulation\    & 248.5 & 2.4  & 0.00    & Lorentzian + DRW                 & 99.2  & 1.3 & 0.03 \\ 
DRW + \matern          & 248.8 & 2.7  & 0.00    & Lorentzian + \granulation\ + Jitter & 99.5 & 1.6 & 0.22 \\
Lorentzian + DRW       & 248.8 & 2.7  & 0.00    & 2$\times$Lorentzian & 101.0 & 3.1 & 0.42 \\
2$\times$DRW           & 249.5 & 3.4  & 0.00    & Lorentzian + DRW + Jitter & 101.6 & 3.7 & 0.03 \\
Lorentzian + 2$\times$DRW & 250.2 & 4.1 & 0.00 & Lorentzian + \granulation\ + DRW & 102.0 & 4.1 & 0.12 \\ 
Lorentzian + Jitter    & 251.6 & 5.5  & 0.03   & Lorentzian + 2$\times$\granulation & 102.4 & 4.4 & 0.08 \\ 
Lorentzian + \granulation& 253.7 & 7.6 & 0.00 & Lorentzian + \granulation + \matern & 102.9 & 5.0 & 0.16 \\ 
Lorentzian + \matern & 261.6 & 11.4 & 0.00 & Lorentzian + \matern\ + DRW & 102.9 & 5.0 & 0.12 \\ 
\matern                &   255.2 & 9.1 & 0.002 &  \matern   & 103.8 & 5.9 & 0.14 \\
\granulation           &   264.9 & 18.8 & 0.02 &  DRW & 104.5 & 6.6 & 0.03       \\
Lorentzian             & 267.9   & 21.8 & 0.03 & Lorentzian  & 104.6 & 6.7 & 0.22 \\
Jitter                 & 310.4  & 64.3 & 0.09  & \granulation  & 104.9 & 7.0 & 0.33 \\   
 \hline \hline 
 \end{tabular} 
 \caption{As per Table~\ref{tab:ngc1365} but now showing the AICc, $\Delta$AICc and $p$-values for the standarised residuals following a Gaussian distribution for the different models tested against the RXTE data of the AGN NGC4945. The left and right values are for the analysis of the full lightcurve and the segment shown in Figure~\ref{fig:NGC4945_segment}, respectively. }
    \label{tab:agn_dbic}
 \end{table*}

We first focus on the analysis of the full lightcurve. Table~\ref{tab:agn_dbic} shows the $\Delta$AICc for the set of models explored, with the DRW yielding the lowest AICc (with $\Delta$AICc = 2.7 over the Lorentzian + DRW). This already suggests the data can be explained under a simpler, stochastic model. 
Performing lightcurve simulations from the DRW posteriors, we find a significance of $\approx$39\% for the Lorentzian component, indicating the addition of the Lorentzian is not supported by the data. However, we note the residuals in all models are narrower than a standard normal distribution ($\sigma = 0.67$), indicating the variability is not well described by a GP. The downward trend in flux around MJD 53,900 may indicate the process is non-stationary over the timescales analysed here. 

We proceed to focus on the segment indicated to the right-hand side of the vertical dashed line in Figure~\ref{fig:NGC4945_segment}, where the authors claimed the QPO significance to be highest. Table~\ref{tab:agn_dbic} lists the models tested against the data in this segment. In this case we find potential evidence for a periodic component, as a model including a Lorentzian (the broadband noise modelled with an \granulation) provides the lowest AICc. The standardised residuals and their ACF are shown in Figure~\ref{fig:NGC4945}. The standardized residuals are compatible with Gaussianity and, therefore, the assumption of a GP is reasonable. We note however the ACF indicates there is still slight variability not captured by the model
potentially indicating that models outside \texttt{celerite} might be more appropriate. Nevertheless, most of the variability is reasonably captured. The central frequency of the Lorentzian is found to be $P = 42_{-3}^{+2}$ days which matches the periodicity reported by \citet{smith_qpo_2020}. 

Using the posteriors from the best-fitting \granulation\ as the null hypothesis, we obtained the significance of the (quasi)-periodic component. To estimate the background contribution for our lightcurve simulations, we assumed the mean source rate was 5\% of the background rate. Although in the average spectra the source contributed 10\% to the total rate, 5\% is both consistent with previous work (\citealt{done_simultaneous_2003}) and we found the simulated lightcurve errorbars matched more closely the data errorbars. As we are assuming a higher background than in the average spectrum, our simulations will be less likely to generate a spurious signal (they will have increased levels of white noise) and the estimated QPO significance will tend to overestimate the true significance, if anything.  

\begin{figure}
    \centering
     \includegraphics[width=0.48\textwidth]{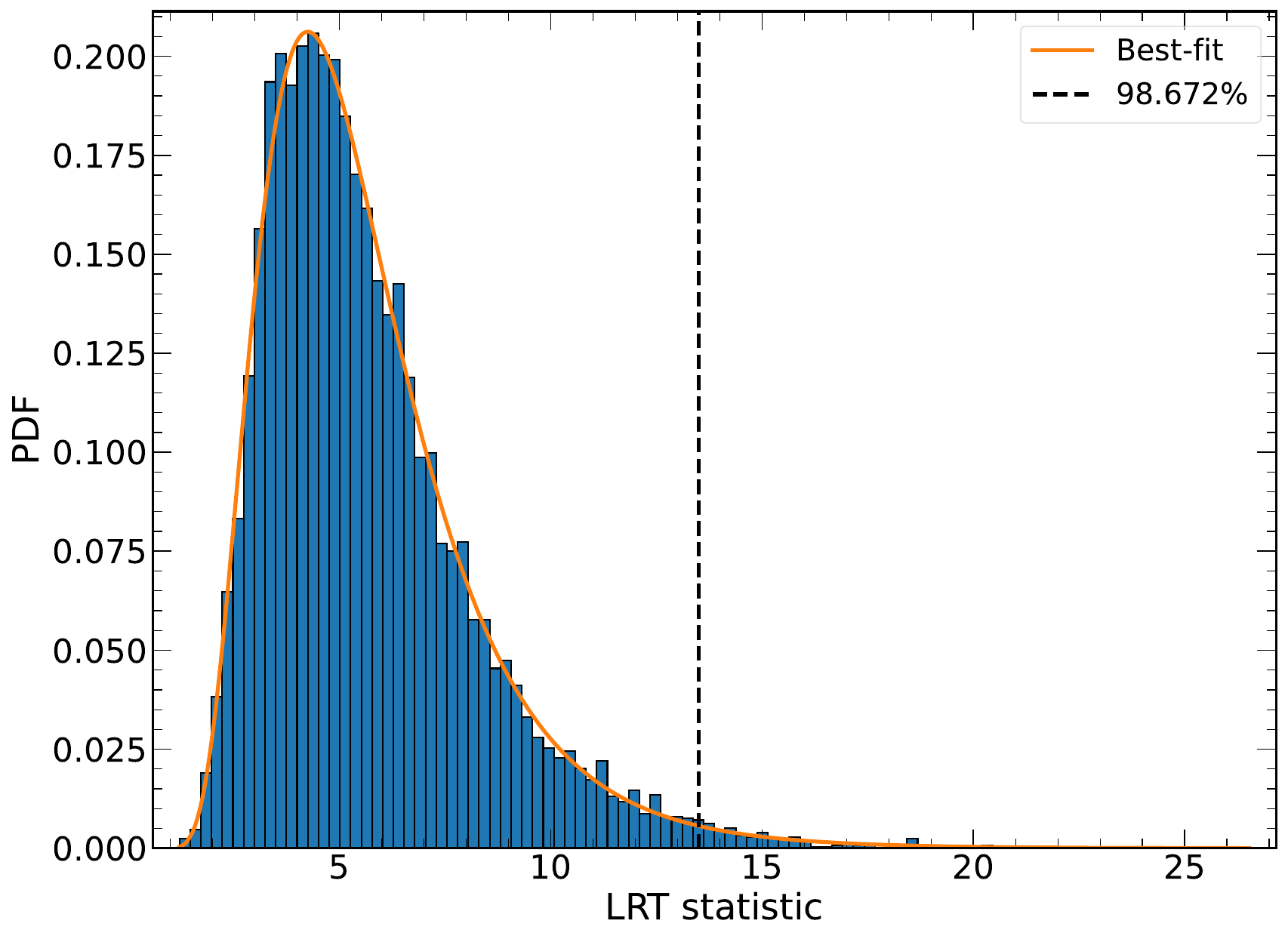}
    \caption{LRT distribution generated from simulated lightcurves from the posteriors of the \granulation\ model (null hypothesis) for NGC~4945. The solid orange line shows a fit to the distribution using a log-normal. The LRT observed in the data is shown as a dashed black line.}
    \label{fig:NGC4945_LRT}
\end{figure}

Figure~\ref{fig:NGC4945_LRT} shows the reference LRT distribution derived from the posteriors of the \granulation\ model. The putative periodicity has a significance of $\sim$98.7\% (i.e. $\approx 2.5\sigma$), which is indeed quite high, but it may not be considered sufficient to claim a detection.


Figure~\ref{fig:NGC4945_segment} shows the mean Lomb-Scargle periodogram of 10,000 lightcurves simulated using the posteriors of the \granulation\ model. Using the process outlined in Section~\ref{sub:goodness} to map the $L_\text{max}^\text{obs}$ to a goodness-of-fit, we find a $p$-value of 0.9 using the \granulation\ model, a deviation of $\sim$1.7$\sigma$ from the mean, indicating the fit is an acceptable description of the data.

Given that the \matern\ provides the lowest AICc when used in isolation (Table~\ref{tab:agn_dbic}), one could argue it represents the best null-hypothesis. Moreover, as discussed in Section~\ref{sub:model_selection}, based on the low $\Delta$AICc $=0.4$ between the Lorentzian + \granulation\ and the Lorentzian + \matern\ models, it may argued these two models offer similar levels of goodness of fit (i.e. we cannot distinguish between the two with the data at hand). We therefore repeated the significance calculation with the posteriors from the \matern-only model, and found a similar value for the the significance ($\sim$96\%). This is consistent with the lower $\Delta$AICc provided by this model when the Lorentzian is added compared to the \granulation\ model (Table~\ref{tab:agn_dbic}) and indicates our results are not strongly dependent on the continuum choice (so as long as it is representative of the data).
\subsubsection{The Blazar B0537-441} \label{sub:qso}
\input{qso}

\subsection{Analysis of the results}\label{sub:analysis_results}

As we have shown, the methodology outlined here is particularly suited for the analysis of irregularly-sampled time series commonly associated with monitoring of systems such as ULXs or AGN \citep[e.g.][]{uttley_measuring_2002}, but its applicability is not restricted to irregularly-sampled time series (Section~\ref{sub:xmm_seyfert}). We have shown its application to the short ($\sim$50 ks) nearly regularly-sampled time series that may be obtained with observatories such as \xmm\ or \nicer, by analysing the QPO recently claimed by \citet{yan_x-ray_2024} in the Seyfert galaxy NGC 1365. These authors found a significance of about 3.6$\sigma$ by comparing periodogram peaks of lightcurves simulated from the continuum-fitted PSD. Using our method we have found instead a much lower significance, of about 1.7$\sigma$ (or 91\%). From their analysis, it is unclear where this discrepancy in the estimate of the significance originates. \citet{yan_x-ray_2024} report fitting the PSD with a bending powerlaw, and then use this model to produce lightcurve simulations to test the significance of the highest peak in the periodogram. However, there is no information regarding the fitting procedure, namely the statistic used to fit the periodogram and whether the appropriateness of the model was taken into account. 
It is also unclear whether the uncertainties on the model were taken into account in the estimation of the false-alarm probability and how the number of trials were considered. It is likely that a combination of these factors can explain the difference in our results. 

We have then applied our method to the ULX in NGC 7793 P13, where slightly dissimilar superorbital periods had been claimed in the sparsely-sampled \swift-UVOT and XRT lightcurves \citep{hu_swift_2017, furst_tale_2018}. As can be seen from Figure~\ref{fig:xrt_P13}, we have found the X-ray period to be 65.6$\pm$0.6 days, which is indeed significantly longer than the period in the UV ($P = 63.9\pm0.4$ days; Figure~\ref{fig:uvot_P13}). As stated above, using time domain methods allows to marginalise over the noise components and obtain accurate uncertainties on the parameters of the periodic component. Thus, we can support earlier assertions that the X-ray period is significantly longer \citep{hu_swift_2017, furst_tale_2018} than the optical/UV period. The high coherence $Q\gtrsim300$ from the Lorentzian components suggest the period amplitude is stable throughout the segment, consistent with the long-term behaviour of the source \citep{furst_long-term_2021}. Regarding the significance of the periodicities, while the third harmonics were marginally significant (at $\sim$95\% and 91\% for the UV and X-ray lightcurves) given that these constitute the harmonics of the same periodicity, these significances likely underestimate the true significance as one could repeat the analysis tying the periods or considering the combined fit improvement provided by the three Lorentzians altogether, but this is beyond the scope of this work.

We have also examined the putative QPO claimed by \citet{smith_qpo_2020} in the AGN NGC 4945. When analysing the whole lightcurve, we have seen that we could not explain the data under a GP, which could indicate deviations from stationarity. Indeed, when analysing the last portion of the whole lightcurve (Figure~\ref{fig:NGC4945_segment}) we have seen the preferred rednoise kernel (\granulation) differed from that obtained for the full lightcurve (DRW), which could support the non-stationarity of the process. Nevertheless, our analysis suggests there is little evidence for periodic variability when analysing the lightcurve as a whole.
In the segment where \citet{smith_qpo_2020} reported the significance of the QPO to be the highest (Figure~\ref{fig:NGC4945_segment} left panel), we have found the putative periodicity to have a significance of
$\sim$98.7\% (i.e. $\approx 2.5\sigma$), much lower than reported by \citet{smith_qpo_2020}. The fact that we are able to produce simulations with comparable fit-improvements as that observed in the data implies our test is well-calibrated and suggests our significance estimate is more plausible. The most obvious discrepancy is that we have correctly accounted for the presence of rednoise. Instead, \citet{smith_qpo_2020} relied on the analytical recipe provided by \citet{horne_prescription_1986}, which may be appropriate in cases where employing white noise as the null hypothesis -- but see \citet{frescura_significance_2008} for caveats on this method.

We note the significances quoted for NGC 4945 may be considered optimistic, as the selection of this segment seems driven by 'a posteriori' analysis of the data \citep[a form of the stopping rule discussed in][]{vaughan_bayesian_2010}, rather than a data-driven decision (such as to avoid a gap in the lightcurve, e.g. Section~\ref{sub:ngc7793}). While this is beyond the scope of this work, one could in principle account for this by simulating lightcurves using the full length of the monitoring (either Figure~\ref{fig:NGC4945} or the entire RXTE history) and then selecting the segment that maximizes the likelihood ratio for each simulation. 

Finally, we have examined the QPO claimed in the Blazar 0537-441 by \citet{tripathi_search_2024} using TESS data. While our analysis supports the presence of a QPO-like feature (at the $\sim$3.7$\sigma$), the identified period is marginally consistent with the 6.5 day QPO reported by \citet{tripathi_search_2024}, although no uncertainties on the claimed periodicity are provided by \citet{tripathi_search_2024}. Differences in our results may be attributed to the different processing of the data and treatment of the underlying noise.
\section{Discussion \& Conclusions} \label{sec:discussion}
The presence of red noise variability, ubiquitously found in accreting systems, makes the detection of periodicities challenging. On the one hand, most periodicity tests are derived for cases of Gaussian white noise, which makes the problem analytically tractable \citep[e.g.][]{scargle_studies_1982}. On the other hand, the presence of red noise increases the likelihood of producing spurious features in the periodogram, particularly because the scatter in the power is proportional to the power itself \citep[e.g.][]{vaughan_simple_2005}. When the data is unevenly sampled, the problem becomes even more profound as stochastic variability can easily be mistaken for periodic behaviour \citep[cf.][]{vaughan_false_2016}.

Extrapolation of tests for periodicities against red noise-like variability were presented in \citet{israel_new_1996} and \citet{vaughan_simple_2005}, who proposed to capture the underlying broadband noise using either a parametric \citep[restricted to PSDs  following a powerlaw;][]{vaughan_simple_2005} or non-parameteric approach \citep{israel_new_1996}, and use these estimates and associated uncertainties to derive the probability of obtaining a spurious signal in the periodogram above a certain level. \citet{vaughan_bayesian_2010} expanded on previous work to model any arbitrary PSD shape using a Bayesian approach which allowed for the inclusions of priors. All these techniques concerned the case where the time series is evenly sampled such that (as discussed in Section~\ref{sec:method}) the periodogram has some well-known statistical properties which allows a well-defined likelihood (and other statistical tests such as goodness-of-fit) to be defined.

Here we have provided a method for periodicity searches in the case of unevenly-sampled data, where constraining the aperiodic variability is considerably more challenging and where it appears preferable to perform the fitting in the time domain (where the probability distribution is known and is generally Gaussian/Poissonian) using GP modelling. Here we have exploited the known likelihood with well-established statistical techniques \citep{protassov_statistics_2002} to estimate the significance of a putative (quasi)periodic component. In a similar manner to the regularly-sampled case, the noise is inferred from the data, allowing a test for the presence of an additional component (e.g. a QPO) by building an empirical \tlrt\ distribution using the method proposed by \cite{protassov_statistics_2002}. Given that the method is entirely data-driven, it is completely generalizable to any system/variability and even choice of mean function (which we haven not exploited here).

If the PSD is of interest, this quantity can be accessed by Fourier-transforming the best-fit GP kernel, rather than the data itself, thereby including the data (heteroscedastic) uncertainties in the final estimate. In doing so, frequency-distorting effects arising from irregular sampling are mitigated, while the data usage is maximized. There is additionally no requirement to rebin the data (so as long as there are enough counts for the data to be Gaussian distributed). A similar approach is discussed in \citet{kelly_flexible_2014} using \textsc{CARMA}, who also advocates for time-domain fitting. The recipe outlined in this work may equally be used employing \textsc{CARMA} kernels and will suffer from the same limitations we discuss below in Section~\ref{sub:limitations}. However, there seem to be certain advantages of using \texttt{celerite} over \textsc{CARMA}. \texttt{celerite} kernels have a more flexible form than \textsc{CARMA} ones \citep{foreman-mackey_fast_2017}. While the PSD of \textsc{CARMA} kernels are restricted to Lorentzian functions, steeper PSDs may be achieved using a single \texttt{celerite} kernel (the \granulation\ kernel being an example used here; Figure~\ref{fig:celerite_models}), which in \textsc{CARMA} may not be straightforward to describe. From a computational point of view, in principle the computational is the same for both \texttt{celerite} and \texttt{carma} implementations \citet{kelly_flexible_2014, foreman-mackey_fast_2017}, scaling as $\mathcal{O}$($NJ^2$), However, \citet{foreman-mackey_fast_2017} showed that in practice \texttt{celerite} seems to perform better computationally.


There are several important advantages of using GP over Lomb-Scargle periodograms. As shown in Section~\ref{sub:ngc7793}, we can not only access more accurately the underlying noise by performing model selection, but also marginalise over the noise parameters, therefore carrying over the full set of uncertainties into our determination of a candidate period's frequency. Instead, both model selection and uncertainties are inaccessible when using Lomb-Scargle periodograms.
\subsection{Limitations and caveats} \label{sub:limitations}
Regardless of its power and improvement over traditional approaches, there remain several limitations of our method, arguably the most pressing being the computational time involved. The computational time of the GP modelling itself scales as $N^3$, which can become intractable if several models need to be tested or for large datasets. Here we have chosen to minimise the compute time using \celerite\ (where the computational time scales as $NJ^2$) at the expense of flexibility, which may not be much more computational expensive than the Lomb-Scargle periodogram.\footnote{The fastest implementation in \texttt{astropy} scales as $\mathcal{O}$($N$log($M$)) where $M$ is the number of frequencies being evaluated.} In addition to the model evaluation, there is the computational time required to perform the simulations for hypothesis testing. This problem is partially mitigated because the likelihood allows us to perform initial model selection (in our case through the $\Delta$AICc) and filter out the most prominent cases. 
Therefore, only in cases with limited signal-to-noise or where the $\Delta$AICc does not provide sufficient indication \citep[e.g.][]{graham_possible_2015}, lightcurve simulations may need to be performed, although having to rely on simulations for hypothesis testing equally applies to regularly-sampled time series \citep[e.g.][]{ashton_searching_2021}. 

Another common drawback of GP modelling is how to choose what kernels to test against the data. A straightforward approach to alleviate this problem is to simply stack basis functions until the minimum of the IC is found \citep{kelly_flexible_2014, foreman-mackey_fast_2017, zhang_search_2023}. Secondly, inspection of the standarized residuals can reveal trends indicative of the model not capturing the full variability (as also illustrated by the ACF).



We have also discussed how to identify cases where the GP might not be a good fit to the data (Section~\ref{sub:agn_qpo}) either due to the process not being a GP or due to the assumption of stationarity not being fullfilled. In the former case, it is still unclear whether our method is still valid. Through our simulations (Section~\ref{sec:lightcurve_simulations}), we have noted that when the lightcurves are produced using a lognormal PDF, the standarized residuals never show compliance with a standard Gaussian distribution, even if the input model parameters are well captured (see Section~\ref{sec:gp_lognormal}). Thus, preliminary tests indicate that the variability is still well-captured even when the flux distribution is not Gaussian. 


Note also that while the assumption of stationarity is another limitation of GPs, the same assumption is inherently made in standard periodograms. In fact, GPs are also more flexible on this regard, as the mean of the time series does not need to be constant. In any instance, in a similar vein as for dynamical periodograms \citep{kotze_characterizing_2012}, one could envision splitting the time series into approximately stationary segments and applying an independent GP modelling to each segment. Then the posteriors of a particular parameter (e.g. the period frequency $P$) could be examined to discern whether a given quantity is varying over the full observation baseline.

Lastly, compared to periodogram fitting, where any functional form may be employed, the fitting process in GP is restricted by the functional form of the kernels. This latter problem may be alleviated at the expense of computational cost, by using kernels outside \texttt{celerite} \citep[e.g.][]{rasmussen_gaussian_2006}. 






\section*{Acknowledgements}
This work made use of data supplied by the UK Swift Science Data Centre at the University of Leicester. This paper includes data collected by the TESS mission. Funding for the TESS mission is provided by the NASA's Science Mission Directorate. This work has made use of lightcurves provided by the University of California, San Diego Center for Astrophysics and Space Sciences, X-ray Group (R.E. Rothschild, A.G. Markowitz, E.S. Rivers, and B.A. McKim), obtained at \url{http://cass.ucsd.edu/\textasciitilde rxteagn/}.
The authors acknowledge support by STFC through grant ST/V001000/1 and the use of the IRIDIS High Performance Computing Facility, and associated support services at the University of Southampton. We would like to thank the referee for their interest in our manuscript and thoughtful comments that helped improve it. A. G\'urpide is also grateful to S. Vaughan, Z. Irving, E. Agol, W. Alston for stimulating discussion on time domain analysis and Gaussian Processes modelling. Software: \texttt{corner} \citep{foreman-mackey_cornerpy_2016}, \texttt{nifty-ls} \citep{garrison_nifty-ls_2024}, \texttt{mind\_the\_gaps} \citep{andres_gurpide_lasheras_2024_14253754}. 

\section*{Data Availability}
All the data used in this paper is publicly available in the corresponding archives. The code used for this manuscript has also been made publicly available.



\bibliographystyle{mnras}
\bibliography{mind_the_gaps} 
\appendix
\input{appendix}

\bsp	
\label{lastpage}
\end{document}

%% file: ngc1365.tex
Using \xmm\ data and employing techniques such as the Lomb-Scargle periodogram, \citet{yan_x-ray_2024} recently reported the detection (significance of 3.6$\sigma$) of a high-frequency ($\sim$4566s) QPO in the Seyfert galaxy NGC1365. We used obsid 0205590301 where \citet{yan_x-ray_2024} reported the detection of the QPO and reanalysed the EPIC-pn and MOS data using tasks \texttt{epproc} and \texttt{emproc} in SAS version 20.0.0. We filtered the lightcurves for particle flaring by first extracting background 10--12 keV lightcurves and then inspected these visually to set a threshold count-rate to reject times of high-background flaring. We applied the standard quality filters and selected {\sc pattern} $\le$ 4 events for pn and \textsc{pattern}$\le$12 events for the MOS cameras. We used \texttt{eregionanalyse}, with the input source coordinates, to select a suitable source region. The circular region as determined by the task contained a fainter source near to the target in some instances, so to avoid contamination we reduced the radius to $\sim$55\arcsec, but keeping the same centroid position. A slightly larger circular region on the same chip, away from the readout region and as close as possible to the source region, was selected for background lightcurve extraction. The final lightcurve was corrected for effects including losses due to vignetting, chip gaps and bad pixels using \texttt{epicclcorr}. Following \citet{yan_x-ray_2024}, the three lightcurves were binned to 200s and their net count-rates combined into a  final lightcurve. Because the asynchronicity of the three instruments can introduce spurious variability \citep{barnard_artificial_2007}, we ensured the start and end times were the same for the three detectors and inspected the individual and combined lightcurves visually.

\begin{figure*}
    \centering
\includegraphics[width=0.49\linewidth]{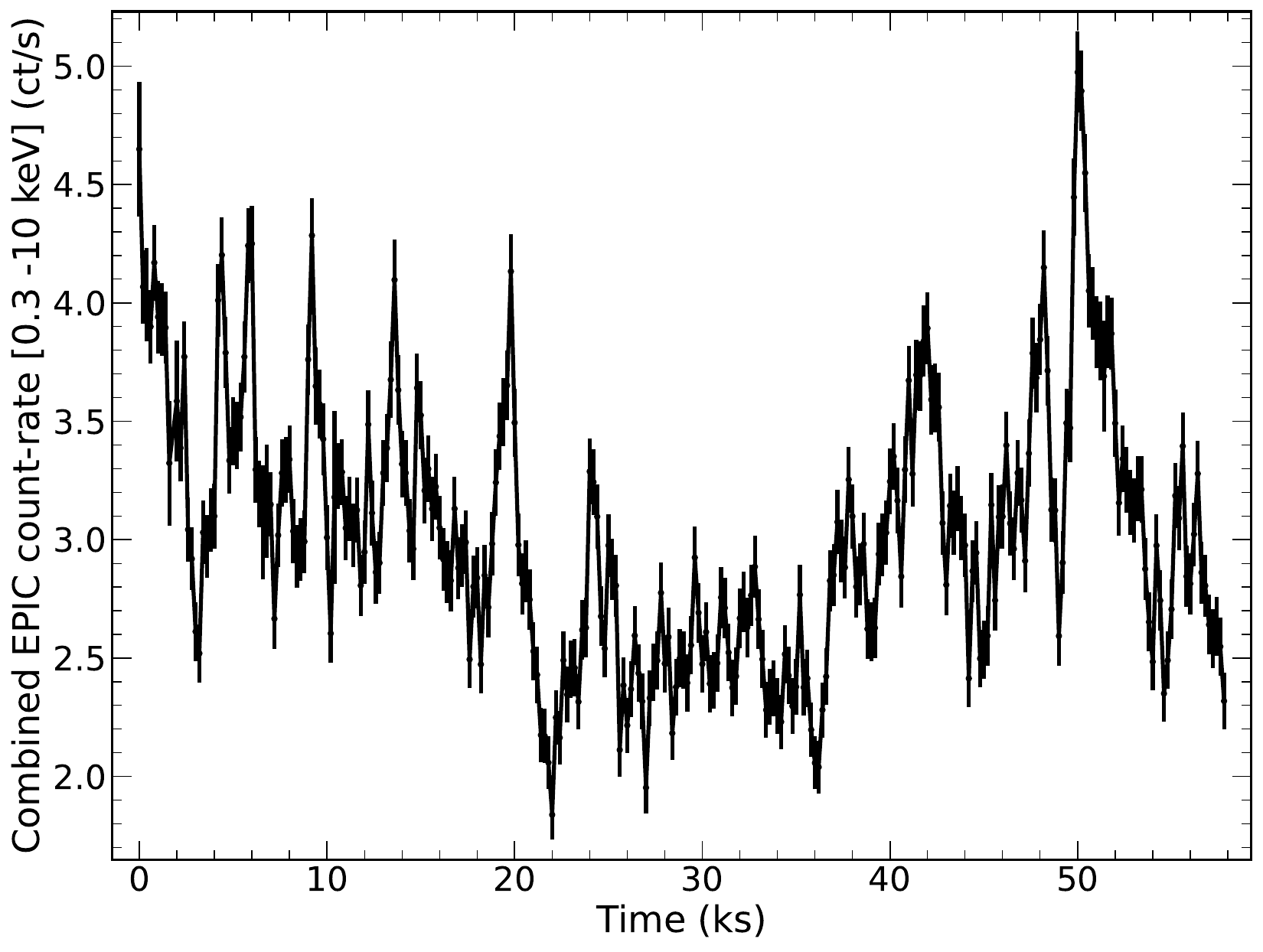}
\includegraphics[width=0.49\linewidth]{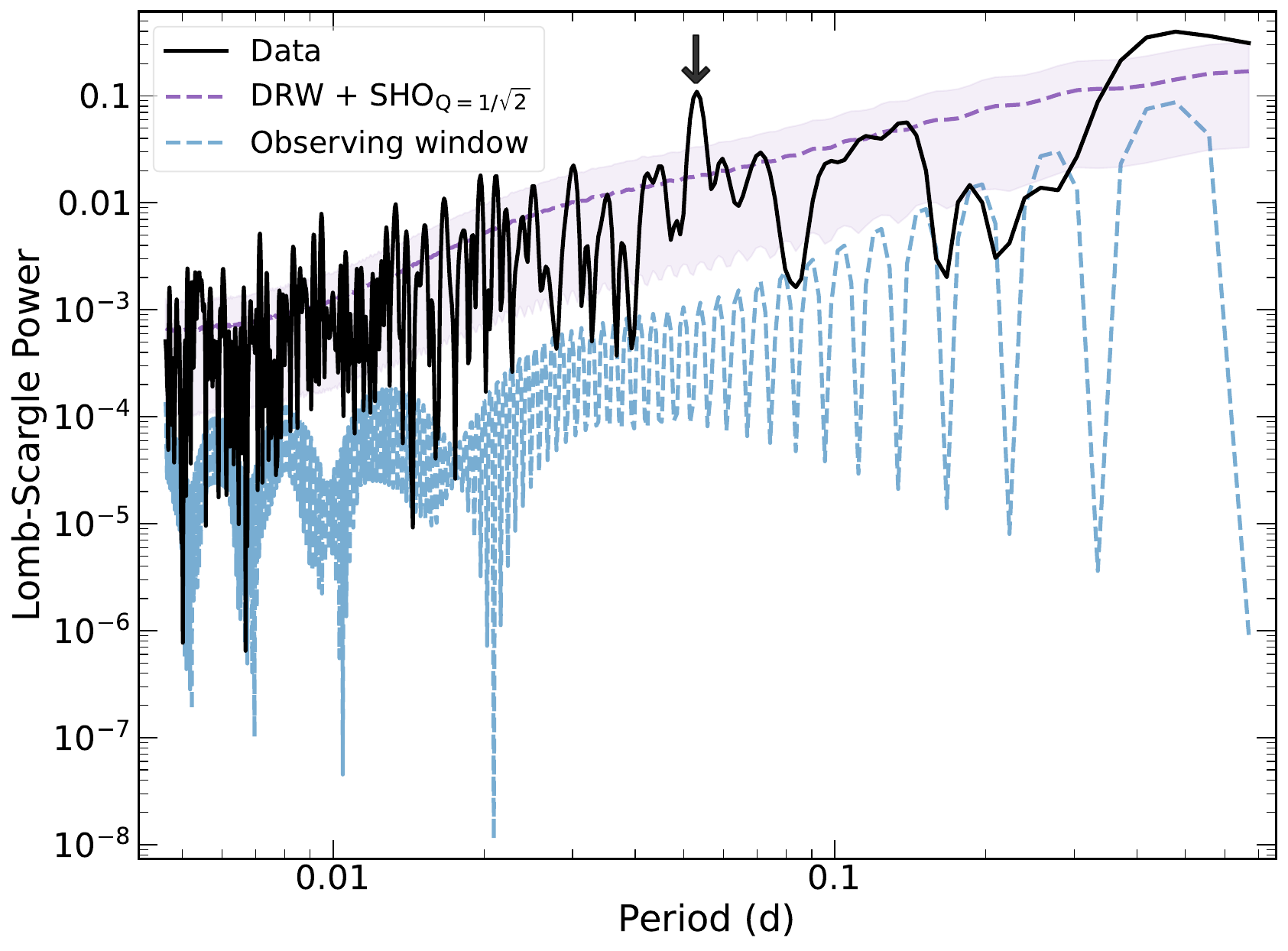}
    \caption{(Left) Combined EPIC 0.3--10 keV lightcurve of the Seyfert galaxy NGC1365. (Right) Corresponding Lomb-Scargle periodogram (oversampled by a factor 5). The pink dashed line shows the mean periodogram of 10,000 lightcurves simulated from the posteriors of the Mat\'ern-3/2 + DRW kernel (best-fit model), with the shaded areas showing the 16\% and 84\% percentiles of the distribution.  The vertical black arrow shows the putative QPO reported by \citet{yan_x-ray_2024}. The dashed blue line shows the power spectrum of the observing window.}
    \label{fig:ngc1365_lc}
\end{figure*}

Figure~\ref{fig:ngc1365_lc} shows the 0.3--10 keV combined EPIC lightcurve of NGC~1365, which comprises 289 datapoints and a duration of 57,800~s. The right-hand panel of Figure~\ref{fig:ngc1365_lc} shows the corresponding periodogram, with an arrow at $\sim$0.05 days indicating the claimed QPO by \citet{yan_x-ray_2024}.

\begin{figure*}
    \centering
    \includegraphics[width=0.49\textwidth]{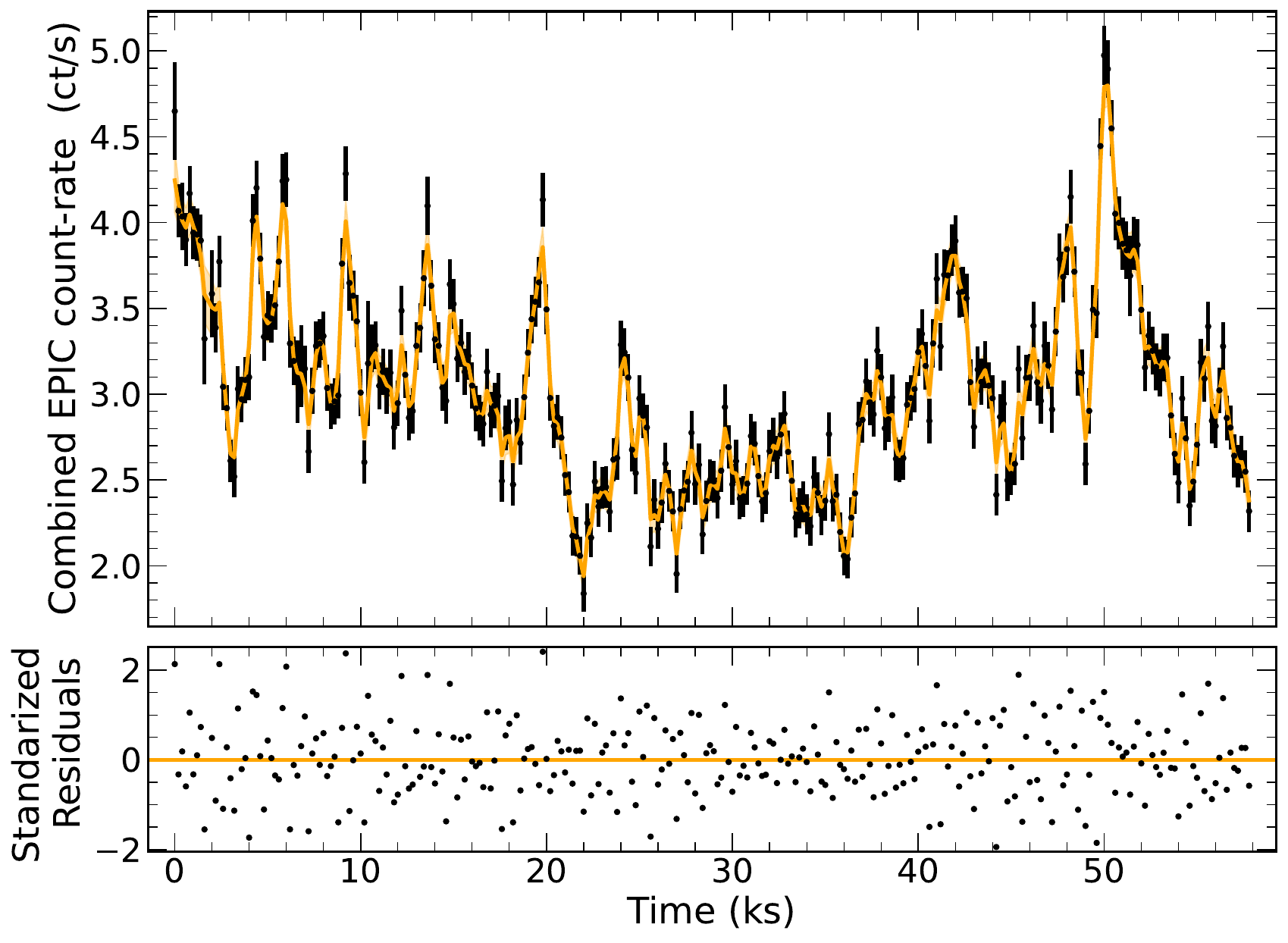}
\includegraphics[width=0.49\textwidth]{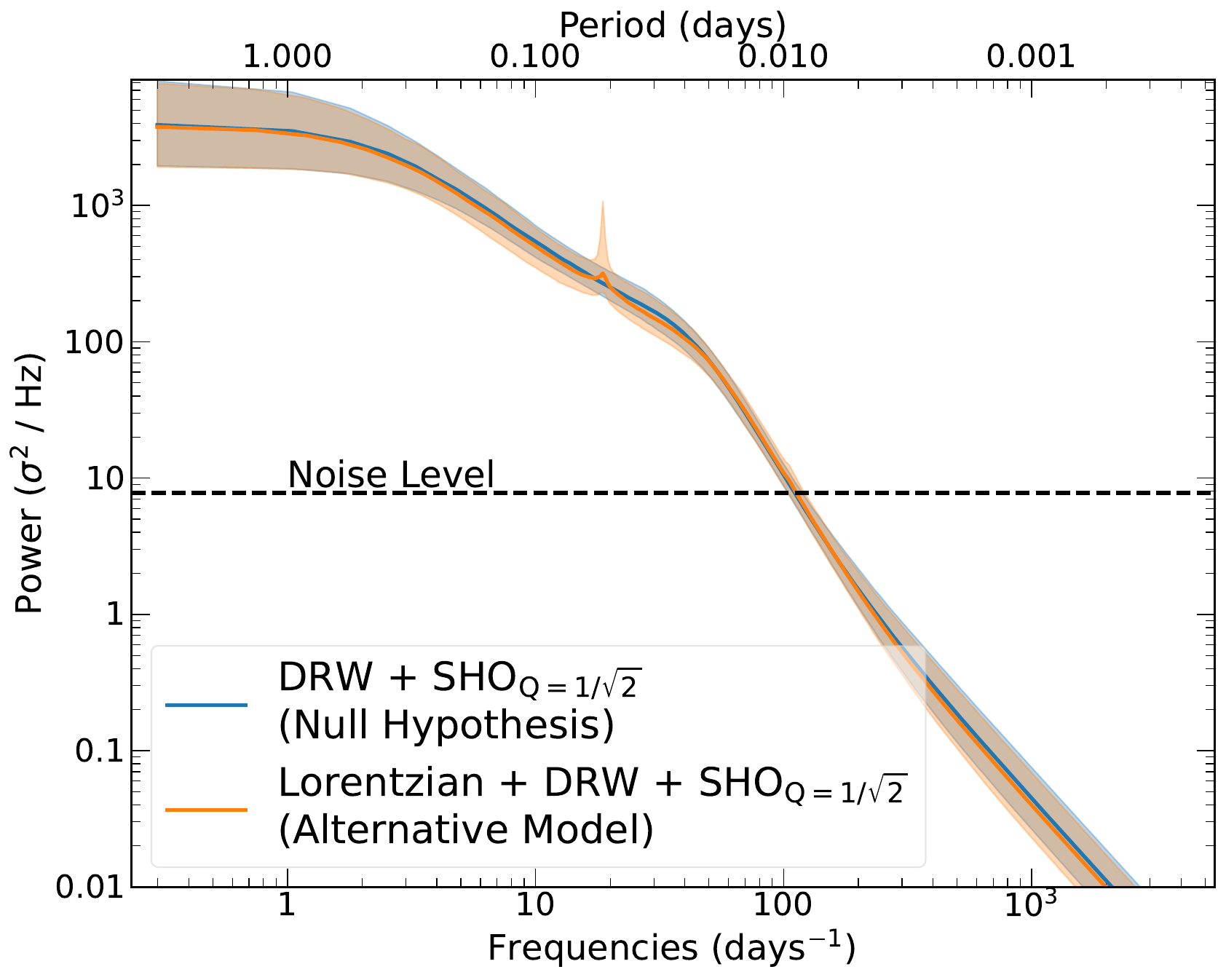}     \includegraphics[width=0.49\textwidth]{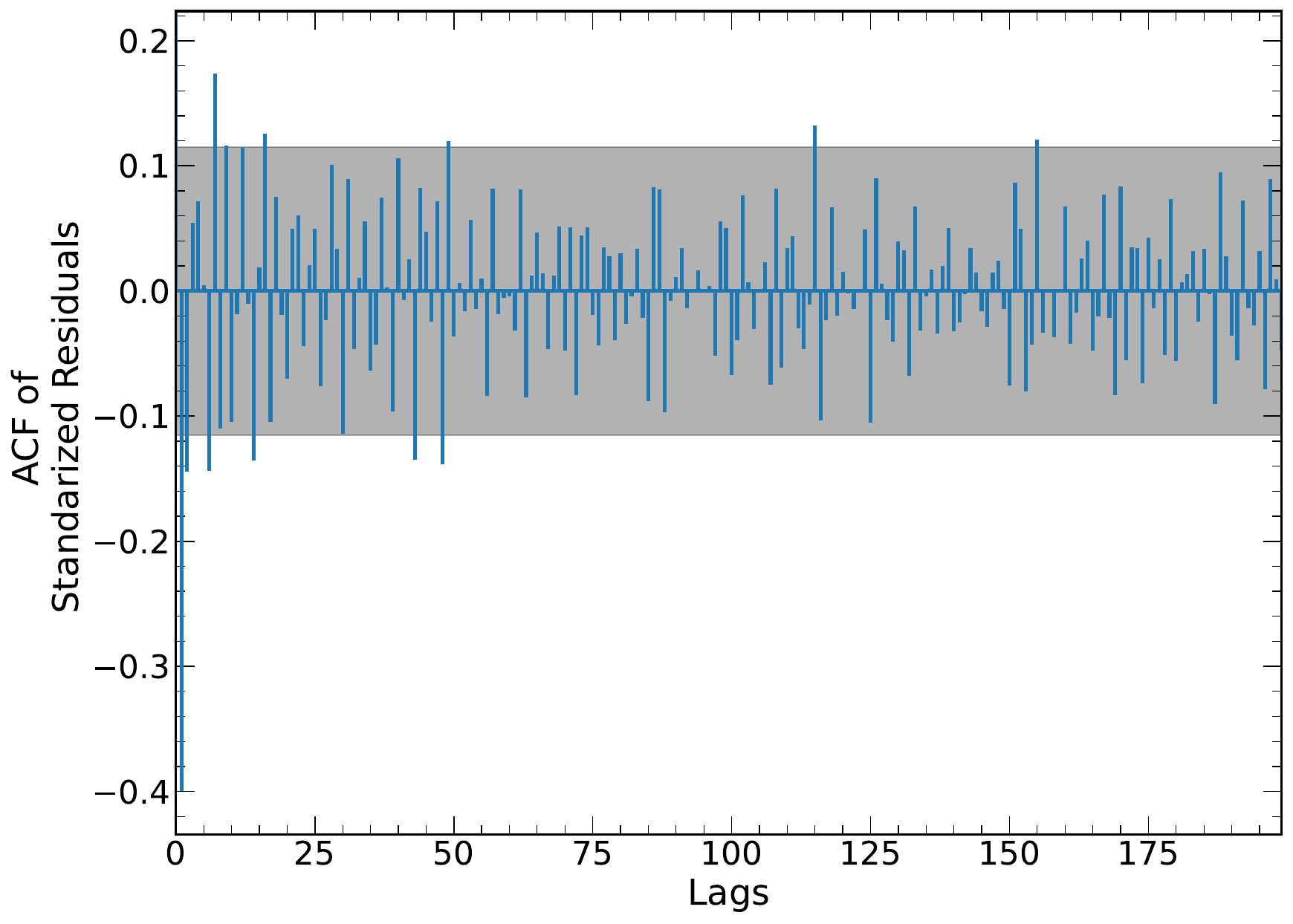}
\includegraphics[width=0.49\textwidth]{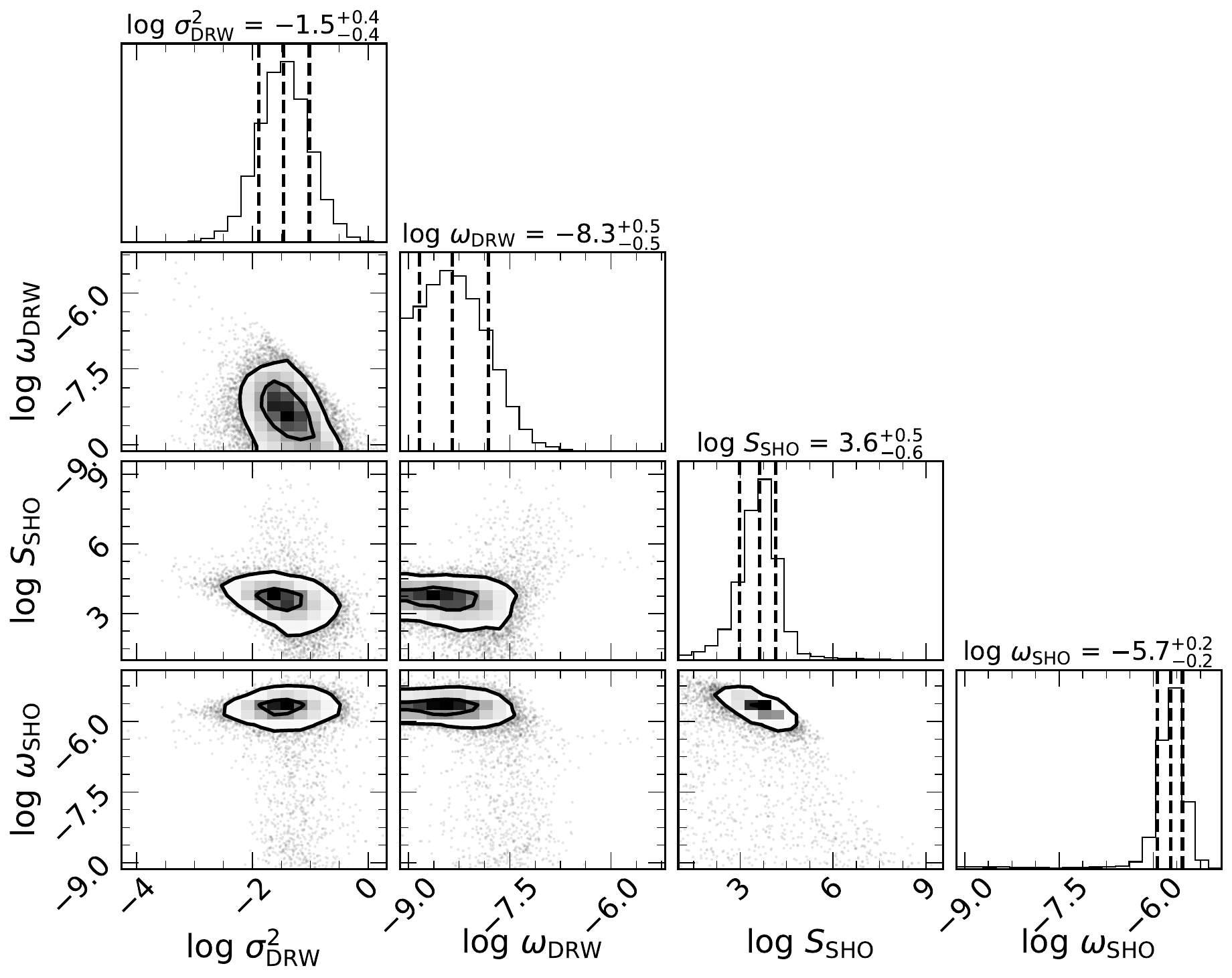}
    \caption{GP modelling results of the combined \xmm\ EPIC data of the Seyfert galaxy NGC 1365 (Figure~\ref{fig:ngc1365_lc}). (Top left) Best-fitting DRW + \granulation\ (solid orange line) and its 1$\sigma$ uncertainties (shaded areas). The bottom panel shows the standarized residuals of the model. (Top right) PSDs derived from the \texttt{celerite} modelling (absolute rms normalization), showing the DRW + \granulation\ (null) and the Lorentzian + DRW + \granulation\ (alternative) models. The solid and shaded areas show the median and 1$\sigma$ uncertainties derived from the posteriors. The dashed horizontal line shows the approximate Poisson level (2 $\tilde{\Delta t} <\sigma_\mathrm{err}^2>$ where $\tilde{\Delta t}$ and $<\sigma_\mathrm{err}^2>$ are the median sampling and the mean square error, respectively). (Bottom left) ACF of the standarized residuals. The shaded areas indicate the 95\% confidence level expected for white noise. (Bottom right) Posteriors of the best-fitting DRW + \granulation\ model. The contours indicate the 2-D, 1 and 2$\sigma$ confidence levels (39 and 86\% respectively) and the dashed lines on the marginalised histograms indicate the 32, 50 and 84\% percentiles (median$\pm1\sigma$). The MCMC run for approximately 64,000 steps until convergence, from which we discarded the first 10,000 as burn-in.}
    \label{fig:ngc1365_posteriors}
\end{figure*}

Table~\ref{tab:ngc1365} lists the models tested to the data, ranked by AICc value. We can see that the best-fit model comprises a Lorentzian (describing the putative QPO) and DRW + \granulation\ kernels to describe the underlying noise. Compared to a DRW + \granulation-only model, the addition of the Lorentzian represents a $\Delta$AICc = 4.5 fit improvement. Figure~\ref{fig:ngc1365_posteriors} shows the best-fit DRW + \granulation\ model, its PSD, the ACF of the standarized residuals and the posteriors. Both models provide an adequate description of the data whilst the ACF (bottom left panel) shows that the variability is approximately well captured by the DRW + \granulation\ model. Therefore, using the posteriors of the DRW + \granulation\ model, we tested whether the addition of the Lorentzian was supported by the data.

\begin{table}
 \begin{tabular}{lcccc} 
 \hline 
 Model & AICc & $\Delta$AICc & $p$-value \\ 
\hline \hline 
Lorentzian + DRW + \granulation & 92.4 & 0.0 & 0.08      \\ 
Lorentzian + \matern + \matern & 92.7 & 0.4 & 0.01       \\ 
Lorentzian + \matern + \granulation & 92.9 & 0.5 & 0.45  \\ 
Lorentzian + \matern + DRW & 93.2 & 0.8 & 0.01           \\ 
\matern + \granulation & 95.7 & 3.4 & 0.18               \\ 
\matern + DRW & 95.8 & 3.4 & 0.06                        \\ 
2$\times$\matern & 96.0 & 3.6 & 0.19                     \\
DRW + \granulation & 96.8 & 4.5 & 0.06                   \\ 
Lorentzian + \matern & 98.7 & 6.3 & 0.70                 \\ 
2$\times$\granulation & 99.3 & 6.9 & 0.27                \\ 
Lorentzian + \granulation & 100.6 & 8.2 & 0.02           \\ 
Lorentzian + DRW & 101.3 & 9.0 & 0.0                     \\ 
\matern & 103.2 & 10.9 & 0.83                            \\ 
DRW & 103.8                 & 11.5 & 0.00                \\ 
Lorentzian + 2$\times$DRW   & 105.3 & 12.9 & 0.00         \\ 
2$\times$DRW & 107.4 & 15.0 & 0.00                        \\ 
3$\times$DRW & 111.6 & 19.2 & 0.00                       \\ 
\granulation & 114.7 & 22.3 & 0.44                       \\ 
Lorentzian + Jitter & 121.3 & 28.9 & 0.00                 \\ 
\hline \hline 
\end{tabular}
 \caption{AICc, $\Delta$AIC and $p$-values for the standarised residuals following a Gaussian distribution for the different models tested on the combined EPIC data of the Seyfert galaxy NGC~1365. $\Delta$AICc refers to the increment in AICc with respect to the first model listed in the Table. Models with lower AICc values are not shown for clarity.} \label{tab:ngc1365}
 \end{table}
 
\begin{figure}
    \centering
    \includegraphics[width=0.48\textwidth]{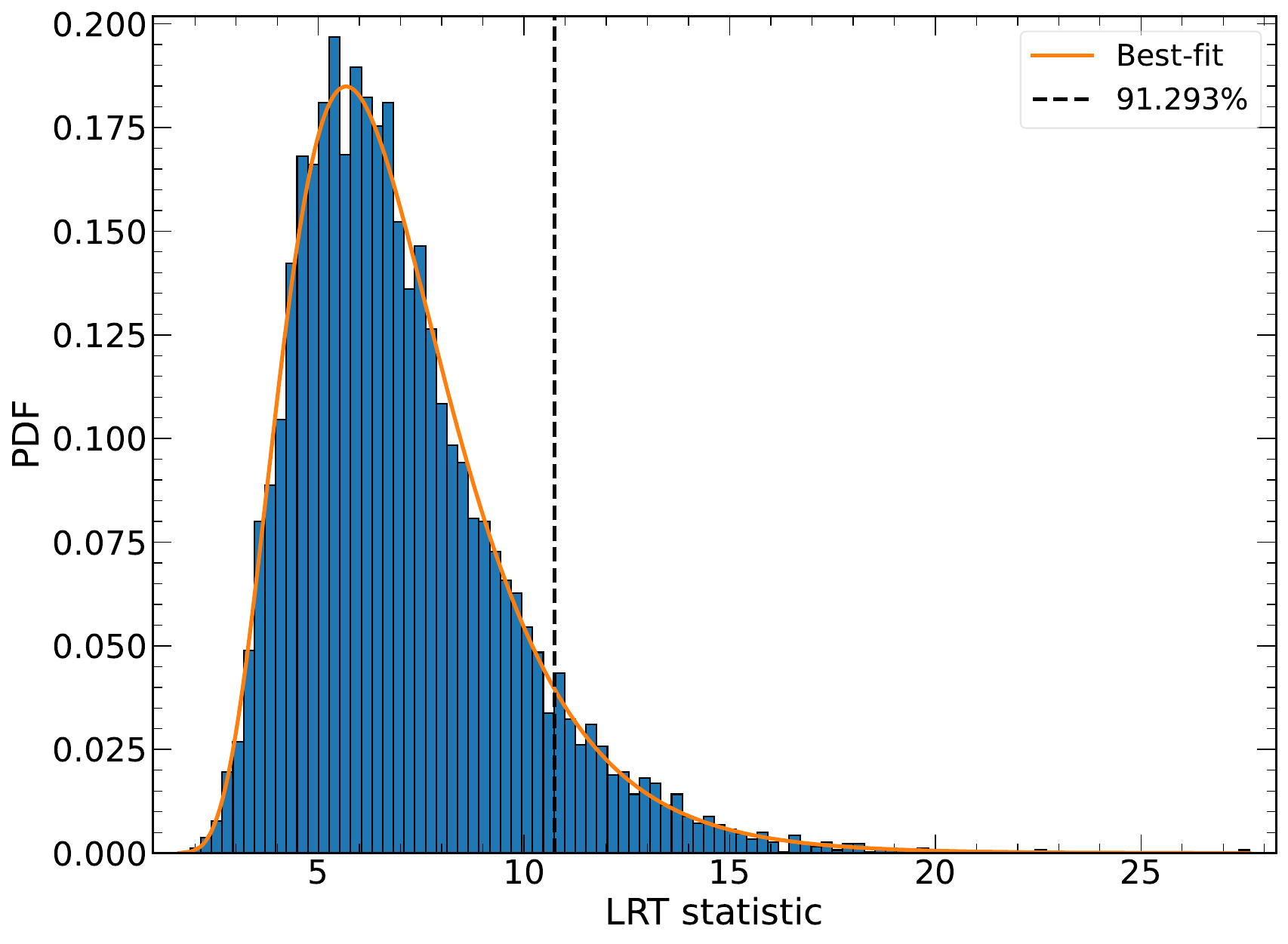}
    \caption{Reference LRT distribution generated from simulated lightcurves from the posteriors of the DRW + \granulation\ model (null hypothesis). The solid orange line shows a fit to the distribution using a log-normal. The $T_\mathrm{LRT}$ observed in the data is shown as a dashed black line.}
    \label{fig:ngc1365_LRT}
\end{figure}

Figure~\ref{fig:ngc1365_LRT} shows the reference LRT distribution derived from lightcurve simulations generated from the posteriors of the DRW + \granulation\ model. As can be seen from the Figure, the addition of the Lorentzian (the QPO component) is significant only at the $\sim$91\% level ($\sim$1.7$\sigma$).

As stated in Section~\ref{sub:model_selection}, owing to the relatively small difference in fit-improvement ($\Delta$AICc = 0.4) with respect to the Lorentzian + 2$\times$\matern, we have repeated the significance calculation with the posteriors of the this other model too. We have found the significance of $\sim$86\%, in line with the lower $\Delta$AICc = 3.6 provided by this model with respect to the null hypothesis. Therefore we do not support the presence of a QPO in this lightcurve of NGC~1365.

%% file: qso.tex
\begin{figure*}
    \centering
    \includegraphics[width=0.49\textwidth]{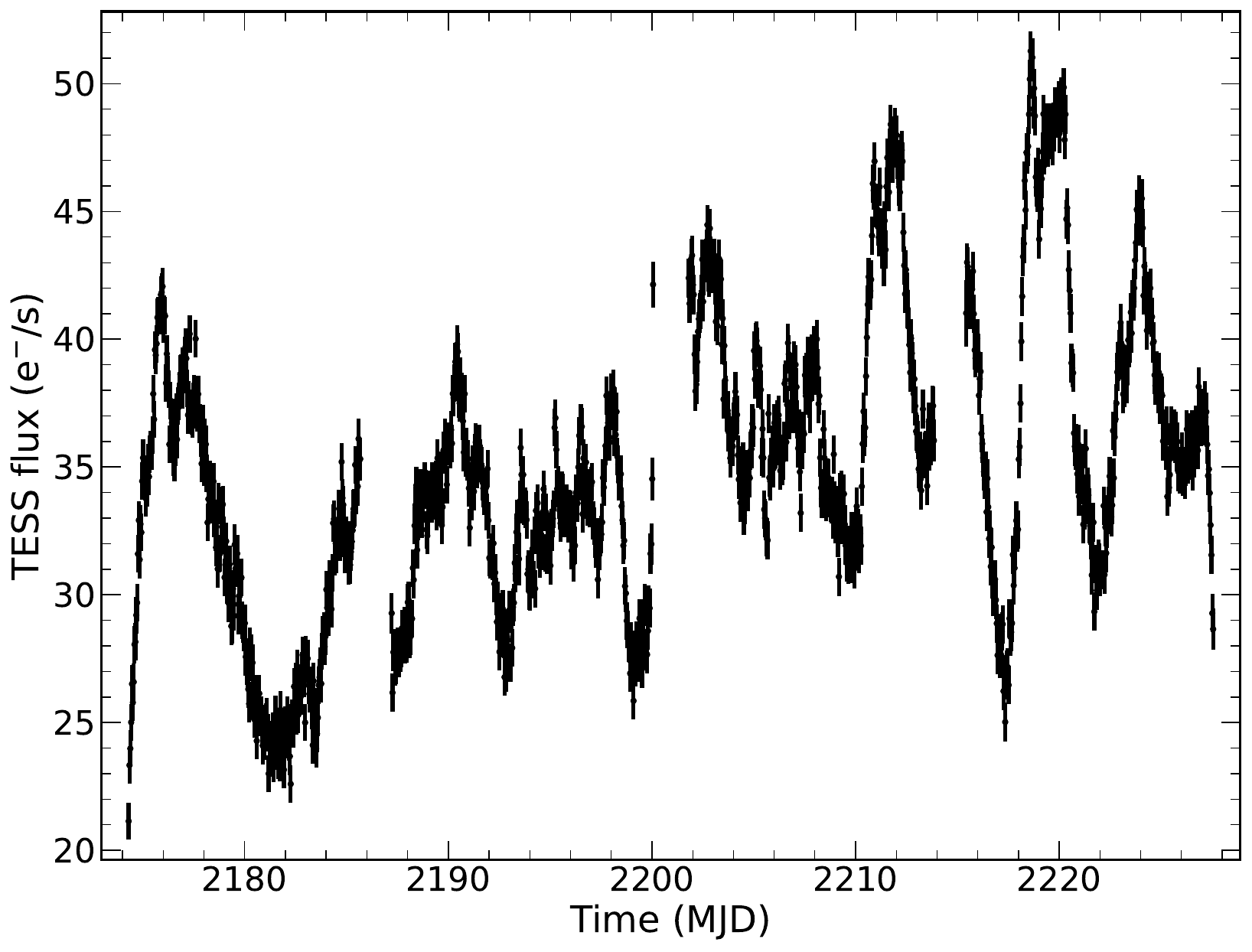}
    \includegraphics[width=0.49\textwidth]{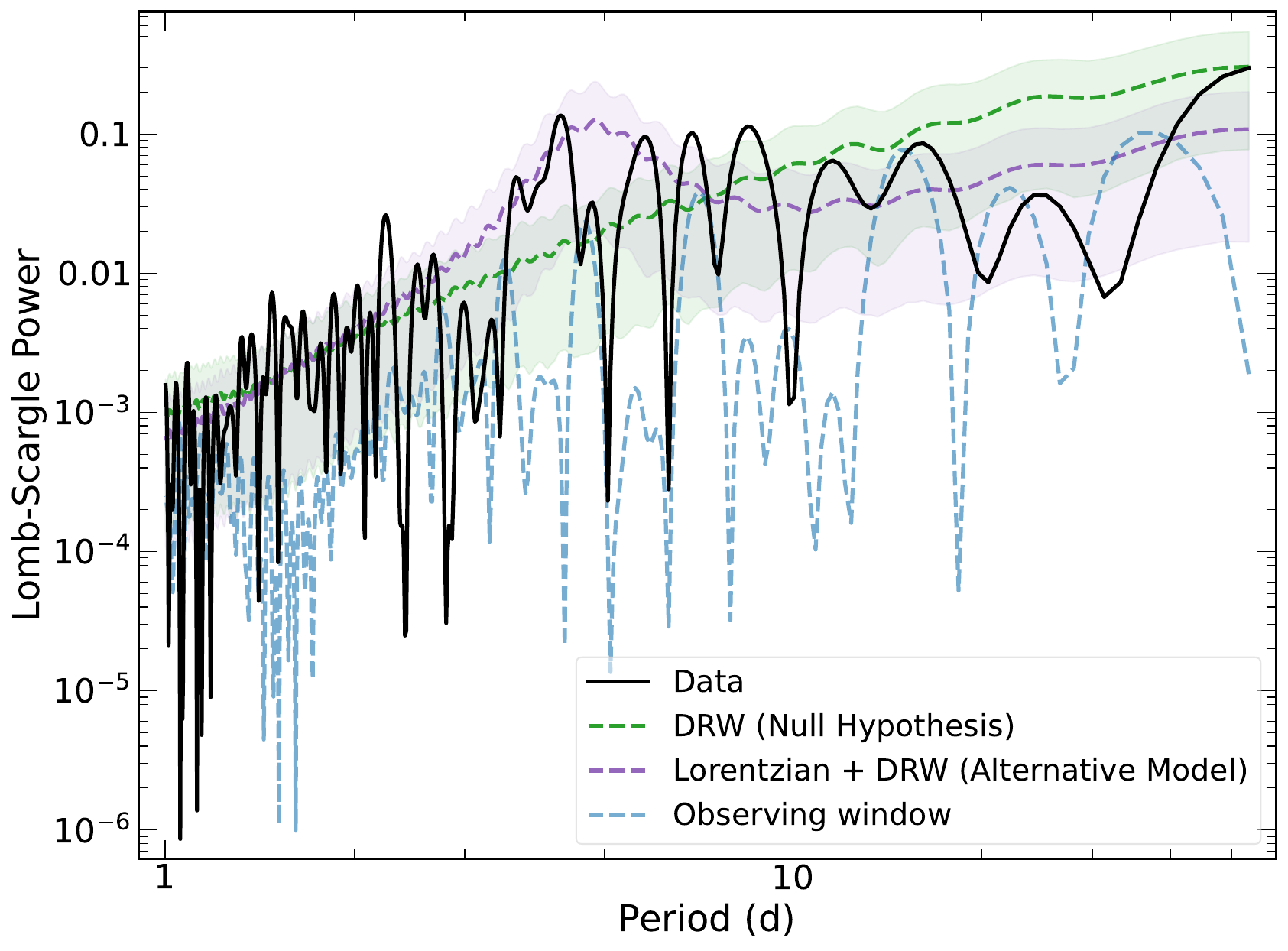}
    \caption{(Left) TESS lightcurve of the Blazar, B0537-441 from sectors 32 \& 33 (cf. Figure 3b in \citealt{tripathi_search_2024}). (Right) Corresponding Lomb-Scargle periodogram (black solid line). The power spectrum of the observing window is shown as per Figure~\ref{fig:uvot_P13_lc}. The dashed purple and green dashed lines show the best-fit Lorentzian + DRW, and DRW-only models (with the shaded areas showing the 16\% and 84\% percentiles of 10,000 simulations; see text for details).}
    \label{fig:qso_lc}
\end{figure*}

\citet{tripathi_search_2024} recently reported the detection of a QPO of $\sim$ 6.5 days in the Blazar B0537-441, using TESS data. We obtained the TESS lightcurves from sectors 32 \& 33 (as analysed by these authors), reduced by the Science Processing Operations Center \citep[SPOC;][]{jenkins_tess_2016} using the python \texttt{lightkurve} package. The lightcurve is extracted using aperture photometry and then corrected with the presearch data conditioning module to remove long-term trends and systematics caused by the spacecraft. The data from sectors 32 \& 33 had been processed with pipeline versions spoc-5.0.21-20210107 and spoc-5.0.22-20210121, respectively. For computational reasons we rebinned the lightcurves to 1h which still allowed us to analyse the variability present and probe the relevant timescales, see Figure~\ref{fig:qso_lc}. The lightcurve had a duration of $\sim$53 days and a total of 1164 datapoints, with two $\sim$1-day gaps at MJD $\sim$2186 and MJD $\sim$2214 due to the satellite's orbit and another gap at MJD $\sim$ 2202 due to the observing strategy of TESS.

\begin{table} 
 \centering 
 \begin{tabular}{lccc} 
 \hline 
 Model & AICc & $\Delta$AICc & $p$-value \\ 
 \hline \hline 
 \matern + \granulation              & 3526.2 & 0.0 & 0.0   \\
 2$\times$\matern                    & 3526.3 & 0.1 & 0.0   \\
 \matern + DRW                       & 3526.5 & 0.3 & 0.0   \\
 Lorentzian + \granulation           & 3527.6 & 1.4 & 0.0   \\ 
 Lorentzian + Mat\'ern-3/2           & 3527.7 & 1.5 & 0.0   \\ 
 2$\times$Lorentzian + \granulation\ & 3527.8 & 1.3 & 0.0   \\
 2$\times$Lorentzian + Mat\'ern-3/2  & 3528.4 & 2.2 & 0.0   \\ 
 \matern                             & 3528.6 & 3.5 & 0.0   \\ 
 \granulation                        & 3536.9 & 10.7 & 0.0  \\ 
 2$\times$Lorentzian + DRW           & 3560.3 & 32.4 & 0.01 \\ 
 Lorentzian + DRW                    & 3573.9 & 47.7 & 0.40 \\
 Lorentzian + 2$\times$DRW           & 3578.1 & 50.5 & 0.35 \\ 
 2$\times$Lorentzian                 & 3578.3 & 52.2 & 0.22 \\ 
 2$\times$Lorentzian + Jitter        & 3581.1 & 53.7 & 0.28 \\
 Lorentzian                          & 3608.0 & 81.8 & 0.94 \\ 
 Lorentzian + Jitter                 & 3610.0 & 83.9 & 0.97 \\ 
 DRW                                 & 3639.9 & 113.8 & 0.27 \\ 
 2$\times$DRW                        & 3644.0 & 117.8 & 0.27 \\ 
 \hline \hline 
 \end{tabular} \caption{As per Table~\ref{tab:ngc1365}, showing the AICc, $\Delta$AICc and $p$-values for the different models tested against the TESS data of the Blazar, B0537-441.}  \label{tab:bic_qso}
 \end{table}
Table~\ref{tab:bic_qso} lists the models tested against this dataset. While some of the single-component models provide seemingly better fits according to the AICc, we can see from their $p$-values that none of these models provide a satisfactory description of the data ($p \lesssim$ 0.05). In such models, the standarized residuals are broader ($\sigma \sim$2) than expected for a standard normal distribution, indicating a deficiency in the fit.

\begin{figure*}
    \centering
\includegraphics[width=0.49\textwidth]{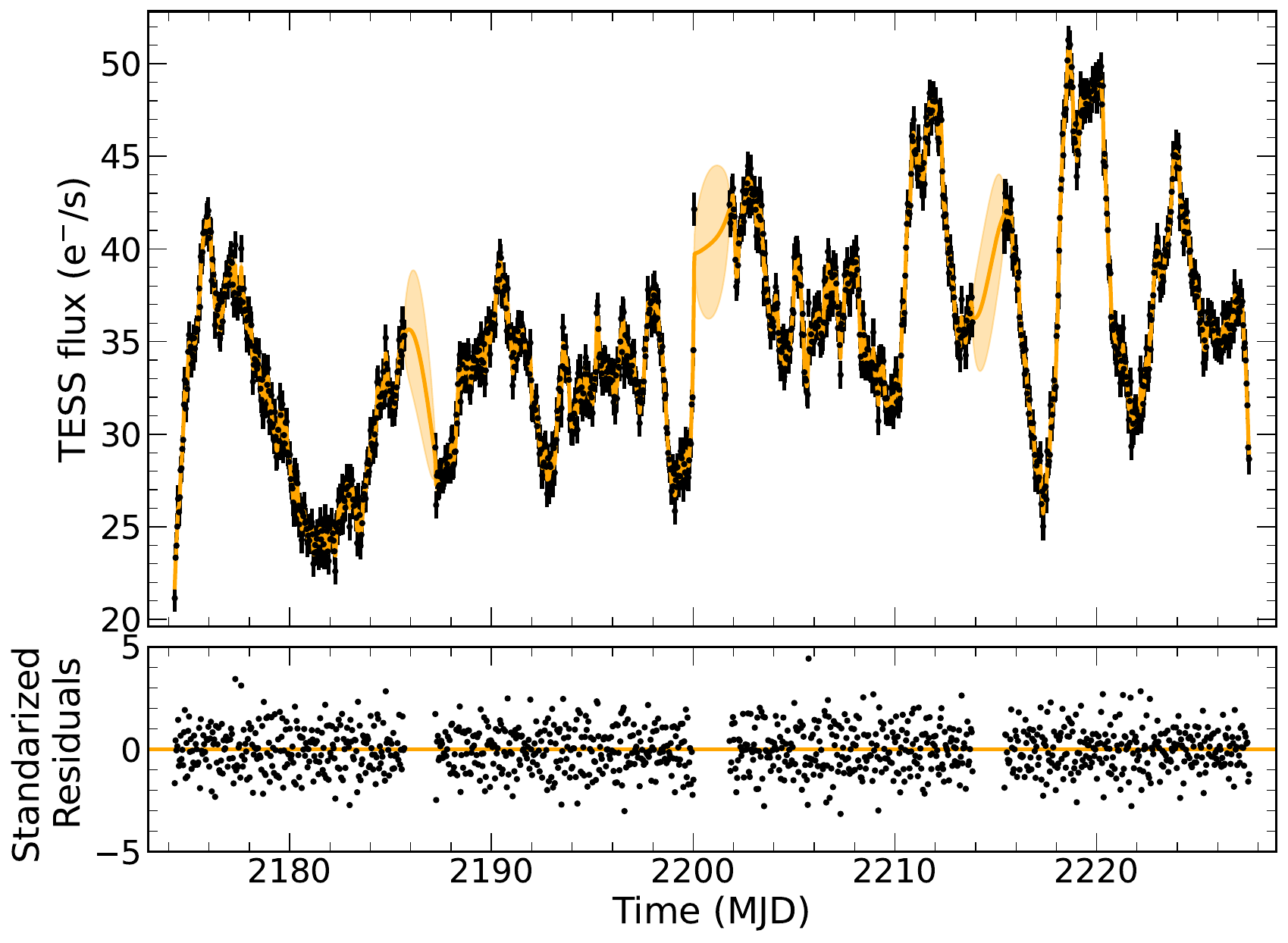}
\includegraphics[width=0.49\textwidth]{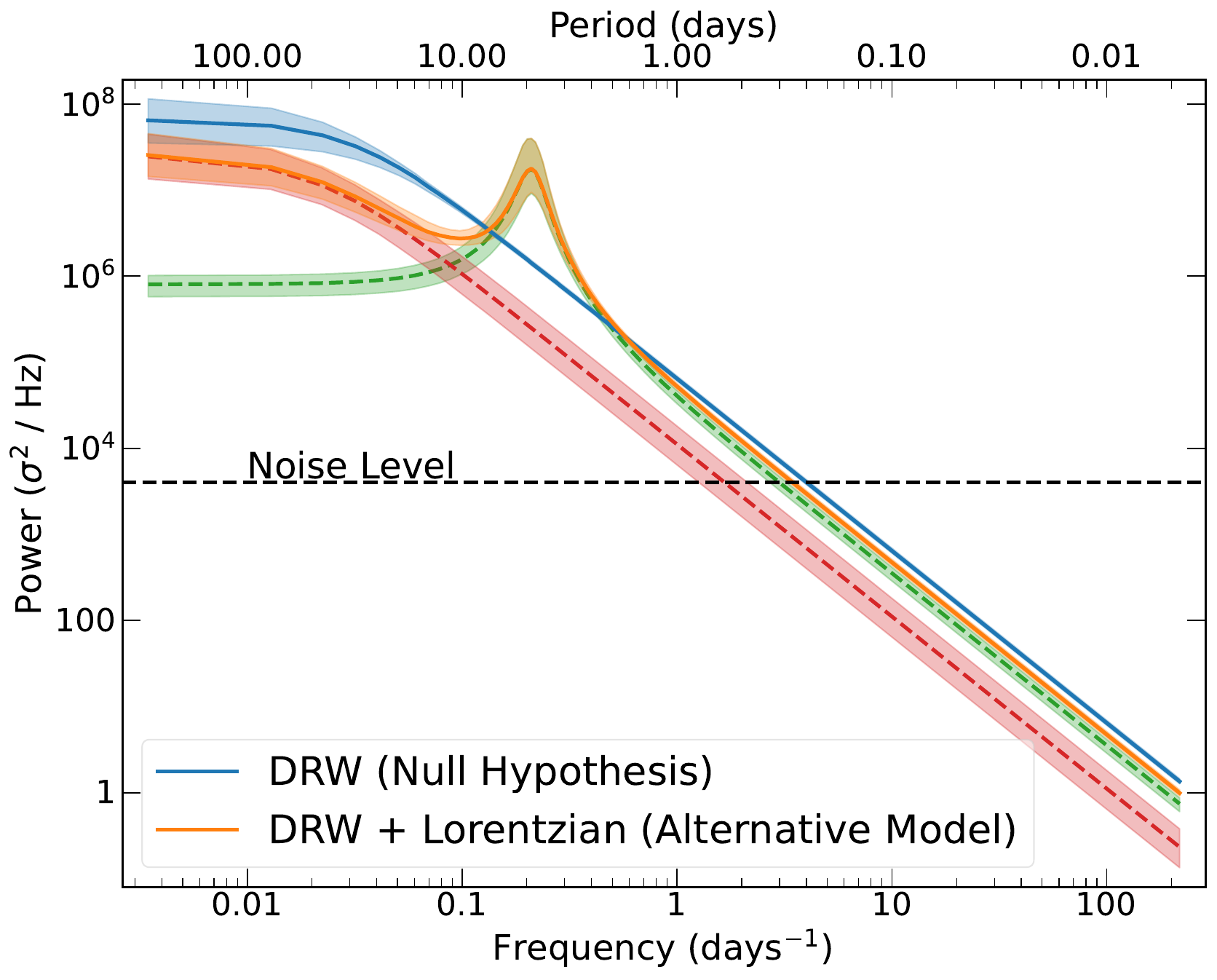}
 \includegraphics[width=0.49\textwidth]{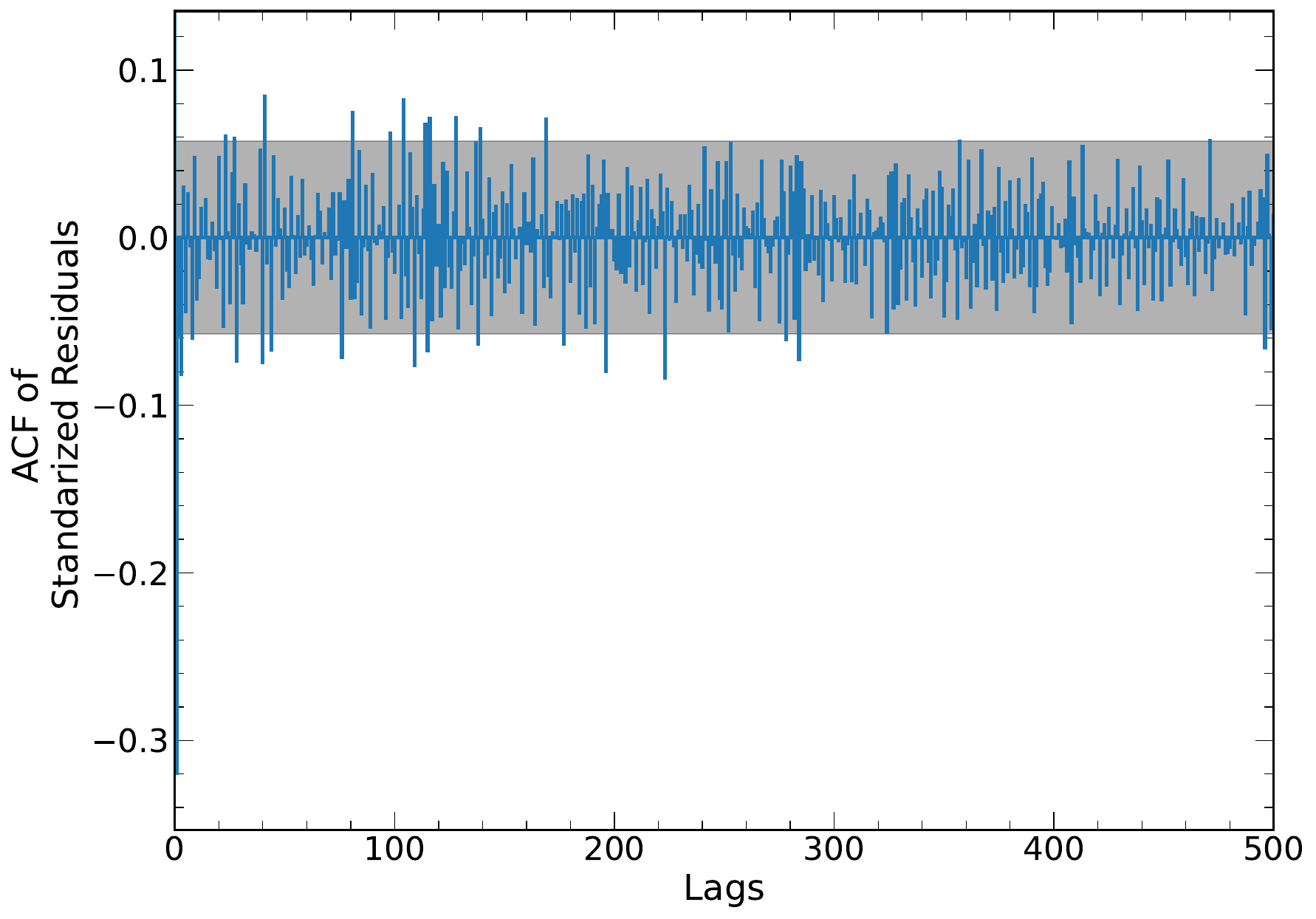}
\includegraphics[width=0.49\textwidth]{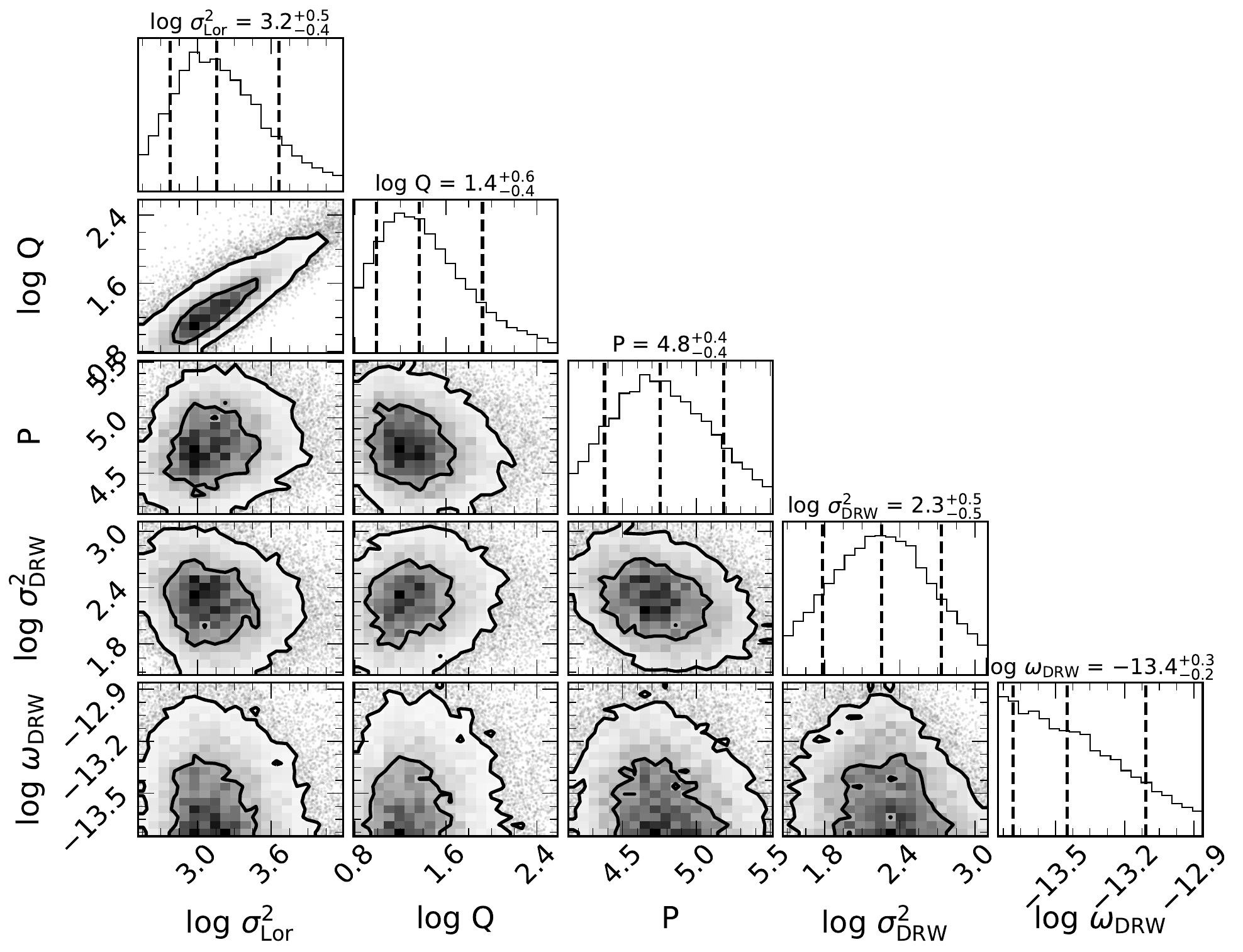}
\caption{GP modelling results of the RXTE data of the TESS lightcurve of the Blazar, B0537-441 shown in Figure~\ref{fig:qso_lc}. (Top Left) Best-fit Lorentzian + DRW model. (Top Right) PSD of the null hypothesis and alternative models. (Bottom left) ACF of the standarized residuals of the Lorentzian + DRW model. (Bottom right) Posterior parameters for the Lorentzian + DRW model. The MCMC run for 17,500 steps until convergence, of which 5,390 were discarded for burn in. Symbols as per Figure~\ref{fig:ngc1365_posteriors}.}
    \label{fig:qso}
\end{figure*}
\begin{figure}
    \centering
    \includegraphics[width=0.48\textwidth]{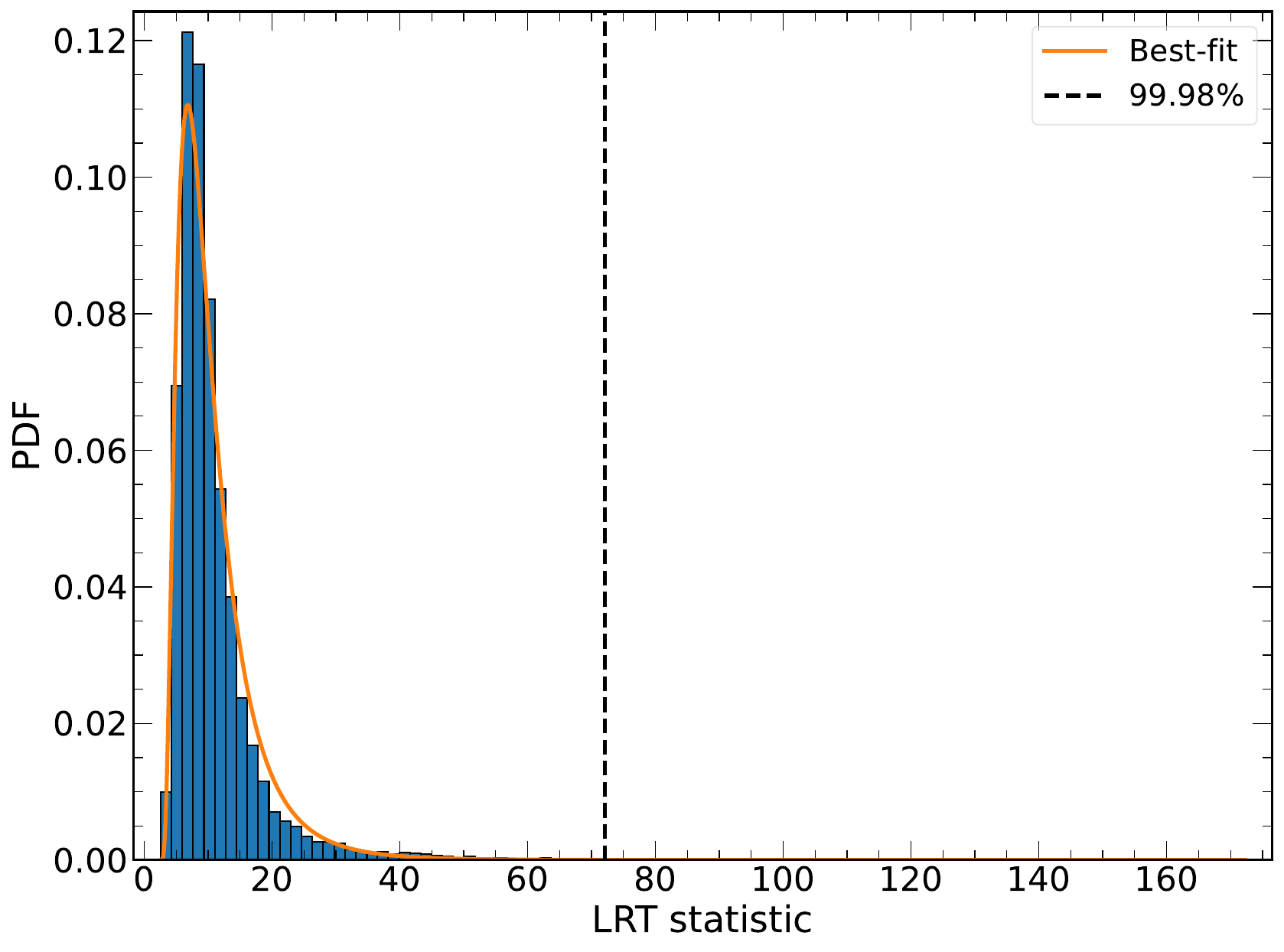}
    \caption{As per Figure~\ref{fig:NGC4945_LRT} but testing the QPO in the TESS lightcurve of the Blazar, B0537-441 using the DRW posteriors as the null hypothesis.}
    \label{fig:qso_sig}
\end{figure}

The first model which provides an adequate description of the data ($p$ = 0.4) is a combination of a Lorentzian and a DRW. The model that maximizes the likelihood, its standarized residuals, ACF and posterior parameters is shown in Figure~\ref{fig:qso}. The best-fit suggests a quasi-periodicity ($Q = 4^{+3}_{-1}$) with $P = 4.8_{-0.4}^{+0.5}$ days. On the other hand, the bend of the DRW is not well constrained, most likely owing to the relatively short baseline ($\approx$54 days) of the data; in our modelling, the DRW mostly acts as a powerlaw with $\beta = -2$. The DRW + Lorentzian model provides an $\Delta$AICc = 66 with respect to the DRW-only model; we proceeded to test whether the $\Delta$AICc was significant using the posteriors of the DRW-only model. We found the Lorenzian component to be significant at the $\sim$99.98\% ($\sim$ 3$\sigma$) level (Figure~\ref{fig:qso_sig}). This is in agreement with the high $\Delta$AICc observed between the Lorentzian + DRW and the DRW-only model. Therefore, we deem the addition of the Lorentzian to be supported by the data. 

Figure~\ref{fig:qso_lc} shows a comparison of the periodogram of the best-fit model (Lorentzian + DRW), the periodogram of the DRW-only model and the Lomb-Scargle of the data. The Lomb-Scargle periodograms of the best-fit models were derived by taking the Lomb-Scargle periodogram of 10,000 lightcurves generated from the posteriors of each of the two models. We can see that the strongest period in periodogram is consistent with our best-fit period of 4.8 days and that the noise is reasonably captured by the DRW.

%% file: appendix.tex
\section{Lightcurve Simulations} \label{sec:lightcurve_simulations}

In order to simulate lightcurves from the kernel PSDs, we have used the method devised by \citet{timmer_generating_1995} and \citet{emmanoulopoulos_generating_2013}. The method proposed by \citet{emmanoulopoulos_generating_2013}, as opposed to the method of \citet{timmer_generating_1995}, which by construction generates Gaussianly distributed data, can generate lightcurves with any flux probability density function (PDF) and PSD model. Therefore, along with having more realistic lightcurves matching more closely the real data and its uncertainties, the issue of negative fluxes is also avoided. 

As stated above, we used the PSD from the GP kernel as input PSD for the method. For the PDF, we used either a Gaussian distribution (in which case we used the \citet{timmer_generating_1995} algorithm), in cases where the observed data was consistent with being Gaussianly distributed (as determined using a KS test) or log-normal distribution if this was not the case (this was only the case for the XRT data of P13; Section~\ref{sub:ngc7793}) -- in which case we reverted to the \citet{emmanoulopoulos_generating_2013} algorithm. In any instance, in practice we have found the PDF used to simulate the lightcurves did not affect the results. The mean of the distribution was set as for the observed data and the variance was determined by integrating the PSD kernel in frequency space from $1 / T$ where $T$ was the duration of the lightcurve, to a pseudo Nyquist frequency defined as 1 /2 min($\Delta t$) where min($\Delta t$) indicates the smallest exposure time in the lightcurve. In this manner we obtained the \textit{intrinsic} variance that generated the lightcurve prior to resampling, as opposed to the observed variance after resampling. The lightcurves were initially generated on a regular grid with a sampling min($\Delta t$) in order to introduce aliasing effects, and a few times longer (typically 5-20 depending on the lightcurve) than the real lightcurve length to introduce red noise leakage. We then drew a random segment matching the duration of the real monitoring and re-sampled it with the same exposure times and cadence as the real observations. We finally added Poisson noise and estimated realistic uncertainties taking into account the background rates and exposure times for each individual snapshot. For the \swift-XRT, as for the real lightcurves, in cases where the simulated source counts dropped below 15, we used instead the posterior probability function derived by \citet{kraft_determination_1991}, which is more suited for the low-count regime and prevents having negative counts.

\section{GP modeling of lognormal lightcurves}\label{sec:gp_lognormal}

It is commonly observed that all accreting systems show a lognormal flux distribution, which translates into the universally-observed linear relationship between the square root of their variance (the rms) and their mean flux \citep[the so-called linear 'rms-flux' relation ;][]{uttley_non-linear_2005}. The implication is that the process generating the flux variations must be multiplicative. A pertinent question to ask is therefore whether the lightcurves of accreting systems can be modelled as a GP, or at the very least, how the retrieved parameters are affected by the lognormality of the fluxes.

\begin{figure*}
    \centering
    \includegraphics[width=0.49\textwidth]{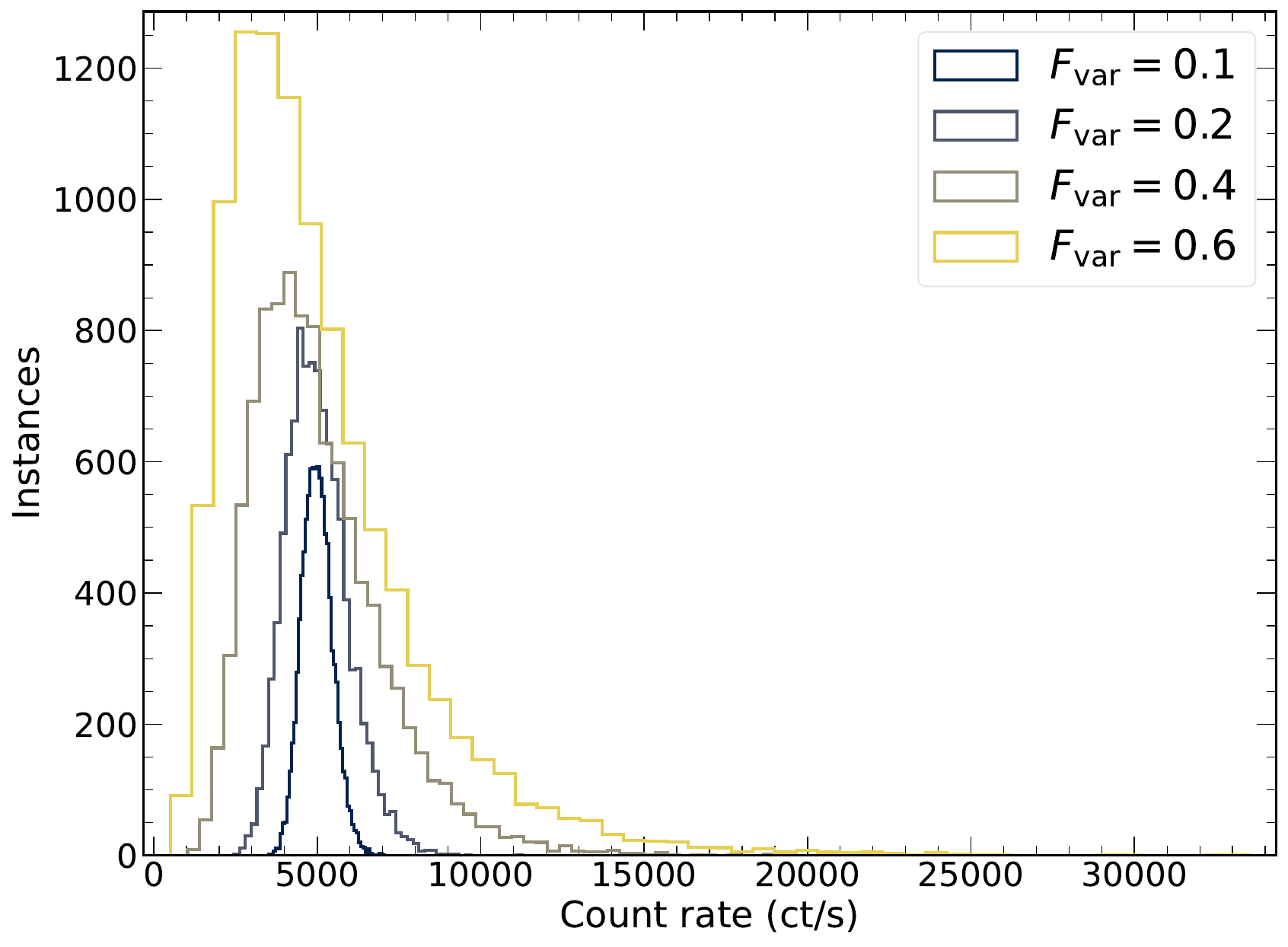}
    \includegraphics[width=0.49\textwidth]{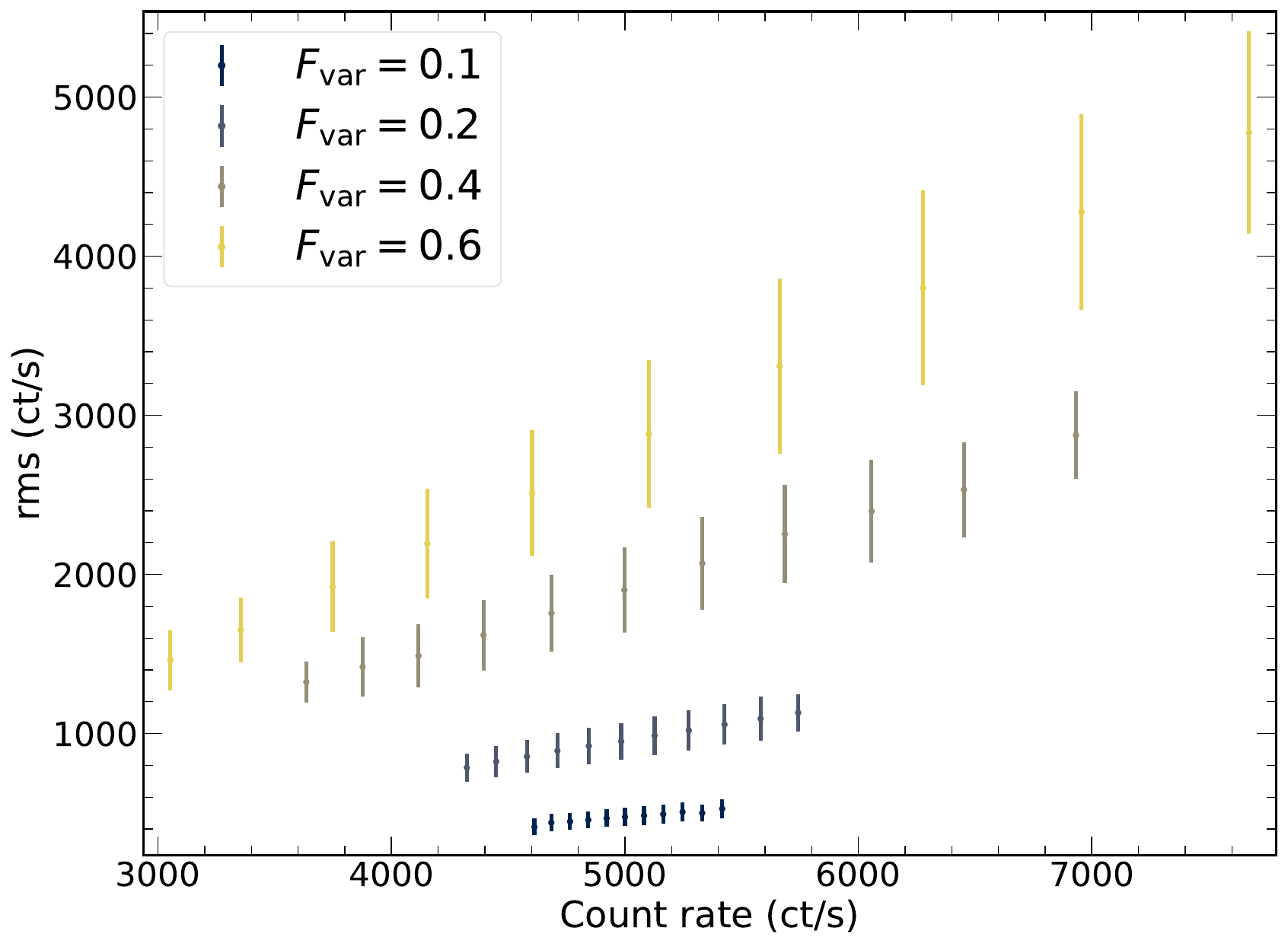}
    \caption{(Left) Example of the PDF of the lognormal lightcurves as a function of \rms. As \rms\ increases, the lognormal deviates more strongly from Gaussianity. (Right) RMS-flux relationship of lightcurves simulated having a lognormal distribution. These were averaged over the ensemble of the 1,000 lightcurves, by averaging the mean and rms calculated using 5000\,s segments.}
    \label{fig:lognormals}
\end{figure*}

The skweness of a lognormal distribution with mean $\mu$ and variance $\sigma^2$ and with Gaussian parameters $\mu_L$ and $\sigma_L$ is given by:
\begin{equation}
    \gamma = (e^{\sigma^2_L} + 2) \sqrt{e^{\sigma^2_L}  - 1} 
\end{equation}
\noindent where $\sigma_L^2 = \ln (1 + \frac{\sigma^2}{\mu^2}) = \ln (1 + F_\mathrm{var}^2)$ and where \rms\ is the fractional rms variability amplitude \citep[$F_\mathrm{var} = \sqrt{\sigma^2 / \mu^2}$;][]{vaughan_characterizing_2003}. This implies that $\gamma = (F_\mathrm{var}^2 + 3) F_\mathrm{var}$ and so for low \rms, the lognormal tends to be symmetric and resembles a Gaussian distribution, but as \rms\ increases, the lognormal distribution becomes more skewed and deviates more strongly from Gaussianity \citep[see also][]{uttley_non-linear_2005}. This is shown in Figure~\ref{fig:lognormals}. 
Naively, we then may expect that GPs might be able to recover the input parameters more readily when the \rms\ is low. Similarly, Gaussian-like lightcurves will show no dependence (or a flat) rms-flux relationship, and as \rms\ increases the rms will show a linear dependence with flux \citep[see also][]{uttley_non-linear_2005}.

In order to inspect any biases introduced by modelling lognormal lightcurves by a GP, we have generated lightcurves possessing a lognormal flux distribution using the method proposed by \citet{emmanoulopoulos_generating_2013} (see Appendix~\ref{sec:lightcurve_simulations}). The lightcurves were generated 10$^6$\,s long, sampled every 10\,s and with exposure times of 1\,s, roughly matching the lightcurve of Cygnus X--1 presented by \citep{uttley_non-linear_2005}. The generative PSD was a DRW, where the bending timescale was set to $\sim$930\,s to ensure it could be well-detected by the choice of sampling. The variance was adjusted to produce lightcurves with a varying degree of \rms\ while the mean count rate was fixed to 5,000 ct/s. In particular, we have tested whether we could recover the input PSD parameters (\omegabend\ and variance $\sigma^2$) using GP modelling of lognormal lightcurves having \rms = 0.1, 0.2, 0.4, 0.6. The lightcurves were produced free of Poisson noise (and the uncertainties were set to zero in the fitting process) as we are only interested in examining any biases introduced by the lognormality of the fluxes.

As can be seen from Figure~\ref{fig:lognormals}, the generated lightcurves naturally follow the observed linear rms-flux relationship. In particular, we can see that for the lowest $\gamma$ (or equivalently, $F_\mathrm{var}$), the relationship is flat, as expected for a Gaussian distribution. As \rms\ increases, we see the linear rms-flux relationship is recovered. 

\begin{figure*}
    \centering
    \includegraphics[width=0.8\textwidth]{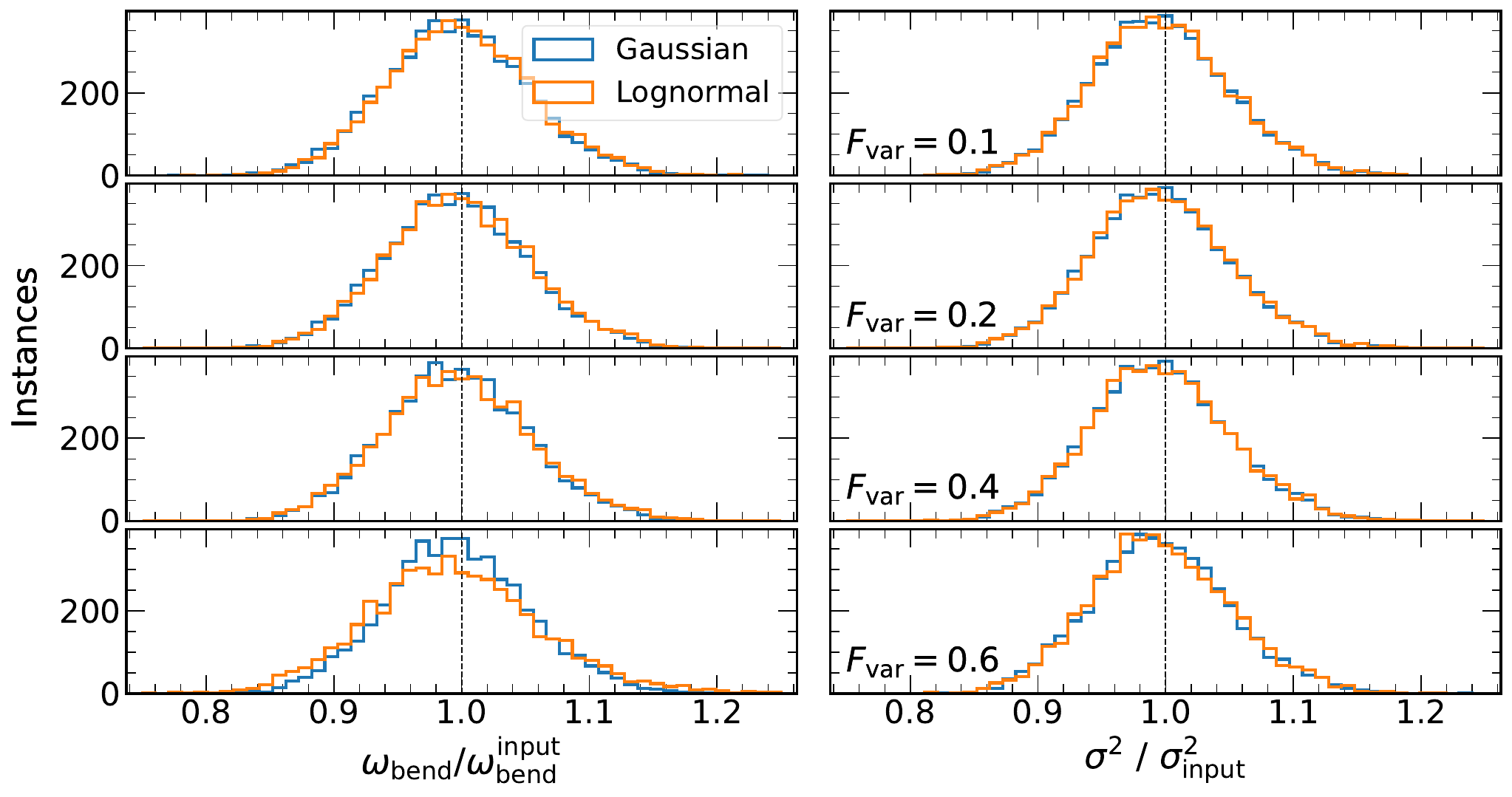}
    \caption{Best-fit \omegabend\ and variance $\sigma^2$ of an ensemble of 5,000 lightcurves generated with varying flux PDFs (as indicated in the legend) and \rms. As can be seen, the recovered parameters are in agreement with the input values, regardless of the PDF used to generated the lightcurves or the \rms.}
    \label{fig:gp_lognormal_fits}
\end{figure*}

Figure~\ref{fig:gp_lognormal_fits} shows histograms of the recovered \omegabend\ and $\sigma^2$ for an ensemble of 5,000 lightcurves.  As can be seen, we do not observe any deviation from the input parameters in the recovered parameters, regardless of \rms, despite the lightcurves following the universal linear rms-flux relationship. 

\begin{figure*}
    \centering
    \includegraphics[width=0.8\textwidth]{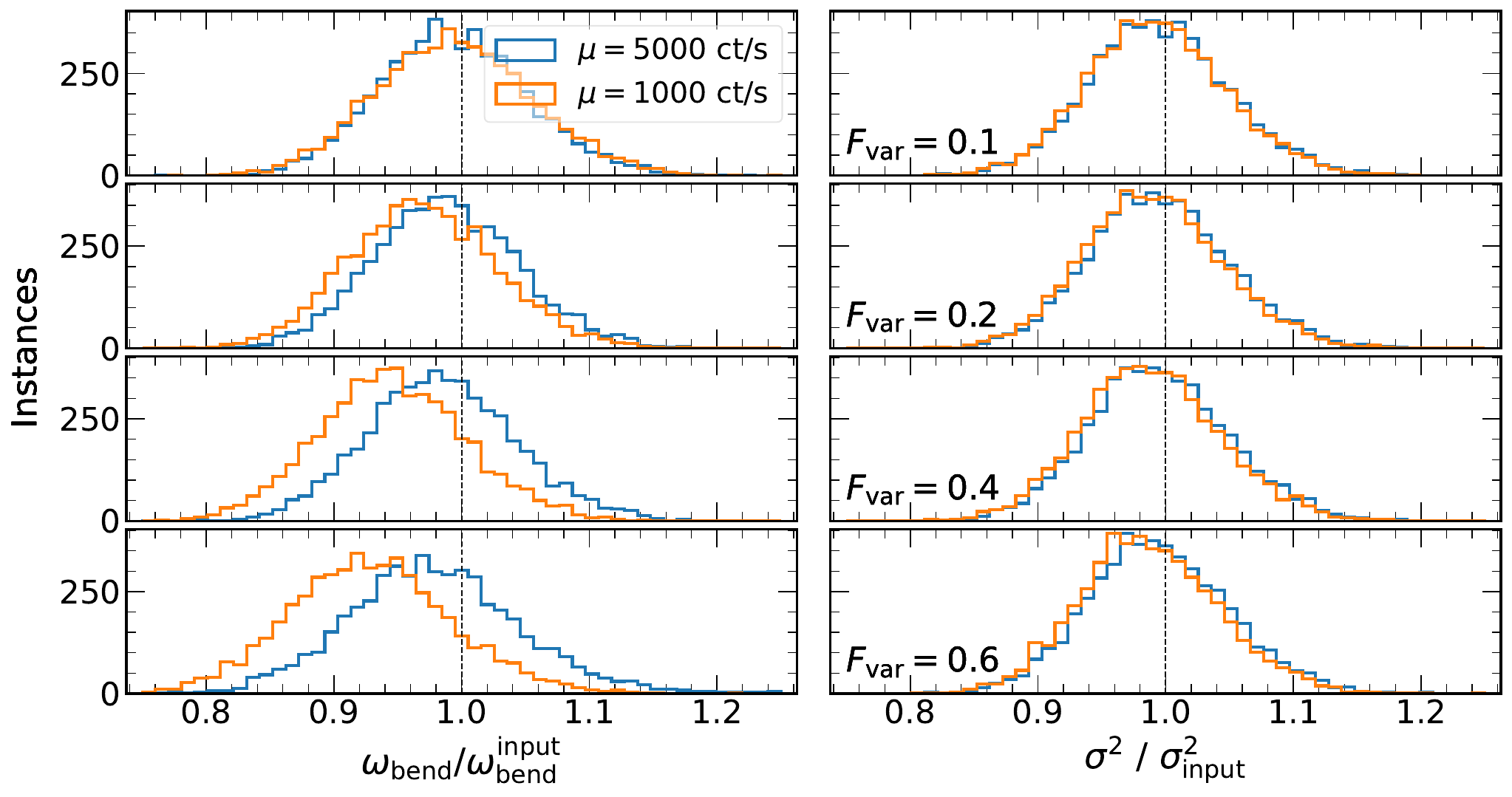}
    \caption{Best-fit \omegabend\ and variance $\sigma^2$ of an ensemble of 5,000 lightcurves generated from a lognormal PDF and varying \rms, but now including Poisson noise and uncertainties in the fitting. The blue color and orange color shows the results for varying mean ($\mu =$ 1,000 and 5,000 ct/s respectively). As can be seen, the bias in the recovered parameters increases with \rms, but even at the highest \rms\ the bias remains small. This suggest GPs have broad applicability.}
    \label{fig:gp_lognormal_fits_err}
\end{figure*}

As a further test, we now incorporate Poisson noise and take into account the (Poissonian) uncertainties in the fitting process. We run this test for the lognormal lightcurves only, as for high \rms\ ($\gtrsim$0.3) the gaussianly distributed lightcurves produce negative counts due to the distribution not being strictly positively defined. Figure~\ref{fig:gp_lognormal_fits_err} shows the histogram of the recovered parameters for an ensemble of 5,000 lightcurves with a lognormal distribution, varying the \rms\ and for two different mean values of $\mu = $1,000 and 5,000 ct/s respectively. As expected, deviations from the input parameters are stronger as \rms\ increases. For $\mu$= 1,000 ct/s and the largest \rms\ values, deviations are at most of the order of $\sim$7\%, affecting more strongly \omegabend. However, we can see that for the higher-mean count-rate case ($\mu = $5,000 ct/s), even at the highest \rms\ of 0.6, biases remain below the order of $\sim$2\%. This suggests that most of the biases we see for $\mu$= 1,000 ct/s are due to Poisson statistics, and that lognormality of the flux has little impact on the recovered parameters. Moreover, since \rms\ values higher than $\gtrsim$0.5 are rarely observed in AGN or X-ray binaries \citep[e.g.][]{breedt_twelve_2010}, this experiment suggests that there is broad applicability of GPs for the recovery of the variability processes in accreting sources.
 
\section{MCMC Sampling} \label{sub:mcmc_sampling}
Here we describe the process for the derivation of the best-fit parameters and their posteriors. These were found by first minimizing the negative log likelihood function using the L-BFGS algorithm. We then applied a small nudge to the best-fit parameters and used Markov-Chain Monte Carlo (MCMC) methods to sample the posterior running 32 independent chains (or walkers) using the \texttt{emceee} \texttt{python} library \citep{foreman-mackey_emcee_2013}. More specifically, after the fitting process the walkers were distributed around the best-fit parameters by drawing from a Gaussian with mean equal to the best-fit values and standard deviation equal to 10\% of their values. 

We adopted fairly uninformative (uniform) priors. Limits on the frequencies of the period and the aperiodic kernel timescales were set based on data constraints. The shortest timescale was set by a pseudo Nyquist frequency ($f_\mathrm{nyq} = $ 1/2$<\Delta t>$ where $<\Delta t>$ was the mean cadence of the lightcurve). The maximum allowed timescale was $T$ for the aperiodic kernels and $T$/2 for the periodic kernel, with $T$ being the lightcurve length. We further imposed $Q \gtrsim 3/2$;  (Figure~\ref{fig:celerite_models}) for the Lorentzian component to force this component to always represent a periodic signal and avoid degeneracy with the aperiodic kernels. The upper bound of $Q$ was effectively unconstrained to allow for cases where the amplitude of the oscillation is not seen to decay.

In order to ensure convergence, the MCMC sampler was run until a) the number of steps reached 100 times the integrated autocorrelation time ($\tau$), which was estimated on the fly every 800 samples, and b) $\tau$ changed less than 1\% compared to the previous estimate. We then discarded the first $30\times\tau$ number of samples (the burn in) and thinned the chains by $\tau$/2 to build the posterior probability density function. We additionally inspected the chains for stationarity and compared the variances within each chain to the variance between chains following \citet{vaughan_bayesian_2010} (and references therein).

%% file: mind_the_gaps.bbl
\begin{thebibliography}{}
\makeatletter
\relax
\def\mn@urlcharsother{\let\do\@makeother \do\$\do\&\do\#\do\^\do\_\do\%\do\~}
\def\mn@doi{\begingroup\mn@urlcharsother \@ifnextchar [ {\mn@doi@}
  {\mn@doi@[]}}
\def\mn@doi@[#1]#2{\def\@tempa{#1}\ifx\@tempa\@empty \href
  {http://dx.doi.org/#2} {doi:#2}\else \href {http://dx.doi.org/#2} {#1}\fi
  \endgroup}
\def\mn@eprint#1#2{\mn@eprint@#1:#2::\@nil}
\def\mn@eprint@arXiv#1{\href {http://arxiv.org/abs/#1} {{\tt arXiv:#1}}}
\def\mn@eprint@dblp#1{\href {http://dblp.uni-trier.de/rec/bibtex/#1.xml}
  {dblp:#1}}
\def\mn@eprint@#1:#2:#3:#4\@nil{\def\@tempa {#1}\def\@tempb {#2}\def\@tempc
  {#3}\ifx \@tempc \@empty \let \@tempc \@tempb \let \@tempb \@tempa \fi \ifx
  \@tempb \@empty \def\@tempb {arXiv}\fi \@ifundefined
  {mn@eprint@\@tempb}{\@tempb:\@tempc}{\expandafter \expandafter \csname
  mn@eprint@\@tempb\endcsname \expandafter{\@tempc}}}

\bibitem[\protect\citeauthoryear{Akaike}{Akaike}{1998}]{parzen_information_1998}
Akaike H.,  1998, in Parzen E.,  Tanabe K.,   Kitagawa G.,  eds, , Selected
  {Papers} of {Hirotugu} {Akaike}.
Springer New York, New York, NY, pp 199--213,
  \mn@doi{10.1007/978-1-4612-1694-0_15}, \url
  {http://link.springer.com/10.1007/978-1-4612-1694-0_15}

\bibitem[\protect\citeauthoryear{Alston, Markeviciute, Kara, Fabian  \&
  Middleton}{Alston et~al.}{2014}]{alston_detection_2014}
Alston W.~N.,  Markeviciute J.,  Kara E.,  Fabian A.~C.,   Middleton M.,  2014,
  \mn@doi [Monthly Notices of the Royal Astronomical Society]
  {10.1093/mnrasl/slu127}, 445, L16

\bibitem[\protect\citeauthoryear{Ashton \& Middleton}{Ashton \&
  Middleton}{2021}]{ashton_searching_2021}
Ashton D.~I.,  Middleton M.~J.,  2021, \mn@doi [Monthly Notices of the Royal
  Astronomical Society] {10.1093/mnras/staa4024}, 501, 5478

\bibitem[\protect\citeauthoryear{Barnard, Trudolyubov, Haswell, Kolb, Osborne
  \& Priedhorsky}{Barnard et~al.}{2007}]{barnard_artificial_2007}
Barnard R.,  Trudolyubov S.,  Haswell C.~A.,  Kolb U.~C.,  Osborne J.~P.,
  Priedhorsky W.~H.,  2007, AIP Conference Proceedings, 924, 691

\bibitem[\protect\citeauthoryear{Belloni, Psaltis  \& Klis}{Belloni
  et~al.}{2002}]{belloni_unified_2002}
Belloni T.,  Psaltis D.,   Klis M. v.~d.,  2002, \mn@doi [ApJ]
  {10.1086/340290}, 572, 392

\bibitem[\protect\citeauthoryear{Bowman \& Dorn-Wallenstein}{Bowman \&
  Dorn-Wallenstein}{2022}]{bowman_photometric_2022}
Bowman D.~M.,  Dorn-Wallenstein T.~Z.,  2022, \mn@doi [A\&A]
  {10.1051/0004-6361/202243545}, 668, A134

\bibitem[\protect\citeauthoryear{Breedt et~al.,}{Breedt
  et~al.}{2010}]{breedt_twelve_2010}
Breedt E.,  et~al., 2010, \mn@doi [Monthly Notices of the Royal Astronomical
  Society] {10.1111/j.1365-2966.2009.16146.x}, 403, 605

\bibitem[\protect\citeauthoryear{Done, Madejski, Mushotzky, Turner, Koyama  \&
  Kunieda}{Done et~al.}{1992}]{done_x-ray_1992}
Done C.,  Madejski G.~M.,  Mushotzky R.~F.,  Turner T.~J.,  Koyama K.,
  Kunieda H.,  1992, \mn@doi [The Astrophysical Journal] {10.1086/171979}, 400,
  138

\bibitem[\protect\citeauthoryear{Done, Madejski, Życki  \& Greenhill}{Done
  et~al.}{2003}]{done_simultaneous_2003}
Done C.,  Madejski G.~M.,  Życki P.~T.,   Greenhill L.~J.,  2003, \mn@doi
  [ApJ] {10.1086/374332}, 588, 763

\bibitem[\protect\citeauthoryear{Emmanoulopoulos, McHardy  \&
  Papadakis}{Emmanoulopoulos et~al.}{2013}]{emmanoulopoulos_generating_2013}
Emmanoulopoulos D.,  McHardy I.~M.,   Papadakis I.~E.,  2013, \mn@doi [Monthly
  Notices of the Royal Astronomical Society] {10.1093/mnras/stt764}, 433, 907

\bibitem[\protect\citeauthoryear{Evans et~al.,}{Evans
  et~al.}{2007}]{evans_online_2007}
Evans P.~A.,  et~al., 2007, \mn@doi [A\&A] {10.1051/0004-6361:20077530}, 469,
  379

\bibitem[\protect\citeauthoryear{Evans et~al.,}{Evans
  et~al.}{2009}]{evans_methods_2009}
Evans P.~A.,  et~al., 2009, \mn@doi [Mon Not R Astron Soc]
  {10.1111/j.1365-2966.2009.14913.x}, 397, 1177

\bibitem[\protect\citeauthoryear{Foreman-Mackey}{Foreman-Mackey}{2016}]{foreman-mackey_cornerpy_2016}
Foreman-Mackey D.,  2016, \mn@doi [Journal of Open Source Software]
  {10.21105/joss.00024}, 1, 24

\bibitem[\protect\citeauthoryear{Foreman-Mackey, Hogg, Lang  \&
  Goodman}{Foreman-Mackey et~al.}{2013}]{foreman-mackey_emcee_2013}
Foreman-Mackey D.,  Hogg D.~W.,  Lang D.,   Goodman J.,  2013, \mn@doi
  [Publications of the Astronomical Society of the Pacific] {10.1086/670067},
  125, 306

\bibitem[\protect\citeauthoryear{Foreman-Mackey, Agol, Ambikasaran  \&
  Angus}{Foreman-Mackey et~al.}{2017}]{foreman-mackey_fast_2017}
Foreman-Mackey D.,  Agol E.,  Ambikasaran S.,   Angus R.,  2017, \mn@doi [AJ]
  {10.3847/1538-3881/aa9332}, 154, 220

\bibitem[\protect\citeauthoryear{Frescura, Engelbrecht  \& Frank}{Frescura
  et~al.}{2008}]{frescura_significance_2008}
Frescura F. A.~M.,  Engelbrecht C.~A.,   Frank B.~S.,  2008, \mn@doi [Monthly
  Notices of the Royal Astronomical Society]
  {10.1111/j.1365-2966.2008.13499.x}, 388, 1693

\bibitem[\protect\citeauthoryear{Fürst et~al.,}{Fürst
  et~al.}{2016}]{furst_discovery_2016}
Fürst F.,  et~al., 2016, \mn@doi [The Astrophysical Journal]
  {10.3847/2041-8205/831/2/L14}, 831

\bibitem[\protect\citeauthoryear{Fürst et~al.,}{Fürst
  et~al.}{2018}]{furst_tale_2018}
Fürst F.,  et~al., 2018, \mn@doi [Astronomy \& Astrophysics]
  {10.1051/0004-6361/201833292}, 616, A186

\bibitem[\protect\citeauthoryear{Fürst et~al.,}{Fürst
  et~al.}{2021}]{furst_long-term_2021}
Fürst F.,  et~al., 2021, \mn@doi [A\&A] {10.1051/0004-6361/202140625}, 651,
  A75

\bibitem[\protect\citeauthoryear{Garrison, Foreman-Mackey, Shih  \&
  Barnett}{Garrison et~al.}{2024}]{garrison_nifty-ls_2024}
Garrison L.~H.,  Foreman-Mackey D.,  Shih Y.-h.,   Barnett A.,  2024, nifty-ls:
  {Fast} and {Accurate} {Lomb}-{Scargle} {Periodograms} {Using} a
  {Non}-{Uniform} {FFT}, \url {http://arxiv.org/abs/2409.08090}

\bibitem[\protect\citeauthoryear{Gehrels et~al.,}{Gehrels
  et~al.}{2004}]{gehrels_swift_2004}
Gehrels N.,  et~al., 2004, \mn@doi [The Astrophysical Journal]
  {10.1086/422091}, 611, 1005

\bibitem[\protect\citeauthoryear{Gierliński, Middleton, Ward  \&
  Done}{Gierliński et~al.}{2008}]{gierlinski_periodicity_2008}
Gierliński M.,  Middleton M.,  Ward M.,   Done C.,  2008, \mn@doi [Nature]
  {10.1038/nature07277}, 455, 369

\bibitem[\protect\citeauthoryear{González-Martín \&
  Vaughan}{González-Martín \& Vaughan}{2012}]{gonzalez-martin_x-ray_2012}
González-Martín O.,  Vaughan S.,  2012, \mn@doi [Astronomy and Astrophysics]
  {10.1051/0004-6361/201219008}, 544, A80

\bibitem[\protect\citeauthoryear{Graham et~al.,}{Graham
  et~al.}{2015}]{graham_possible_2015}
Graham M.~J.,  et~al., 2015, \mn@doi [Nature] {10.1038/nature14143}, 518, 74

\bibitem[\protect\citeauthoryear{Gúrpide \& Mangham}{Gúrpide \&
  Mangham}{2025}]{andres_gurpide_lasheras_2024_14253754}
Gúrpide A.,  Mangham S.,  2025, {andresgur/mind\_the\_gaps: Release for
  paper}, \mn@doi{10.5281/zenodo.14600069}, \url
  {https://doi.org/10.5281/zenodo.14600069}

\bibitem[\protect\citeauthoryear{Horne \& Baliunas}{Horne \&
  Baliunas}{1986}]{horne_prescription_1986}
Horne J.~H.,  Baliunas S.~L.,  1986, \mn@doi [The Astrophysical Journal]
  {10.1086/164037}, 302, 757

\bibitem[\protect\citeauthoryear{Hu, Li, Kong, Ng  \& Lin}{Hu
  et~al.}{2017}]{hu_swift_2017}
Hu C.-P.,  Li K.~L.,  Kong A. K.~H.,  Ng C.-Y.,   Lin L. C.-C.,  2017, \mn@doi
  [ApJ] {10.3847/2041-8213/835/1/L9}, 835, L9

\bibitem[\protect\citeauthoryear{Hurvich \& Tsai}{Hurvich \&
  Tsai}{1989}]{hurvich_regression_1989}
Hurvich C.~M.,  Tsai C.-L.,  1989, \mn@doi [Biometrika]
  {10.1093/biomet/76.2.297}, pp 297--307

\bibitem[\protect\citeauthoryear{Hübner, Huppenkothen, Lasky, Inglis, Ick  \&
  Hogg}{Hübner et~al.}{2022}]{hubner_searching_2022}
Hübner M.,  Huppenkothen D.,  Lasky P.~D.,  Inglis A.~R.,  Ick C.,   Hogg
  D.~W.,  2022, \mn@doi [The Astrophysical Journal] {10.3847/1538-4357/ac7959},
  936, 17

\bibitem[\protect\citeauthoryear{Ingram \& Done}{Ingram \&
  Done}{2011}]{ingram_physical_2011}
Ingram A.,  Done C.,  2011, \mn@doi [Monthly Notices of the Royal Astronomical
  Society] {10.1111/j.1365-2966.2011.18860.x}, 415, 2323

\bibitem[\protect\citeauthoryear{Israel \& Stella}{Israel \&
  Stella}{1996}]{israel_new_1996}
Israel G.~L.,  Stella L.,  1996, \mn@doi [The Astrophysical Journal]
  {10.1086/177697}, 468, 369

\bibitem[\protect\citeauthoryear{Israel et~al.,}{Israel
  et~al.}{2017}]{israel_discovery_2017}
Israel G.~L.,  et~al., 2017, \mn@doi [Mon Not R Astron Soc Lett]
  {10.1093/mnrasl/slw218}, 466, L48

\bibitem[\protect\citeauthoryear{Jenkins et~al.,}{Jenkins
  et~al.}{2016}]{jenkins_tess_2016}
Jenkins J.~M.,  et~al., 2016, in Software and {Cyberinfrastructure} for
  {Astronomy} {IV}. SPIE, pp 1232--1251, \mn@doi{10.1117/12.2233418}

\bibitem[\protect\citeauthoryear{Jiang et~al.,}{Jiang
  et~al.}{2022}]{jiang_tick-tock_2022}
Jiang N.,  et~al., 2022, Tick-{Tock}: {The} {Imminent} {Merger} of a
  {Supermassive} {Black} {Hole} {Binary}, \mn@doi{10.48550/arXiv.2201.11633},
  \url {http://arxiv.org/abs/2201.11633}

\bibitem[\protect\citeauthoryear{Jordán, Eyheramendy  \& Buchner}{Jordán
  et~al.}{2021}]{jordan_state-space_2021}
Jordán A.,  Eyheramendy S.,   Buchner J.,  2021, \mn@doi [Res. Notes AAS]
  {10.3847/2515-5172/abfe68}, 5, 107

\bibitem[\protect\citeauthoryear{Kaastra}{Kaastra}{2017}]{kaastra_use_2017}
Kaastra J.~S.,  2017, \mn@doi [A\&A] {10.1051/0004-6361/201629319}, 605, A51

\bibitem[\protect\citeauthoryear{Kelly, Sobolewska  \& Siemiginowska}{Kelly
  et~al.}{2011}]{kelly_stochastic_2011}
Kelly B.~C.,  Sobolewska M.,   Siemiginowska A.,  2011, \mn@doi [The
  Astrophysical Journal] {10.1088/0004-637X/730/1/52}, 730, 52

\bibitem[\protect\citeauthoryear{Kelly, Becker, Sobolewska, Siemiginowska  \&
  Uttley}{Kelly et~al.}{2014}]{kelly_flexible_2014}
Kelly B.~C.,  Becker A.~C.,  Sobolewska M.,  Siemiginowska A.,   Uttley P.,
  2014, \mn@doi [ApJ] {10.1088/0004-637X/788/1/33}, 788, 33

\bibitem[\protect\citeauthoryear{Khan \& Middleton}{Khan \&
  Middleton}{2023}]{khan_long-term_2023}
Khan N.,  Middleton M.~J.,  2023, \mn@doi [Monthly Notices of the Royal
  Astronomical Society] {10.1093/mnras/stad2071}, 524, 4302

\bibitem[\protect\citeauthoryear{Kotze \& Charles}{Kotze \&
  Charles}{2012}]{kotze_characterizing_2012}
Kotze M.~M.,  Charles P.~A.,  2012, \mn@doi [Monthly Notices of the Royal
  Astronomical Society] {10.1111/j.1365-2966.2011.20146.x}, 420, 1575

\bibitem[\protect\citeauthoryear{Kraft, Burrows  \& Nousek}{Kraft
  et~al.}{1991}]{kraft_determination_1991}
Kraft R.~P.,  Burrows D.~N.,   Nousek J.~A.,  1991, \mn@doi [The Astrophysical
  Journal] {10.1086/170124}, 374, 344

\bibitem[\protect\citeauthoryear{Lomb}{Lomb}{1976}]{lomb_least-squares_1976}
Lomb N.~R.,  1976, \mn@doi [Astrophysics and Space Science, Volume 39, Issue 2,
  pp.447-462] {10.1007/BF00648343}, 39, 447

\bibitem[\protect\citeauthoryear{Markowitz}{Markowitz}{2010}]{markowitz_x-ray_2010}
Markowitz A.,  2010, \mn@doi [The Astrophysical Journal]
  {10.1088/0004-637X/724/1/26}, 724, 26

\bibitem[\protect\citeauthoryear{Motch, Pakull, Soria, Grisé  \&
  Pietrzyński}{Motch et~al.}{2014}]{motch_mass_2014}
Motch C.,  Pakull M.~W.,  Soria R.,  Grisé F.,   Pietrzyński G.,  2014,
  \mn@doi [Nature] {10.1038/nature13730}, 514, 198

\bibitem[\protect\citeauthoryear{Mueller \& Madejski}{Mueller \&
  Madejski}{2009}]{mueller_parameter_2009}
Mueller M.,  Madejski G.,  2009, \mn@doi [The Astrophysical Journal]
  {10.1088/0004-637X/700/1/243}, 700, 243

\bibitem[\protect\citeauthoryear{Pasham et~al.,}{Pasham
  et~al.}{2019}]{pasham_loud_2019}
Pasham D.~R.,  et~al., 2019, \mn@doi [Science] {10.1126/science.aar7480}, 363,
  531

\bibitem[\protect\citeauthoryear{Pasham et~al.,}{Pasham
  et~al.}{2024}]{pasham_lense-thirring_2024}
Pasham D.~R.,  et~al., 2024, Lense-{Thirring} {Precession} after a
  {Supermassive} {Black} {Hole} {Disrupts} a {Star},
  \mn@doi{10.48550/arXiv.2402.09689}, \url {http://arxiv.org/abs/2402.09689}

\bibitem[\protect\citeauthoryear{Protassov, van Dyk, Connors, Kashyap  \&
  Siemiginowska}{Protassov et~al.}{2002}]{protassov_statistics_2002}
Protassov R.,  van Dyk D.~A.,  Connors A.,  Kashyap V.~L.,   Siemiginowska A.,
  2002, \mn@doi [The Astrophysical Journal] {10.1086/339856}, 571, 545

\bibitem[\protect\citeauthoryear{Rasmussen \& Williams}{Rasmussen \&
  Williams}{2006}]{rasmussen_gaussian_2006}
Rasmussen C.~E.,  Williams C. K.~I.,  2006, Gaussian processes for machine
  learning.
Adaptive computation and machine learning, MIT Press, Cambridge, Mass

\bibitem[\protect\citeauthoryear{Ricker et~al.,}{Ricker
  et~al.}{2014}]{ricker_transiting_2014}
Ricker G.~R.,  et~al., 2014, in Space {Telescopes} and {Instrumentation} 2014:
  {Optical}, {Infrared}, and {Millimeter} {Wave}. SPIE, pp 556--570,
  \mn@doi{10.1117/12.2063489}

\bibitem[\protect\citeauthoryear{Scargle}{Scargle}{1982}]{scargle_studies_1982}
Scargle J.~D.,  1982, \mn@doi [The Astrophysical Journal] {10.1086/160554},
  263, 835

\bibitem[\protect\citeauthoryear{Smith, Robles  \& Perlman}{Smith
  et~al.}{2020}]{smith_qpo_2020}
Smith E.,  Robles R.,   Perlman E.,  2020, \mn@doi [ApJ]
  {10.3847/1538-4357/abb593}, 902, 65

\bibitem[\protect\citeauthoryear{Stella, Arlandi, Tagliaferri  \&
  Israel}{Stella et~al.}{1994}]{stella_continuum_1994}
Stella L.,  Arlandi E.,  Tagliaferri G.,   Israel G.~L.,  1994, Continuum
  {Power} {Spectrum} {Components} in {X}-{Ray} {Sources}: {Detailed}
  {Modelling} and {Search} for {Coherent} {Periodicities},
  \mn@doi{10.48550/arXiv.astro-ph/9411050}, \url
  {http://arxiv.org/abs/astro-ph/9411050}

\bibitem[\protect\citeauthoryear{Timmer \& Koenig}{Timmer \&
  Koenig}{1995}]{timmer_generating_1995}
Timmer J.,  Koenig M.,  1995, Astronomy and Astrophysics, 300, 707

\bibitem[\protect\citeauthoryear{Tripathi, Smith, Wiita  \& Wagoner}{Tripathi
  et~al.}{2024}]{tripathi_search_2024}
Tripathi A.,  Smith K.~L.,  Wiita P.~J.,   Wagoner R.~V.,  2024, Search for
  {Quasi}-{Periodic} {Oscillations} in {TESS} light curves of bright {Fermi}
  {Blazars}, \url {http://arxiv.org/abs/2402.04352}

\bibitem[\protect\citeauthoryear{Uttley, McHardy  \& Papadakis}{Uttley
  et~al.}{2002}]{uttley_measuring_2002}
Uttley P.,  McHardy I.~M.,   Papadakis I.~E.,  2002, \mn@doi [Monthly Notices
  of the Royal Astronomical Society] {10.1046/j.1365-8711.2002.05298.x}, 332,
  231

\bibitem[\protect\citeauthoryear{Uttley, McHardy  \& Vaughan}{Uttley
  et~al.}{2005}]{uttley_non-linear_2005}
Uttley P.,  McHardy I.~M.,   Vaughan S.,  2005, \mn@doi [Monthly Notices of the
  Royal Astronomical Society] {10.1111/j.1365-2966.2005.08886.x}, 359, 345

\bibitem[\protect\citeauthoryear{VanderPlas}{VanderPlas}{2018}]{vanderplas_understanding_2018}
VanderPlas J.~T.,  2018, \mn@doi [ApJS] {10.3847/1538-4365/aab766}, 236, 16

\bibitem[\protect\citeauthoryear{Vasilopoulos, Lander, Koliopanos  \&
  Bailyn}{Vasilopoulos et~al.}{2020}]{vasilopoulos_m51_2020}
Vasilopoulos G.,  Lander S.~K.,  Koliopanos F.,   Bailyn C.~D.,  2020, \mn@doi
  [Monthly Notices of the Royal Astronomical Society] {10.1093/mnras/stz3298},
  491, 4949

\bibitem[\protect\citeauthoryear{Vaughan}{Vaughan}{2005}]{vaughan_simple_2005}
Vaughan S.,  2005, \mn@doi [A\&A] {10.1051/0004-6361:20041453}, 431, 391

\bibitem[\protect\citeauthoryear{Vaughan}{Vaughan}{2010}]{vaughan_bayesian_2010}
Vaughan S.,  2010, \mn@doi [Monthly Notices of the Royal Astronomical Society]
  {10.1111/j.1365-2966.2009.15868.x}, 402, 307

\bibitem[\protect\citeauthoryear{Vaughan \& Uttley}{Vaughan \&
  Uttley}{2005}]{vaughan_where_2005}
Vaughan S.,  Uttley P.,  2005, \mn@doi [Monthly Notices of the Royal
  Astronomical Society] {10.1111/j.1365-2966.2005.09296.x}, 362, 235

\bibitem[\protect\citeauthoryear{Vaughan, Edelson, Warwick  \& Uttley}{Vaughan
  et~al.}{2003}]{vaughan_characterizing_2003}
Vaughan S.,  Edelson R.,  Warwick R.~S.,   Uttley P.,  2003, \mn@doi [Monthly
  Notices of the Royal Astronomical Society]
  {10.1046/j.1365-2966.2003.07042.x}, 345, 1271

\bibitem[\protect\citeauthoryear{Vaughan, Uttley, Markowitz, Huppenkothen,
  Middleton, Alston, Scargle  \& Farr}{Vaughan
  et~al.}{2016}]{vaughan_false_2016}
Vaughan S.,  Uttley P.,  Markowitz A.~G.,  Huppenkothen D.,  Middleton M.~J.,
  Alston W.~N.,  Scargle J.~D.,   Farr W.~M.,  2016, \mn@doi [Mon. Not. R.
  Astron. Soc.] {10.1093/mnras/stw1412}, 461, 3145

\bibitem[\protect\citeauthoryear{Waller, Smith, Childs  \& Real}{Waller
  et~al.}{2003}]{waller_monte_2003}
Waller L.~A.,  Smith D.,  Childs J.~E.,   Real L.~A.,  2003, \mn@doi
  [Ecological Modelling] {10.1016/S0304-3800(03)00011-5}, 164, 49

\bibitem[\protect\citeauthoryear{Yan, Zhang, Liu, Chang, Liu, Yan  \& Zeng}{Yan
  et~al.}{2024}]{yan_x-ray_2024}
Yan Y.,  Zhang P.,  Liu Q.,  Chang Z.,  Liu G.,  Yan J.,   Zeng X.,  2024, An
  {X}-{Ray} {High}-{Frequency} {QPO} in {NGC} 1365, \url
  {http://arxiv.org/abs/2405.16187}

\bibitem[\protect\citeauthoryear{Zhang, Yang  \& Dai}{Zhang
  et~al.}{2023}]{zhang_search_2023}
Zhang H.,  Yang S.,   Dai B.,  2023, \mn@doi [ApJ] {10.3847/1538-4357/acbe37},
  946, 52

\bibitem[\protect\citeauthoryear{van~der Klis}{van~der
  Klis}{1988}]{van_der_klis_fourier_1988}
van~der Klis M.,  1988, NATO Advanced Study Institutes Series. Series C,
  Mathematical and Physical Sciences

\makeatother
\end{thebibliography}
